\documentclass[letterpaper, 10 pt, journal]{ieeeconf}

\usepackage{cite}
\usepackage{tikz}
\usepackage[latin1]{inputenc}
\usepackage{graphicx,subfig,float,dblfloatfix}
\usepackage{amssymb,amsmath,dsfont}
\usepackage{url}
\usepackage{tcolorbox,lipsum,tikz,pstricks}
\usepackage{multicol}
\usepackage{blindtext}
\usepackage{pst-node,graphicx}
\usepackage{enumerate}
\SpecialCoor
\usepackage{soul}

\usepackage{wrapfig,color}
\usepackage{pgfplots}
\usepackage{tikz}
\usetikzlibrary{external}
\usepackage{algorithm}
\usepackage[noend]{algpseudocode}

\tcbuselibrary{breakable}

\makeatletter
\def\BState{\State\hskip-\ALG@thistlm}
\makeatother
\usepackage{color, colortbl}
\usepackage{hyperref} 

\newcommand{\real}{\mathbb{R}}

\newcommand{\integernonnegative}{\mathbb{Z}_{\ge 0}}

\newcommand{\bigCI}{\mathrel{\text{\scalebox{1.07}{$\perp\mkern-10mu\perp$}}}}

\newcommand{\R}{\mathbb{R}} 
\newcommand{\N}{\mathbb{N}}  

\newcommand{\G}{\mathcal{G}} 
\newcommand{\V}{\mathcal{V}} 
\newcommand{\E}{\mathcal{E}} 
\newcommand{\Sc}{\mathcal{S}} 

\usepackage{xcolor}

\definecolor{airforcecarmine}{rgb}{0.36, 0.54, 0.66}
\definecolor{carmine}{rgb}{0.59, 0.0, 0.09}
\newcommand{\X}{\mathcal{X}} 

\newcommand{\e}{\textup{e}} 

\newcommand{\1}{\mathds{1}} 

\newcommand{\diag}{\operatorname{diag}} 

\newcommand{\argmin}[1]{\underset{#1}{\mathrm{argmin\,}}}

\newtheorem{definition}{Definition}
\newtheorem{example}{Example}

\newcommand{\Mc}{\mathcal{M}}

\newcommand{\Ab}{\mathbf{A}}
\newcommand{\Bb}{\mathbf{B}}

\newcommand{\Sb}{\mathbf{S}}
\newcommand{\Lb}{\mathbf{L}}
\newcommand{\Mb}{\mathbf{M}}
\newcommand{\Db}{\mathbf{D}}
\newcommand{\Wb}{\mathbf{W}}
\newcommand{\Vb}{\mathbf{V}}
\newcommand{\Ub}{\mathbf{U}}

\newcommand{\Adj}{\mathbf{Adj}}

\newcommand{\xb}{\mathbf{x}}
\newcommand{\yb}{\mathbf{y}}

\newcommand{\ub}{\mathbf{u}}
\newcommand{\zb}{\mathbf{z}}
\newcommand{\wb}{\mathbf{w}}

\newcommand{\chiara}[1]{\textcolor{cyan}{\@Fabrizio: #1}}

\usepackage[normalem]{ulem}

\newcommand{\cl}[1]{{{#1}}}
\newcommand{\killmath}[1]{\ifmmode\text{\sout{\ensuremath{\textcolor{blue}{#1}}}}\else\sout{#1}\fi}

\newcommand{\ant}[1]{{#1}}

\tcbset{breakable}

\usepackage{tikz}

\definecolor{LightGray}{gray}{0.96} 
\definecolor{DarkGray}{gray}{0.8}

\title{Learning hidden influences in large-scale dynamical social networks\\\medskip \large A data-driven sparsity-based approach}

\author{
Chiara Ravazzi\thanks{C.\ Ravazzi and F.\ Dabbene are with the National Research Council of Italy, CNR-IEIIT, c/o Politecnico di Torino, Corso Duca degli Abruzzi 14, 10129, Turin, Italy.}, \IEEEmembership{Member, IEEE},  Fabrizio Dabbene, \IEEEmembership{Senior Member, IEEE},  \\Constantino Lagoa, \IEEEmembership{Member, IEEE}\thanks{C. Lagoa is with
         The Pennsylvania State University, University Park, PA 16802.},
  Anton V. Proskurnikov, \IEEEmembership{Senior Member, IEEE}\thanks{A.V.\ Proskurnikov is with Politecnico di Torino, Corso Duca degli Abruzzi 14, 10129, Turin, Italy, and also with the Institute for Problems in Mechanical Engineering of the Russian Academy of Sciences, St. Petersburg, Bolshoy pr. V.O. 61, Russia.}}
\usepackage{siunitx}
\usepackage{xcolor}
\usepackage{booktabs,colortbl, array}
\usepackage{pgfplotstable}
\pgfplotsset{compat=1.8}

\definecolor{dgr}{RGB}{0,0,100} 

\definecolor{rulecolor}{RGB}{0,71,171}
\definecolor{tableheadcolor}{gray}{0.92}
\newcommand{\topline}{ %
        \arrayrulecolor{rulecolor}\specialrule{0.1em}{\abovetopsep}{0pt}%
        \arrayrulecolor{tableheadcolor}\specialrule{\belowrulesep}{0pt}{0pt}%
        \arrayrulecolor{rulecolor}}
\newcommand{\midtopline}{ %
        \arrayrulecolor{tableheadcolor}\specialrule{\aboverulesep}{0pt}{0pt}%
        \arrayrulecolor{rulecolor}\specialrule{\lightrulewidth}{0pt}{0pt}%
        \arrayrulecolor{white}\specialrule{\belowrulesep}{0pt}{0pt}%
        \arrayrulecolor{rulecolor}}
\newcommand{\bottomline}{ %
        \arrayrulecolor{white}\specialrule{\aboverulesep}{0pt}{0pt}%
        \arrayrulecolor{rulecolor} %
        \specialrule{\heavyrulewidth}{0pt}{\belowbottomsep}}%

\pgfplotstableset{normal/.style ={%
        header=true,
        string type,
        font=\addfontfeature{Numbers={Monospaced}}\small,
        column type=l,
        every odd row/.style={
            before row=
        },
        every head row/.style={
            before row={\topline\rowcolor{tableheadcolor}},
            after row={\midtopline}
        },
        every last row/.style={
            after row=\bottomline
        },
        col sep=&,
        row sep=\\
    }
}

\newcommand{\boxref}[1]{{\bf \textcolor{carmine!50}{``#1''}}}
\newcommand{\boxrefb}[1]{{\bf \textcolor{airforcecarmine!100}{``#1''}}}

\begin{document}
\onecolumn
\maketitle
\tableofcontents

\clearpage
\begin{multicols}{2}

\section{Towards a unified system theory of opinion formation and social influence}

Processes of information diffusion over social networks, e.g. opinions {spread} and beliefs {formation}, are attracting substantial interest of various disciplines, from behavioral sciences to mathematics and engineering.
Since opinions and behaviors of each individual are influenced by interactions with others, understanding
the structure of interpersonal influences is a key ingredient to predict, analyze and, possibly, control information and decisions~\cite{Tang:2009:SIA:1557019.1557108}.

With rapid increase of many social media platforms that provide instant messaging, blogging and other social networking services -- see the box \boxrefb{Online Social Networks} (OSN) -- people can nowadays easily share news, opinions and preferences. Information can reach a broad audience much faster than before, and opinion mining and sentiment analysis is now becoming a key challenge in the modern society \cite{Metaxas472}. The first anecdotal evidence of this fact is probably the use that the Obama campaign made of social networks during the US 2008 American Presidential Elections~\cite{Lai2008}. More recently, several news outlets, according to the study in \cite{Allcott2017}, stated that Facebook users played a major role in feeding into fake news that might have influenced the outcome of the US 2016 American Presidential Election. This can be explained by the phenomena of homophily and biased assimilation~\cite{mcpherson2001birds,DandekarPNAS2013,Singla:2008:YCS:1367497.1367586} in social networks, that {corresponds to} the tendency of people to follow the behaviors of their friends and {to} establish friendship with like-minded individuals.

The inference of social ties from empirical data becomes of central interest in political organizations and business firms due to its potential impact in decision making and action planning. According to the report published by McKinsey \& Company~\cite{MK10}, {\em ``Marketing-induced consumer-to-consumer word of mouth generates more than twice the sales of paid advertising''}.
The influence analysis {is becoming} a key input into sophisticated recommendation engines that identify potential customers, exploiting similarities amongst several users to predict preferences. {The same} report~\cite{MK10} {estimates} that $35\%$ of revenue of Amazon company and $75\%$ of what users watch on Netflix came from product recommendations. The study of structures in networks (such as e.g. community detection and computing the node's centralities) has been the main concern of \textit{social network analysis} (SNA)~\cite{wasserman_faust_1994}, embraced now by the multidisciplinary field of Network Science~\cite{BarabasiLinked,Newman:2003,EasleyKleinberg}. {On a parallel line of research}, many works {have been} published in physical, mathematical and engineering literature that focus on dynamical models of opinion diffusion (see~\cite{Castellano:2009,Friedkin:2015,ProTempo:2017-1,ProTempo:2018,DongZhanKou:18-survey,ABID2018,Mastroeni:2019} and references therein).

There are numerous gaps between SNA and opinion dynamics modeling, and the relations between structures of social influence and information spread mechanisms are far from being well studied. This paper takes a step towards filling {these gaps}, and in the direction of deriving a unified theory, describing the intricate relations between structural and dynamical properties of social systems. In this new area, the methods of systems and control should play a key role.

Our aim, as explained in the \boxref{Summary}, is to provide a general overview of the main concepts, algorithmic tools, results, and open problems in the systematic study of learning interpersonal influence in networked systems.


\begin{tcolorbox}[title= \sf \textbf{\color{black}Summary}, colframe=carmine!10,
colback=carmine!10,
coltitle=black,
]\sf \small
Interpersonal influence estimation from empirical data is a central challenge in the study of social structures and dynamics. {\em Opinion dynamics} theory is a young interdisciplinary science that studies opinion formation in social networks and has a huge potential in applications, such as marketing, advertisement and recommendations.
\medskip

 The term {\em social influence} refers to the behavioral change of individuals due to the interactions with others in a social system, e.g. organization, community, or society in general.
 \medskip

 The advent of {the} Internet has made a huge volume of data easily available that can be used to measure social influence over large populations. Here, we aim at qualitatively and quantitatively infer social influence from data using a {\em systems and control viewpoint}. First, we introduce some definitions and models of opinions dynamics and review some structural constraints of online social networks, based on the notion of sparsity. Then, we review the main approaches
to infer the network's structure from a set of observed data. Finally, we present some algorithms that exploit the introduced models and structural constraints, focusing on the sample complexity and computational requirements.
\end{tcolorbox}

As summarized in the box \boxref{Opinion dynamics in a nutshell}, we can group the main research lines in {this field into} three broad categories: modeling, analysis, and control. \emph{Modeling} aims at finding a coherent mathematical description of the social interactions. To build a mathematical model, one has to define (a) the interaction {protocol}, e.g. the times of interactions, that can be discrete or continuous, the contact modes (deterministic or random), and the frequency of interactions among social network members; (b) the {dynamical} mechanism of {social interactions, or ties}, which can be described by linear or nonlinear functions~\cite{Abelson:1967,HunterDanesCohenBook_1984}. In the simplest situation, each social tie is described by a single scalar, treated as the ``influence weight'' one individual assigns to the other~\cite{Friedkin:2015}.
\emph{Analysis} of social networks is usually focused on {the} study of the qualitative and quantitative properties of the opinion dynamics: asymptotic convergence or oscillations, eventual consensus or disagreement etc.
It is also important to extract low-dimensional features of the network, e.g. to identify communities or the most influential leaders.
A long-standing goal in the study of social networks is to \emph{control}
\ant{the final distribution of opinions}~\cite{hegselmann2015optimal,Masuda_2015,ZhaoLiuWang:2016,Grabisch:2018}.

\end{multicols}
\begin{tcolorbox}[title= \sf \textbf{Online Social Networks},colframe=airforcecarmine!20,colback=airforcecarmine!20,coltitle=black,]\small\sf
\begin{multicols}{2}
\begin{minipage}{1\columnwidth}
The words \textbf{Online Social Network} (OSN) or Techno-Social Networks refer to a group of individuals or organizations that use the new communication technologies (i.e. Internet, Mobiles, etc.) as a communication medium, forming a social structure described by particular relations \cite{cruz2011handbook}. The study of OSN has increasingly attracted the attention of the scientific community. In fact, online services, such as Facebook, Twitter, or Instagram, play an increasingly important role in the dissemination of opinions and in the emergence of certain behaviors. They facilitate social interactions, helping individuals to find other people with common interests, {to} establish a forum for discussion, and {to} exchange information \cite{Cheung2011}, \cite{Helmond2009}. 

\medskip

The Special Digital Report 2020 \cite{we_are_social_2020} states that digital, mobile and social media are a fundamental part of people's daily lives around the world. According to \url{Statista.com}, in 2019 the social penetration rate of OSN reached 70\% in East Asia and North America, followed by North Europe at 67\%, leading to a global social penetration rate of 45\%.
Moreover, since the COVID-19 outbreak was declared a Public Health Emergency of International Concern on 30 January 2020, social media usage has reportedly increased significantly.
On the 24th March  2020, Facebook recorded a 50\% increase in total messages in many of the countries most affected by the virus, with a 70\% increase in the time users have spent on social media since the beginning of the pandemic \cite{napoleoncat}.
\medskip

In the figure below the total number of active users of the most popular social media networks are shown \cite{Digitalinformationworld}.
\medskip

\begin{center}
  \includegraphics[trim={0cm 0cm 0cm 0cm},width=1\columnwidth]{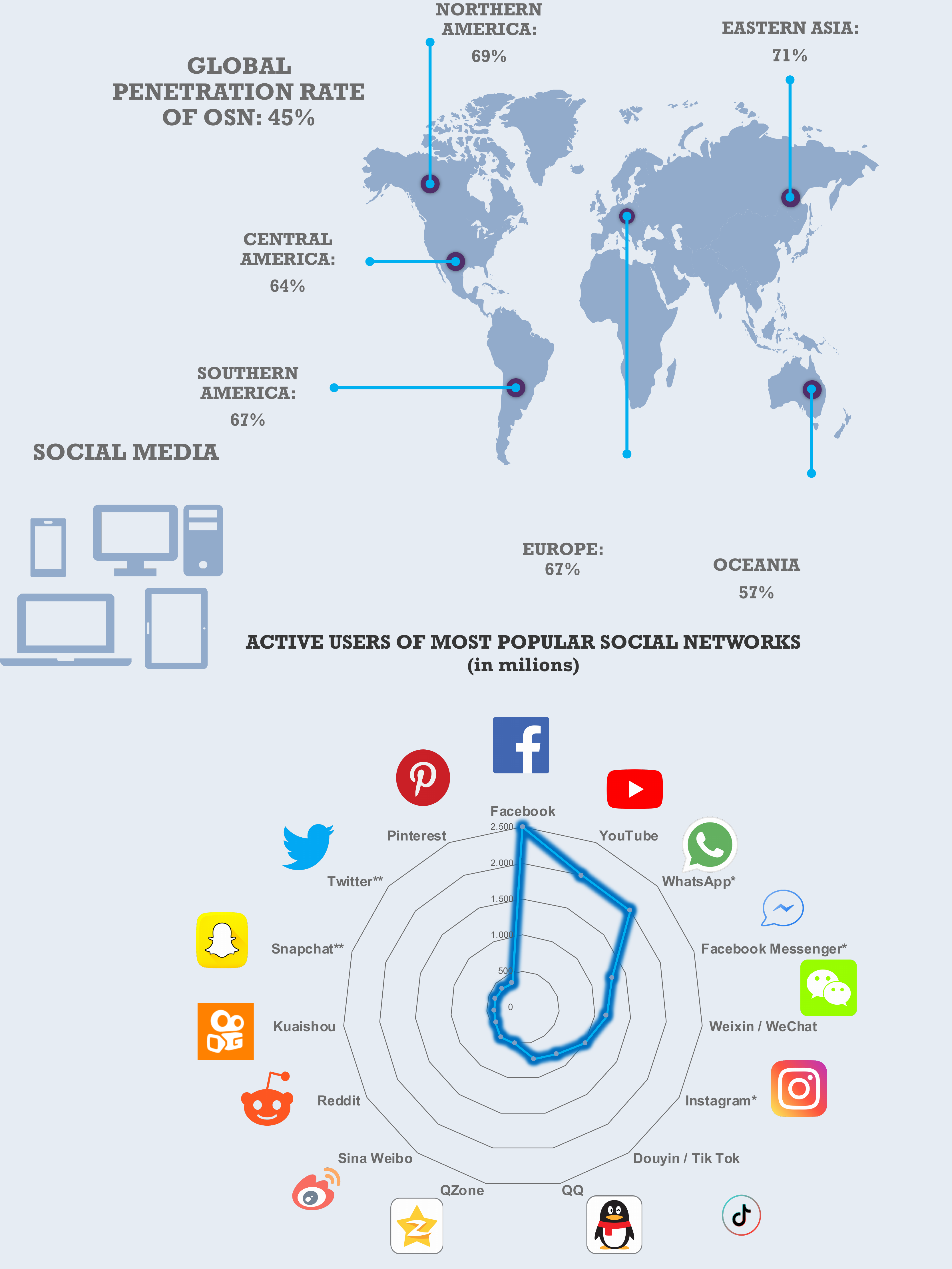}
\end{center}

\end{minipage}

\begin{minipage}{1\columnwidth}
A downside of this high social penetration rate is the spread of \emph{fake news} \ant{whose detection is becoming a serious problem}~\cite{Fake_news_Science,Bovet2019}.
The figure shows the share concerned about what is real and fake on the internet when it comes to news. The data are updated until in 2019 \cite{STATS_FAKE_NEWS}. Moreover, around 43\% adults in USA get news from Facebook, according to a survey conducted in July and August 2018 \cite{STATS_FAKE_NEWS2}. This \ant{proportion} is much higher than the percentage of adults who get news through YouTube (21\%), Twitter (only 12\%) and other platforms.

\begin{center}
  \includegraphics[trim={0cm 0cm 0cm 0cm},width=1\columnwidth]{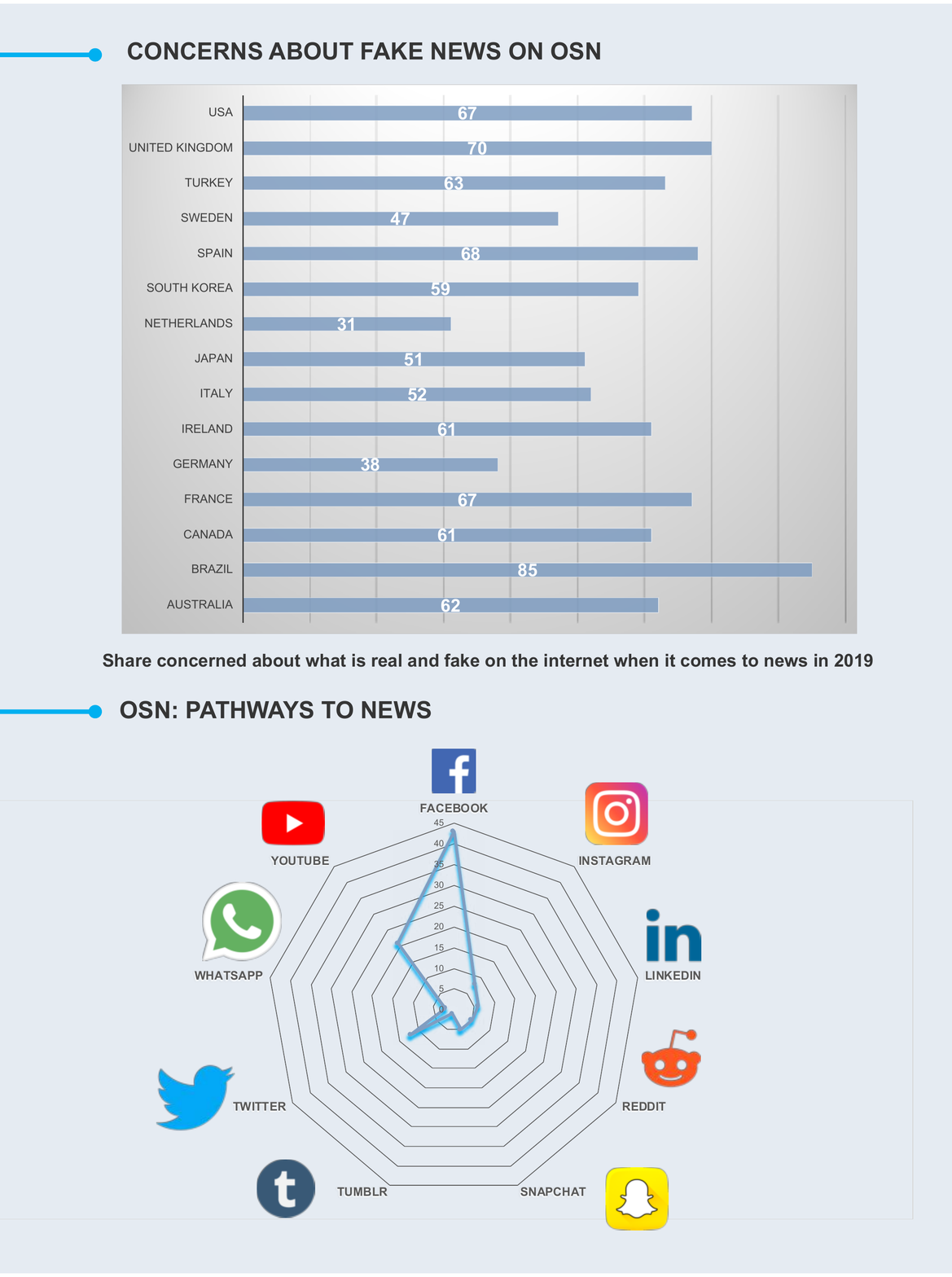}
\end{center}

Very recent contributions of researchers in the systems and control field focus on the problem of modeling the dynamics of the spread of misinformation \cite{Del_Vicario201517441}.
\ant{In particular,} it has been shown~\cite{Pierri2020TopologyCO} that in Twitter diffusion networks misleading content spreads deeper than mainstream news with a small number of followers, and communities sharing fake news are more connected and clustered.
Then structural properties of Twitter diffusion networks, such as the number and the size of Weakly Connected Components, Average Clustering Coefficient, Diameter of the Largest Weakly Connected Components, to mention just a few, can effectively be used to identify misleading and harmful information \cite{Pierri2020TopologyCO}.
Inferring the networks structure and learning global properties from partial information becomes a central question \cite{GomezRodriguez:2010:IND:1835804.1835933,GomezRodriguezLBS2014}.


\end{minipage}
\end{multicols}

\end{tcolorbox}
\begin{multicols}{2}



\begin{tcolorbox}[title= \sf \textbf{Opinion dynamics over networks in a nutshell},colframe=carmine!10,
colback=carmine!10,
coltitle=black,
]\small{\sf{
{\textbf{Assumptions}:}
\begin{itemize}
\item Population of individuals (or actors)
\item Individuals interact and exchange their opinions
\item As a result, their opinions evolve
\end{itemize}

\begin{center}
  \includegraphics[width=0.7\columnwidth]{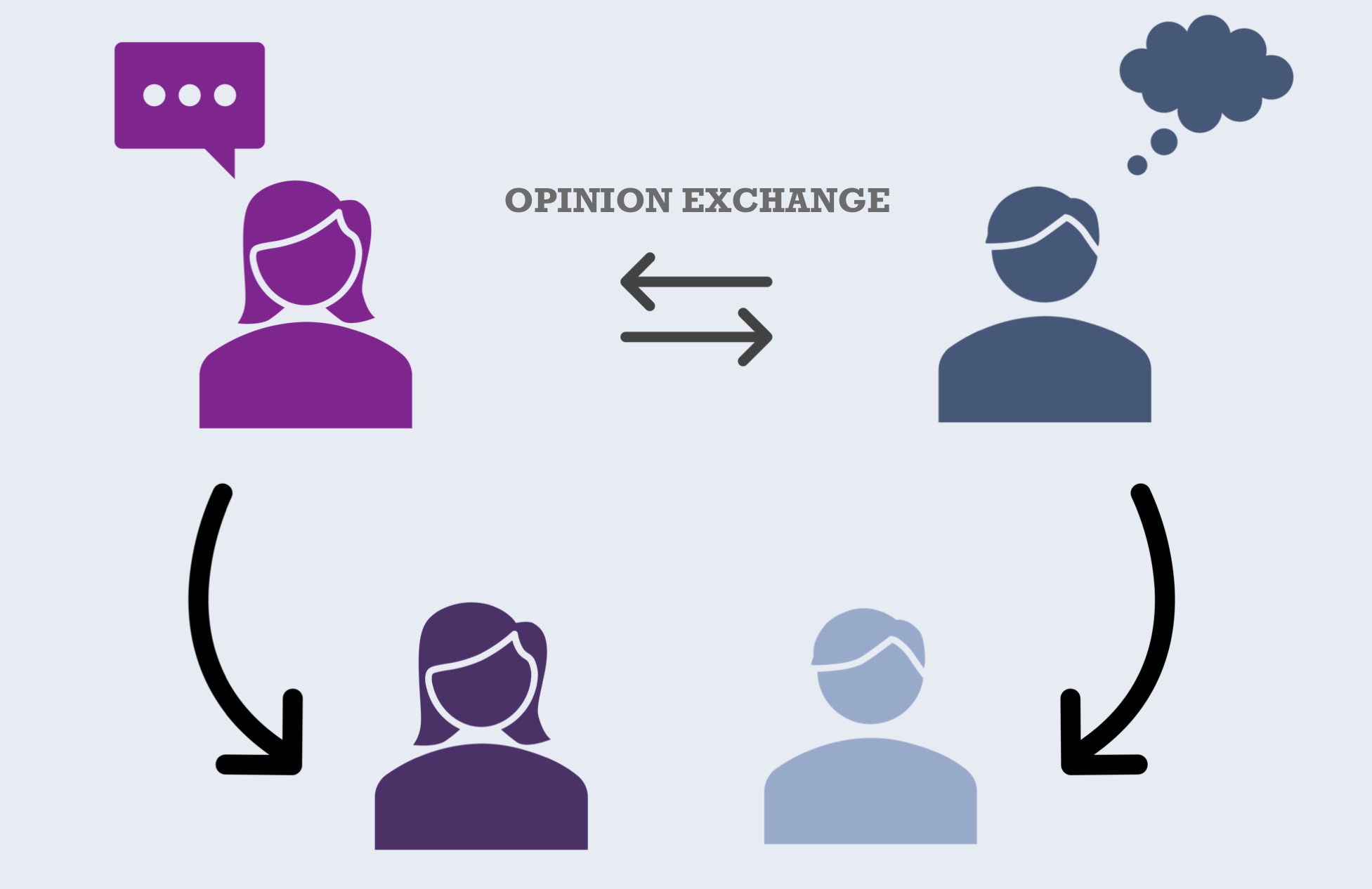}
\end{center}

{\textbf{Three research directions:}}
\begin{enumerate}
\item \textbf{Mathematical modeling} of social behaviors
{
\begin{itemize}
\item {\em{interaction {protocol}}}: discrete/continuous-time, deterministic/random {contacts}
\item {\em{social ties}}: linear (parameterized by scalar ``weights'') or nonlinear functions
\end{itemize}}
\item \textbf{Analysis}: sociology + computer science
\begin{itemize}
\item   the \em{evolution} of opinions (convergence / oscillations, consensus / disagreement)
\item   \em{low-dimensional features} (communities, opinion leaders)
\end{itemize}
\item \textbf{Control}: design mechanisms to provide the desired behavior of the opinion profile
\begin{itemize}
\item induce qualitative changes (e.g. consensus)
\item induce quantitative changes (e.g. drive the individual's opinions to a desired value)
\end{itemize}
\end{enumerate}
}}
\end{tcolorbox}

The role of the systems and control community in the area of social networks has increased with
the introduction of \emph{dynamical models}, that {are able to capture}
phenomena  observed in sociology and thus enable to understand social processes and interactions.
The workflow {we envisage} for a systematic study of opinion formation  and the networks' structural properties is shown in the diagram in the box \boxref{Data-driven systems and control approach}.
All research lines Data Collection, Design of Mathematical framework (Modeling, Analysis and Control), Algorithms development, and Design of Large-Scale experiments must proceed in parallel
 with continuous interactions among the blocks.

Data collection and processing constitute the backbone of \ant{computational social science}~\cite{Lazer721} and
require a careful systematization. More precisely, one needs to define what kind of information can be acquired, the frequency of the samples, and subsequently, to explore how much information is contained in each sample.
In this context, the first issue is to encode people's opinions, sentiments, and preferences from written language into a formal language or numerical representation that can be processed with numerical techniques. To this regard, several methods for sensing opinions based on sentiment analysis have been proposed in recent years~\cite{Liu:2012:SAO:3019323}. Efficient sampling procedures of graph signals require only a few nodes in a networks to be directly observed and sensed~\cite{7480396,Tremb15} and remove irrelevant information in order to improve the performance of processing.

In analysis, we need to identify the best suitable models that accurately characterize the social system, and to select the evaluation metrics to quantify the interpersonal influence in the network. Then, efficient algorithms and new control mechanisms of centrality measures (see "Centrality measures") must be designed to improve social network interconnectivity and resilience. We stress out that, in our opinion, the control models should be {\em data-driven}, i.e.  simulation based on data collected from real social networks must be performed to validate and refine the dynamic models, predict and control the opinion diffusion over the network. Since parameters of system dynamics models are subject to uncertainty, a sensitivity analysis is crucial to explore the effects of parameter uncertainty on the behavior patterns.
\end{multicols}

\begin{tcolorbox}[title= \sf \textbf{Data-driven systems and control approach}, colframe=carmine!10,
colback=carmine!10,
coltitle=black,
]
The {ultimate goal} is the development of a \ant{theoretical framework, which is based on systems theory and data-driven control,  able to predict the processes of opinion formation and information} spread in \ant{large-scale} social networks and provides well grounded tools to quantify and control the impact of specific actions.
The workflow for a systematic study of network structures and dynamics is \ant{shown} in the diagram below.
\begin{center}
\includegraphics[trim={1cm 5cm 1cm 3cm},clip,width=0.68\columnwidth]{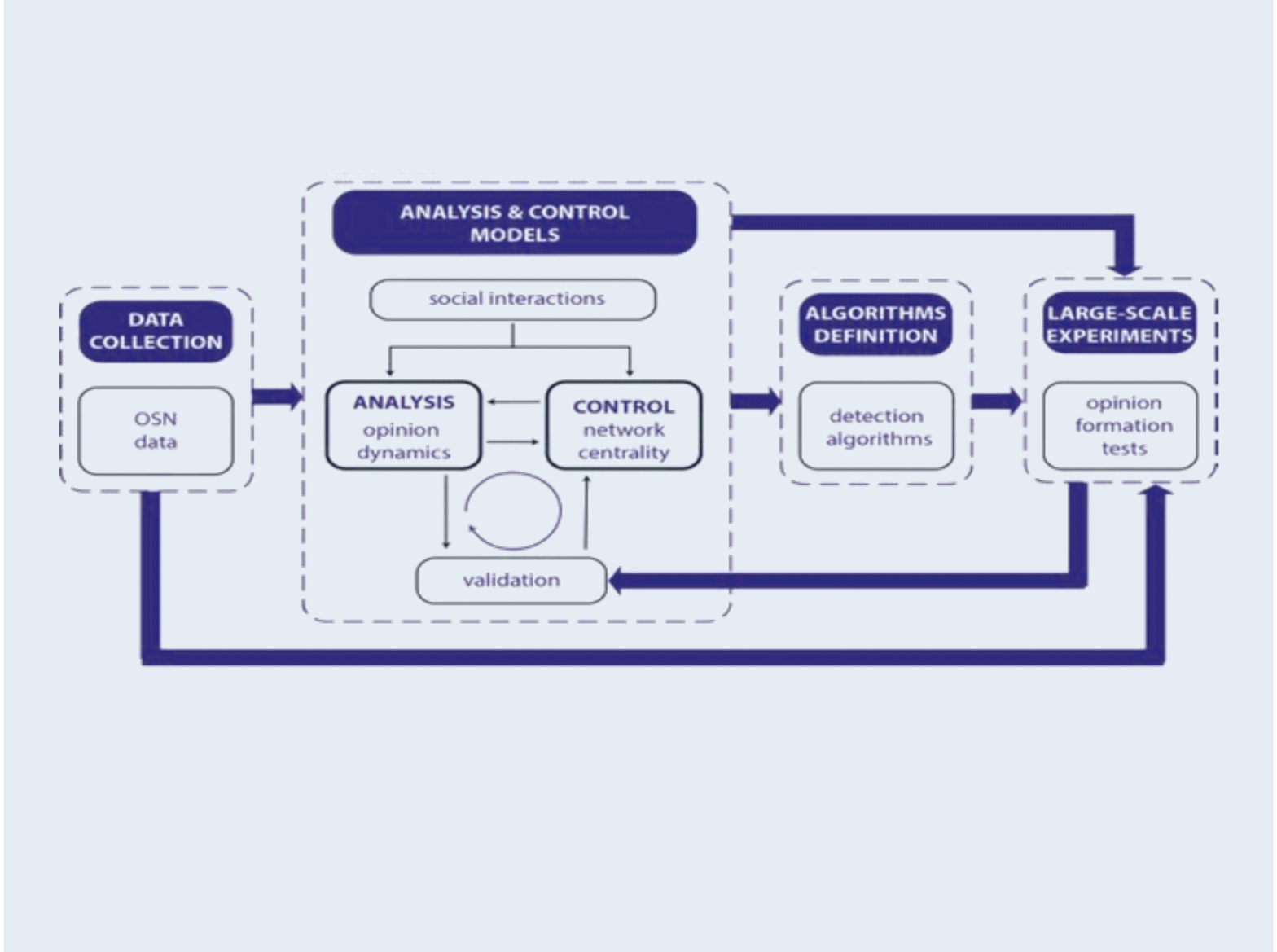}
\end{center}
\end{tcolorbox}

\begin{tcolorbox}[breakable,title= \sf \textbf{Friedkin and Johnsen experiment\cite{Friedkin:Johnsen:1999,Friedkin:Johnsen:2011}},colframe=airforcecarmine!20,colback=airforcecarmine!20,coltitle=black,]
\sf\small\label{exp.chips}
A seminal example of coupling the theory with empirical research can be found in \cite{Friedkin:Johnsen:2011}, where Friedkin and Johnsen model \cite{Friedkin:Johnsen:1999} for single-issue opinion dynamics is validated for small and medium size groups of individuals
\ant{(social actors)}.\\

\medskip

\begin{minipage}{0.49\columnwidth}
 \begin{center}
\includegraphics[width=0.8\columnwidth]{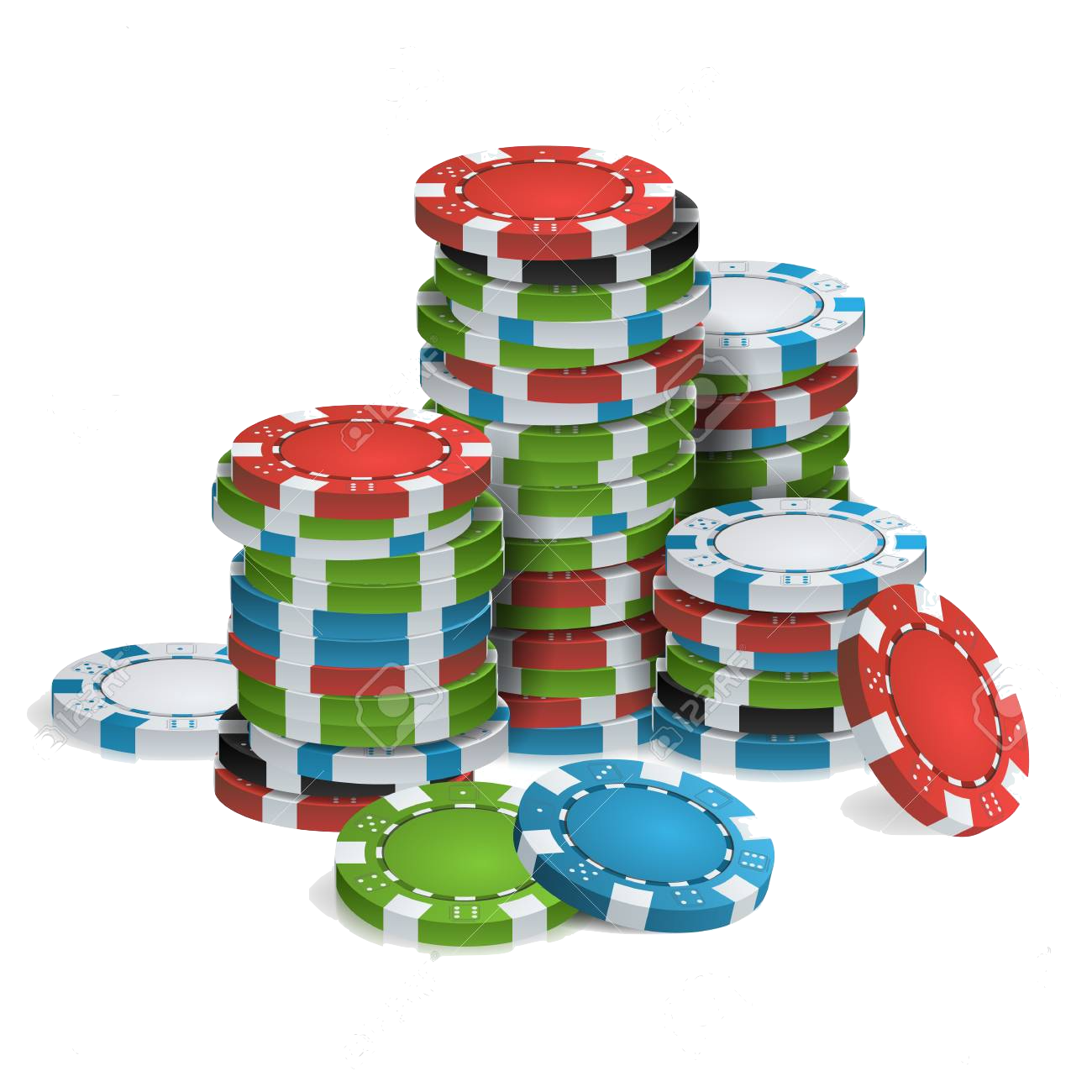}
\end{center}
\begin{center}
\sf  \it The influence is measured experimentally by asking the participants to distribute ``chips" to measure the influence weights each participant has assigned to him/herself and to the others during the decision making process.
\end{center}
\end{minipage}
\begin{minipage}{0.49\columnwidth}
\begin{center}
{\em{Experiment design:}}
\begin{itemize}
\item 30 groups
\item 4 individuals
\item 15 issues (Risk Choice Dilemmas)\end{itemize}
\end{center}
\medskip
 \begin{center}
{\em{''The issues involve opinions on the {{minimum level of confidence}}, that is a value in the [0, 1] interval, required to {{accept a risky option with a high payoff}} over a {less risky option with a low payoff}.'' }}
\\
\end{center}
\bigskip

{\small{Example:}}\medskip
\begin{center}
{\em{''\underline{Investment Choice}: Imagine you want to invest some money you recently inherited. You may invest in secure low return securities (small risk), or alternatively in more risky securities that offer the possibility of large gains (great risk).''}}
\end{center}
\medskip
\begin{itemize}
\item {Initial opinions} of actors recorded
\item Fifteen minutes of discussion
\item Actors {distribute ''chips''} to quantify the influence
\item {FinaI actors'  opinions} recorded
\end{itemize}
\end{minipage}
\vskip 6mm

\noindent
30 groups consisting of 4 people participate in the experiment.
The participants are asked to  express their opinions on 15 different issues selected uniformly at random without replacement from a set of risk choice dilemmas.
Risk choice dilemmas are hypothetical life decision situations that are used to measure the willingness to assume a risk. More precisely, the agents are asked to express their minimum level of confidence, i.e., a scalar value in the range $[0, 1]$, required to accept a risky option with a high payoff over a less risky option with a low payoff. Individuals in the group record their initial opinions on the issue, then a 15 minutes discussion is opened, then the final opinions are recorded.\\[1mm]

\textit{Influence determination:}
To estimate social influence, in the post-discussion people are asked to distribute ``chips'' between the actors they interacted with, as a subjective measure of influence exercised by other group members. The model validation is then performed by showing that the opinions predicted through the model are close to those recorded.
\end{tcolorbox}

\begin{multicols}{2}
In order to validate and examine the models, social ties between the individuals have to be quantified. In a small-size group \ant{participating in a round-table discussion, individuals can estimate the influence of themselves and the others on the formation of their opinions (\boxrefb{Friedkin-Johnsen experiment}). However, this approach is inapplicable to large-scale groups and online social networks whose structures of influence relations can only be inferred from data}.

The rapid development of the Internet, on one hand, makes a large volume of data easily available for analysis. On the other hand, it poses new challenges. Data size is getting larger and larger, and information collected becomes heterogeneous and more complex. In fact, the massive data in OSN consist of linked data, mainly in the form of graphic structures, describing the communications between any two entities, text, images, audio, and video, that must be processed. Hence efficient analytic tools and algorithms to reconstruct social influence mechanisms are required.

These considerations motivate the present work, which aims to present a unified overview onto the two main aspects of the interpersonal influence estimation: i) the social network sensing problem and ii) network reconstruction algorithms with a particular focus on sample complexity and computational requirements. The main challenge is to guarantee efficiency and scalability of the algorithms in the face of big data produced by OSN.
It is shown that the interpersonal influence estimation problem can leverage a mature technical background and strong mathematical foundations and can be tackled efficiently using modern techniques.
The main studies performed on this subject are highly innovative, blending learning tools with high dimensional data analysis, including \ant{principal component analysis} \cite{PCA}, compressed sensing \cite{eldar2012compressed} and graph analytics \cite{series/sbcs/YanTC17}, and \ant{encompassing various fields of research, e.g.}
\begin{enumerate}
\item graph theory and linear algebra;
\item control theory techniques: stability, controllability, system identification, optimal and robust control;
\item signal processing, \ant{statistics} and machine learning for big data analysis; 
\item efficient \ant{optimization-based} algorithms for sampling and reconstruction of graph signals.
\end{enumerate}

The main core of the paper is  based on previous works \cite{FrascaTempo:2013,FrascaIshiiTempo:2015,FrascaTempo:2015,Ravazzi2018,8619770,8796302,Parsegov2017TAC,tutorial}. \ant{We also refer an interested reader to the additional literature, cited} throughout the paper in order to gain a deeper insight on this subject. \ant{The material of this survey has been partly presented on the tutorial section ``Control and Learning for Social Sciences: Dynamical Networks of Social Influence'' of IFAC World Congress 2020.}

\medskip
\noindent

\section{Defining influence in social networks}
 As defined in~\cite{friedkin_1998, doi:10.1177/0049124193022001006}, the interpersonal influence is a {\em ``causal effect of one actor on another''}, {such as a change in opinions and behaviors of the influenced actor~\cite{Sun2011}.
Quantification and measurement of social ties are long-standing problems that have been studied
since 1950s~\cite{March57,Simon53,French:1956,FrenchRaven:1959}.
{One principal difficulty is} to separate direct and indirect influence:} {\em ``if the opinion change has occurred within a system of
influences involving other actors, then these other actors may have induced the observed opinion
difference or change''}~\cite{friedkin_1998}. Another problem is the {co-evolution of social ties and the individuals' behaviors.} 
On one hand, people modify their behaviors to align them with the behaviors of their friends (social influence), but on the other hand, people tend to form friendship with others like themselves ({social} selection). Opinions and other mutable characteristics of people are thus formed by the \emph{interplay} between social selection and influence~\cite{EasleyKleinberg}.

{Several research lines on interpersonal influence exist in the literature, among which three {main} directions prevail. The first direction of research is developing the seminal ideas of Granovetter~\cite{Granovetter73}, defining the strength of a social tie between two individuals as a function of their \emph{positions} in the social group: for instance, the more common friends actors A and~B have, the stronger is {the} tie among them~\cite{Granovetter73,Sun2011}. Social influence introduced in this way thus depends only on the structure of a social network. A large amount of available data from real-world social networks and the existence of efficient tools for their analysis makes this approach very attractive for both behavioral and computer sciences.}

{The second line of research  relates the social influence to temporal (dynamical) mechanisms, modifying some numerical attributes of social actors, such as e.g. opinions or some quantities related to them. The influence (or power) of actor B over actor A is a parameter of the corresponding mechanism, measuring A's sensitivity to the opinion of B or the level of trust in B's opinions. This idea has been elaborated in the Friedkin-Johnsen theory of social influence~\cite{friedkin_1998,Friedkin:Johnsen:1999,Friedkin:Johnsen:2011,Friedkin:2015}. The fundamental results reported in~\cite{friedkin_1998} establish interrelations between {the structural and the dynamical} approaches to social influence. Namely, in networks of scientific collaborations social positions {(``opinions'')} of individual researchers can be encoded by multidimensional vectors. Two opinions are close if the researchers have similar (in some sense) sets of collaborators. The evolution of these opinions is predicted by the Friedkin-Johnsen model of opinion formation (see Section~IV) whose parameters can be constructed via a structural analysis.}

{The third direction of research on influence in complex networks (not necessarily social) is concerned with \emph{statistical} (learning-based) methods of network reconstruction. Similar to the second approach, it assumes that the actors at a network's nodes are endowed with some numerical values that are supposed to be random. Unlike the second approach, an existence of a temporal mechanism modifying the values is not stipulated. A tie between two nodes corresponds to \emph{statistical correlation} between their values, and the strength of this tie is naturally measured by the correlation coefficient. In other words, a network is considered as a probabilistic \emph{graphical model}~\cite{jordan2004,Airoldi} and is analyzed by methods of statistics and statistical learning theory.}

\cl{This paper develops the second and third lines of research. In Section~\ref{sec:graphID}, we consider statistical estimation of social influence (the third direction). Sections~\ref{sec:FJ} through~\ref{sec:SIENNA} deal with identification of dynamic mechanisms of opinion formation, namely, the Friedkin-Johnsen model.} Both approaches are aimed in the reconstruction of a \emph{weighted directed graph}, whose nodes have some numerical attributes (considered as opinions of social actors), the arcs represent social ties whose strengths are described by weights.
A natural question arises how the estimates of these weights can be used to study the structure of a social network, e.g. exploring communities? The remainder of this section is devoted to this problem and introduces important characteristics of a weighted graph.

\subsection{Influence related measures}

A social network {consists of the two main components: i) social actors (individuals or organizations)} ii)  the dependency, influence or similarity relations. {Each actor has a numerical attribute (standing e.g. for an opinion)}.

The social network can be mathematically described by a directed  weighted graph (see box \boxref{A Glossary on Graphs} for graph-related definitions).

At the local level, the social influence is a {\em directional effect from node $i$ to node $j$, and is related to the edge strength $(i,j)\in\E$} \cite{Sun2011}. The social influence can then be encoded in the \textit{social influence matrix} $\Wb=[w_{ij}]$,
 which is adapted to the graph (if $(i,j)\notin\E$ then the corresponding entry $w_{ij}$ is zero). In the dynamic models of social influence~\cite{Friedkin:2015} this matrix is typically normalized to be row stochastic.


{At the} global level, some nodes can be more influential than others {due to the network interconnections. Several} global measures have been introduced to identify the most relevant entities in the network. These global measures can refer to nodes or edges and can be defined in several ways according to the specific context and application, leading to different notions of {\em centrality} (a node's/edge importance) measure.
Various measures on centrality are defined in the box  \boxref{Centrality measures in weighted graphs}.
\end{multicols}

\medskip
\begin{tcolorbox}[title= \sf \textbf{A Glossary on Graphs },colframe=carmine!10,colback=carmine!10,coltitle=black,
]

\sf\small

An \textbf{unweighted graph} $\G$ is represented by the couple $(\V,\E)$, where
\begin{itemize}
\item $\V$ is the set of \textbf{nodes} (corresponding e.g.\ to agents in the network), indexed as $1,\ldots,n$.
\item $\E\subseteq\V\times\V$ is a set of ordered pairs of nodes describing the relationships, e.g. if $(i,j) \in \E$ then $j$ is influenced by $i$. We refer to the couples $(i,j)$ as the \textbf{edges} of the graph.\\
\end{itemize}

Given an unweighted graph, one can define its \textbf{adjacency matrix}
$\Adj$, with ${ij}$ entry $[\Adj]_{ij}=1$
if $(i,j)\in\E$ and zero otherwise.\\

A \textbf{weighted graph} $\G$ is represented by a triple $(\V,\E,\Wb)$, where $\V$ and $\E$ are the nodes and edges of the graph, and  $\Wb=[w_{ij}]$ is the weighted adjacency matrix, known as \textbf{influence matrix}, whose {entry $w_{ij}$ defines} the \textbf{weight} of the edge $(i,j)$ {and $w_{ij}=0$ if $(i,j)\not\in\E$, that is, $i$ and $j$ are not connected. Each square matrix $\Wb=(w_{ij})_{i,j\in\V}$ can be associated with a graph $\G[\Wb]=(\V,\E,\Wb)$, where $\E=\{(i,j):w_{ij}\neq 0\}$.} \\

A matrix $\Mb$ is said to be \textbf{adapted to the graph} $\G$ {if $\G[\Mb]=\G$}. 
By construction, the adjacency matrix $\Adj$ and {every influence matrix $\Wb$ of a graph $\G$} are adapted to {$\G$}.\\

\begin{center}\begin{tikzpicture} [scale=.7,every node/.style={circle,fill=airforcecarmine!50,inner sep=6pt}]
  \node (n4) at (8,7) {};
  \node (n5) at (10,9)   {};
  \node (n1) at (13,8) {};
  \node (n2) at (13,6)   {};
  \node (n3) at (10,5) {};

    \tikzset{every node/.style={}}
  \node (n10) at (10,9)   {${x_u}$};
  \node (n11) at (13,6)   {$x_v$};

    \foreach \from/\to in {n4/n5,n1/n5,n1/n2,n2/n3,n3/n4}
    \path (\from) edge[->,bend right=3] (\to);
     \tikzset{mystyle/.style={->}}
\tikzset{every node/.style={fill={carmine!10}}}

      \path (n2) edge[->,bend right=3] node { ${w_{ij}}$}(n5);

\end{tikzpicture}
\end{center}

The matrix $\Wb$ is said to be \textbf{row stochastic} if its rows sum up to one, i.e.\ $\sum_i w_{ij}=1$. In a compact form, we can write $\Wb\1=\1$, with $\1\doteq [1\,\cdots\, 1]^\top$.
Similarly, matrix $\Wb$ is  \textbf{column stochastic} if \ $\sum_j w_{ij}=1$, i.e. $\1^\top\Wb=\1^\top$.

For unweighted graphs, we say that $\G=(\V,\E)$ is an \textbf{undirected graph} if $(i,j) \in\E$ implies that $(j,i)$ is also an edge in $\E$. For a weighted graph $\G=(\V,\E,\Wb)$, we also require that 
the {weights of} edges $(i,j)$ and $(j,i)$ {coincide}: $\Wb=\Wb^\top$.\\

The \textbf{Laplacian} matrix of a weighted graph \ant{(possibly, directed)} is defined as
\[
\Lb\doteq\Db-\Wb,
\]
where $\Db\doteq\mathrm{diag}(d_1,\ldots,d_n)$ is the {weighted} degree matrix, {where $d_i=\sum_{j}w_{ij}$.}

For each {node} $i\in\V$, we denote its \textbf{neighborhood} with $\mathcal{N}_i \doteq  \{j\in \V : (i,j) \in\E\}$.\\

A sequence of edges $(i,i_1),(i_1,i_2),\ldots,(i_{m-1},j)$  without repeated vertices forms a \textbf{path} from $i$ to $j$.
A graph is said to be \textbf{strongly connected} if there exists a path between any pair of nodes.\\

\end{tcolorbox}

\begin{multicols}{2}
The simplest and most popular definition of centrality is the \textit{degree centrality}, i.e. the number of neighbors of a node. This measure  can be interpreted as a measure of the immediate risk of a node of catching (in-degree) or spreading (out-degree) some information.
\ant{A more general concept is \emph{$K$-path} centrality~\cite{SADE:1989}, defined as the number of paths of length $K$ starting from a node.
Both degree and $K$-path centrality definitions are local. To measure the importance of a node for the graph as a whole, other alternative centrality measures have been considered}. Among them, we discuss briefly  the closeness, betweenness, and eigenvector centrality.

The {\em closeness centrality} is a measure of how much a node is close to most of the other nodes~\cite{LF:79} and gives an insight on how long it will take to spread information from $i$ to all other nodes in the network.


\end{multicols}

\begin{tcolorbox}[title= \sf \textbf{Centrality measures in weighted graphs},colframe=carmine!10,
colback=carmine!10,
coltitle=black,
]

\sf\small
Consider a weighted graph $\G=(\V,\E,\Wb)$, where
$\V$ is the set of agents in the network, $\E\subseteq\V\times\V$ is a set links describing the interpersonal influences and $\Wb\in[0,1]^{\V\times\V}$ is the \textit{social influence matrix}, which is adapted to the graph.\\

A \textbf{centrality measure} is a nonnegative scalar measuring \ant{the importance of a node or an arc in the graph. We illustrate alternative definitions of centrality on} a simple directed network, known as the Football Dataset~\cite{Dagstuhl2001}. The network records 35 soccer teams which participated in the World Championship in Paris, 1998. Every edge records the number of national team players of one country who play in the league of another country.

\begin{minipage}{0.49\columnwidth}
\begin{center}
\textbf{Degree centrality}
\end{center}

\begin{center}
\includegraphics[trim={6cm 10cm 5cm 10cm},clip,width=0.7\columnwidth]{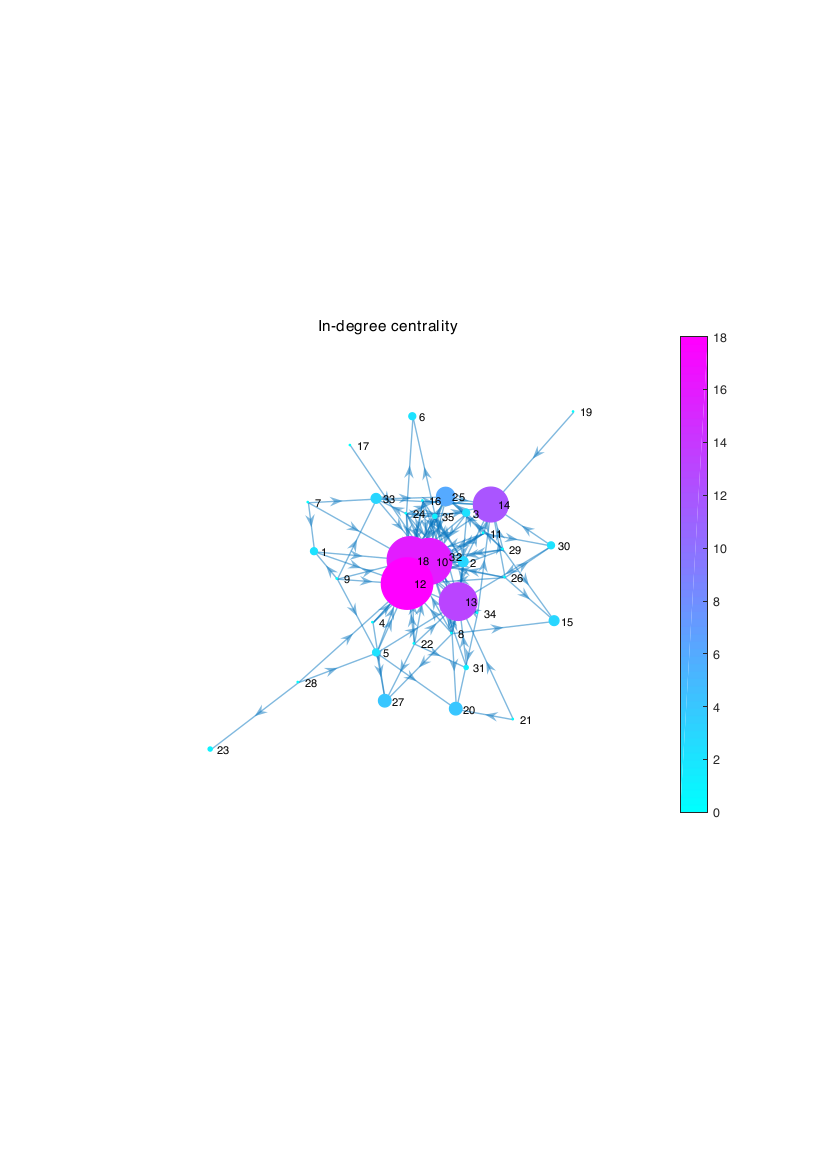}
\end{center}

The in/out-degree of $i\in\V$ are defined as follows {\color{blue}
\begin{align*}
\text{in-deg}(i)&=|\{j\in\V:w_{ij}\neq0\}|,\\
\text{out-deg}(i)&=|\{j\in\V:w_{ji}\neq0\}|,
\end{align*}}
respectively, where $|\X|$ denotes the cardinality of the set $\X$.
In social systems, the degree corresponds to the number of paths of length 1 starting from a node.
The \textit{weighted} in/out-degree are defined as the sum of weights when analyzing weighted networks
\begin{align*}
\text{in-deg}_\Wb(i)&=\sum_{j\in{\mathcal{V}}} w_{ij},\quad
\text{out-deg}_\Wb(i)&=\sum_{j\in{\mathcal{V}}} w_{ji}.
\end{align*}
\end{minipage}
\quad
\begin{minipage}{0.47\columnwidth}

 \begin{center}
\textbf{Closeness centrality}
\end{center}
 \begin{center}
\includegraphics[trim={6cm 10cm 5cm 9.8cm},clip,width=0.7\columnwidth]{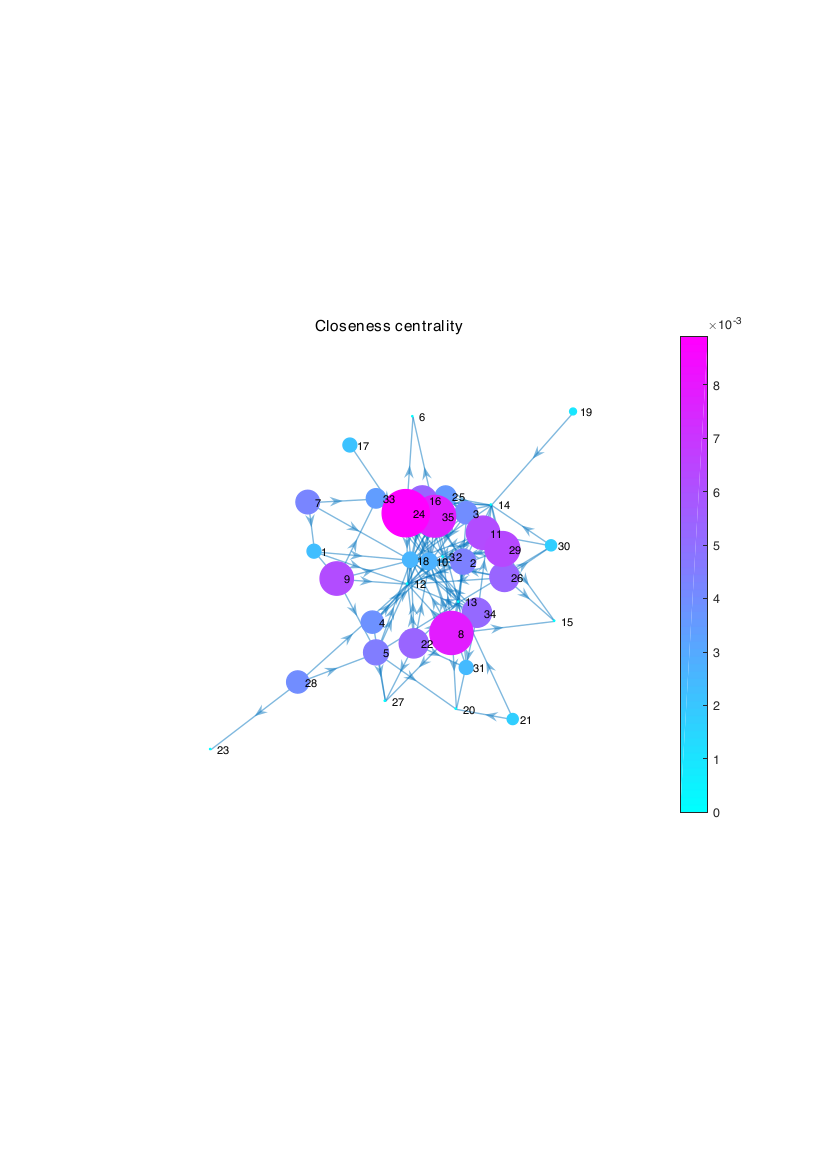}
\end{center}
The \textit{closeness centrality} of node $i$ is defined as
\[
c_i=\frac{1}{\sum_{j\in\V\setminus\{i\}}d_{ij}}
\]
where $d_{ij}$ denotes the length of the shortest path between $i$ and $j$. This notion can be modified using other definitions of distances, as considered in~\cite{F77,DRW-SPB:94}.
In particular, closeness centrality for weighted graphs can be defined introducing  ``weighted distance $d_{ij}$'', {that is,
the minimal weight of all paths that connect $i$ to $j$. The weight of a path is naturally defined as the sum of the weights on the traversed edges.}
\bigskip
\end{minipage}
\quad

\begin{minipage}{0.49\columnwidth}
\begin{center}
\textbf{Betweenness centrality}
\end{center}
 \begin{center}
\includegraphics[trim={6cm 10cm 5cm 9.8cm},clip,width=0.9\columnwidth]{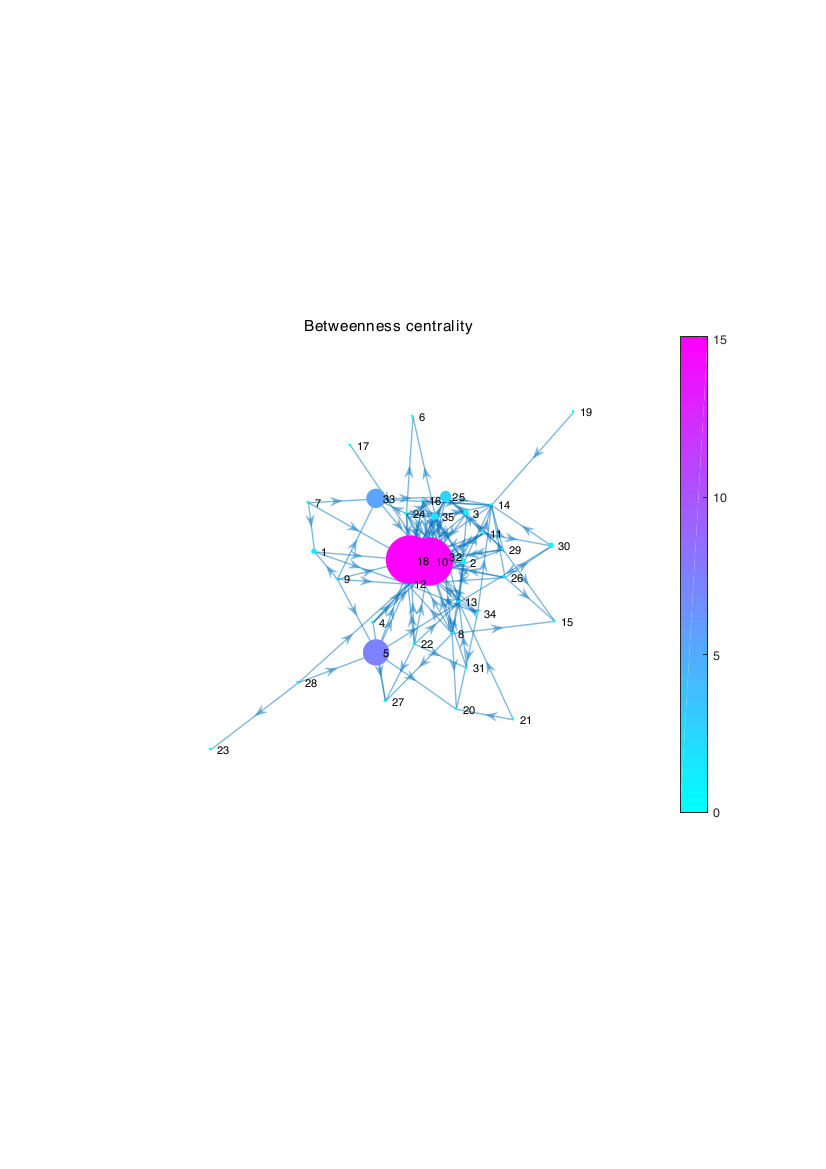}
\end{center}

The  \textit{betweenness} of node $i$ is defined as
\[
 b_i=
\sum_{j,k\in \V, j\neq k\neq i} \!\!\frac{|\Sc_i(j,k)|}{|\Sc(j,k)|}
\]
where $\Sc(j,k)$ denotes the set of shortest paths from $j$ to $k$, and
$\Sc_i(j,k)$  the set of shortest paths from $j$ to $k$
 that contain the node $i$.
 For weighted graphs, the length of each edge forming the paths in $\Sc(j,k)$ and $\Sc_i(j,k)$ is measured through the entries of the influence matrix $\Wb$.
\end{minipage}
\quad
\begin{minipage}{0.47\columnwidth}
 \begin{center}
\textbf{Eigenvalue centrality}
\includegraphics[trim={6cm 10cm 5cm 9.8cm},clip,width=0.7\columnwidth]{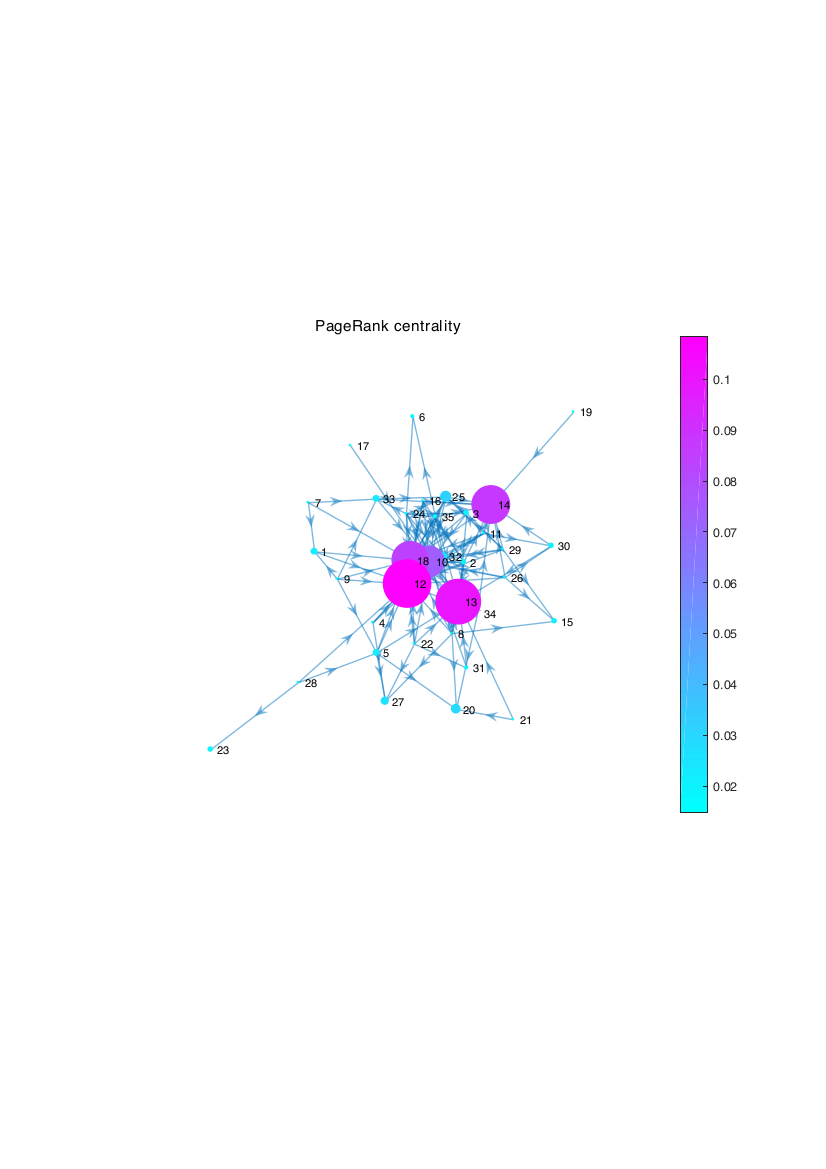}
\end{center}
The idea of \emph{eigenvector centrality} is based on a simple principle: a node is important if it is connected to other important nodes. This centrality measure is determined by the dominating (Perron-Frobenius) eigenvector $x^{\star}$ of some properly defined nonnegative matrix $\Ab$, compatible with the graph. Formally,
\[
\Ab x^{\star}=\lambda x^{\star},\quad \1_n^{\top}x^{\star}=1,\quad x^*_i\geq 0\,\forall i.
\]
where $\lambda=\rho(\Ab)$ is the maximal positive eigenvalue (being also the spectral radius) of $\Ab$. In the standard definition of eigenvector centrality~\cite{Newman:2003}, $\Ab=\Adj$ is the standard adjacency matrix. A more general construction
\[
\Ab(\mathbf{M})=(1-m)\mathbf{M}+\frac{m}{n}\1_n\1_n^{\top},
\]
where $\mathbf{M}$ is a \emph{column stochastic} matrix and $m\in (0,1)$, arises in the definition of PageRank centrality~\cite{Broder2006}.

\end{minipage}

\end{tcolorbox}

\begin{multicols}{2}

Another relevant measure is represented by the node \textit{betweenness} \cite{F77,DRW-SPB:94}. Nodes with a high betweenness occupy critical positions in the network and are bridges between two groups of vertices within the network, since many paths in different groups must pass through this node.

The {\em eigenvector centrality} of a node is a function of its neighbors and the relevance is assigned according to the entries of the leading eigenvector $x^{\star}$ of a suitable weighted adjacency matrix of the network.
Contrary to the degree centrality, this notion does not depend on the number of neighbors but takes into account the relevance of its neighbors. In this way, a node with a few influent neighbors has larger eigenvector centrality than a node with various neighbors of limited influence. The most famous eigenvalue centrality measure is the PageRank centrality~\cite{SB-LP:98}, which was introduced in the context of ranking of webpages. {Many other centrality measures, such as e.g.}
Katz centrality \cite{Katz1953}, Bonacich centrality~\cite{B87,Bonacich:01}
and  harmonic influence centrality~\cite{Friedkin:1991,Friedkin:2015,ProTempo:2017-1}, {naturally arise as extensions of the eigenvector centrality and PageRank}.
\medskip

\section{Sparsity structure of social network}\label{subsec:Structure_OSN}

{A} systematic study of the structure of social networks {offers} several metrics and algorithms 
for extracting low-dimensional {network's} features. 
Metrics {can} quantify global or local structural properties. The {\em network density} is an aggregate network metric defined as the ratio {$|\E|/n^2$} of the number of social relationships observed in the network to the total number of possible relationships that could exist among nodes {(that is, } the proportion of ties within the network{)}.

A collection of large social network datasets is made available by
Stanford Network Analysis Platform (SNAP, \cite{leskovec2016snap}) and can be visualized using the software GraphViz \cite{Ellson01graphviz}. In the box \boxref{Sparsity structure in OSN'} the table reports the type, the number of nodes and the network density of some of these social networks.
If you look at these data, you will realize that OSN have some common features:
\begin{itemize}
\item they are massive networks with a number of nodes $n=|\V|$ ranging from tens thousands to millions;
\item they are not dense, in the sense that number of edges is not close to $n^2$, i.e. the maximal number of possible edges and is at most of order of the size of the network.
\end{itemize}

\end{multicols}

\begin{tcolorbox}[title= \sf \textbf{Sparsity structure in OSN},colframe=carmine!10,
colback=carmine!10,
coltitle=black,
breakable
]

\begin{minipage}{0.49\columnwidth}
 \begin{center}
\includegraphics[trim={5cm 5cm 5cm 5cm},clip,width=0.8\columnwidth]{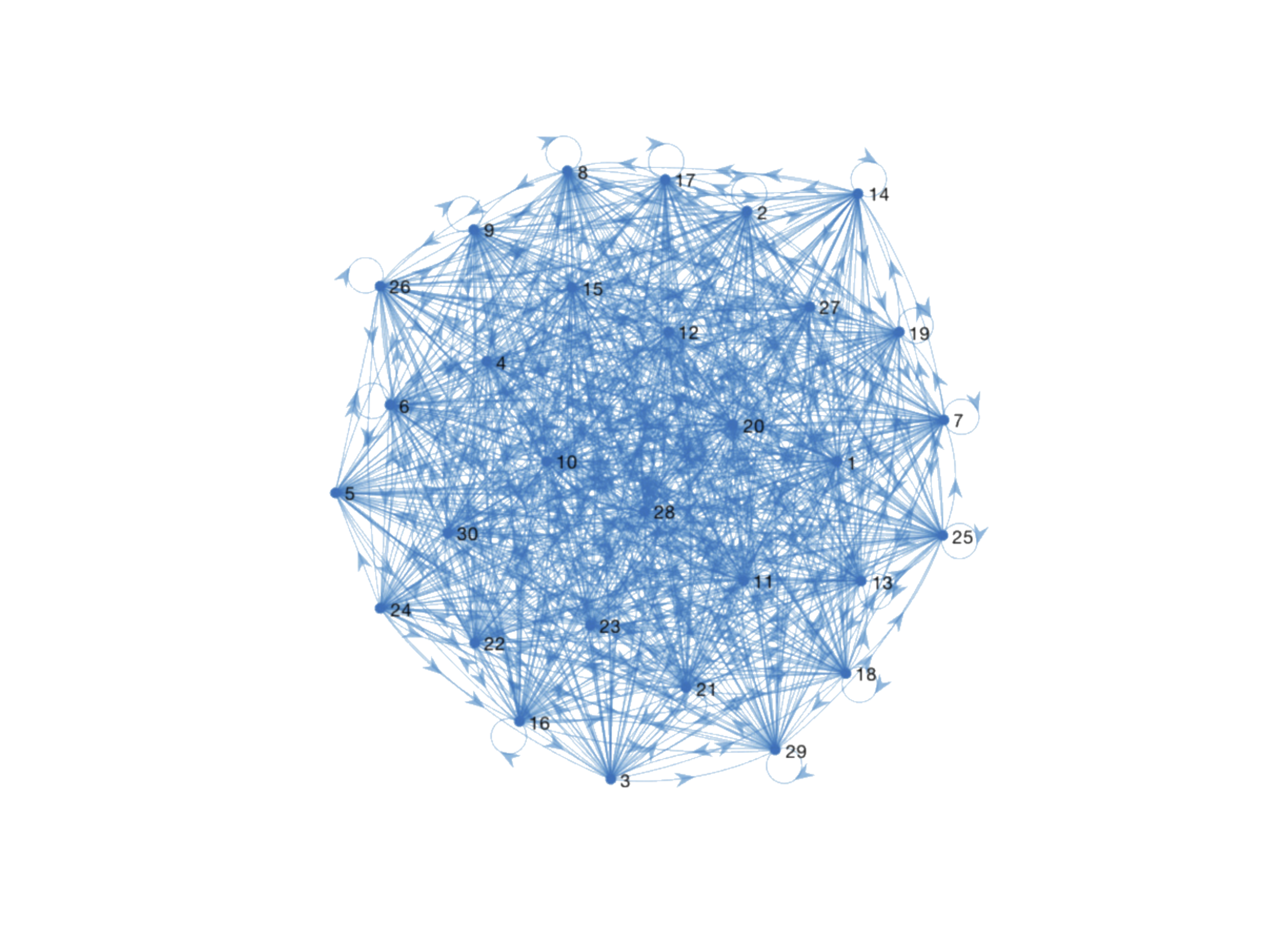}
\end{center}
\begin{center}
\sf Dense network
\end{center}
\end{minipage}
\begin{minipage}{0.49\columnwidth}
\begin{center}
\includegraphics[trim={5cm 5cm 5cm 5cm},clip,width=0.8\columnwidth]{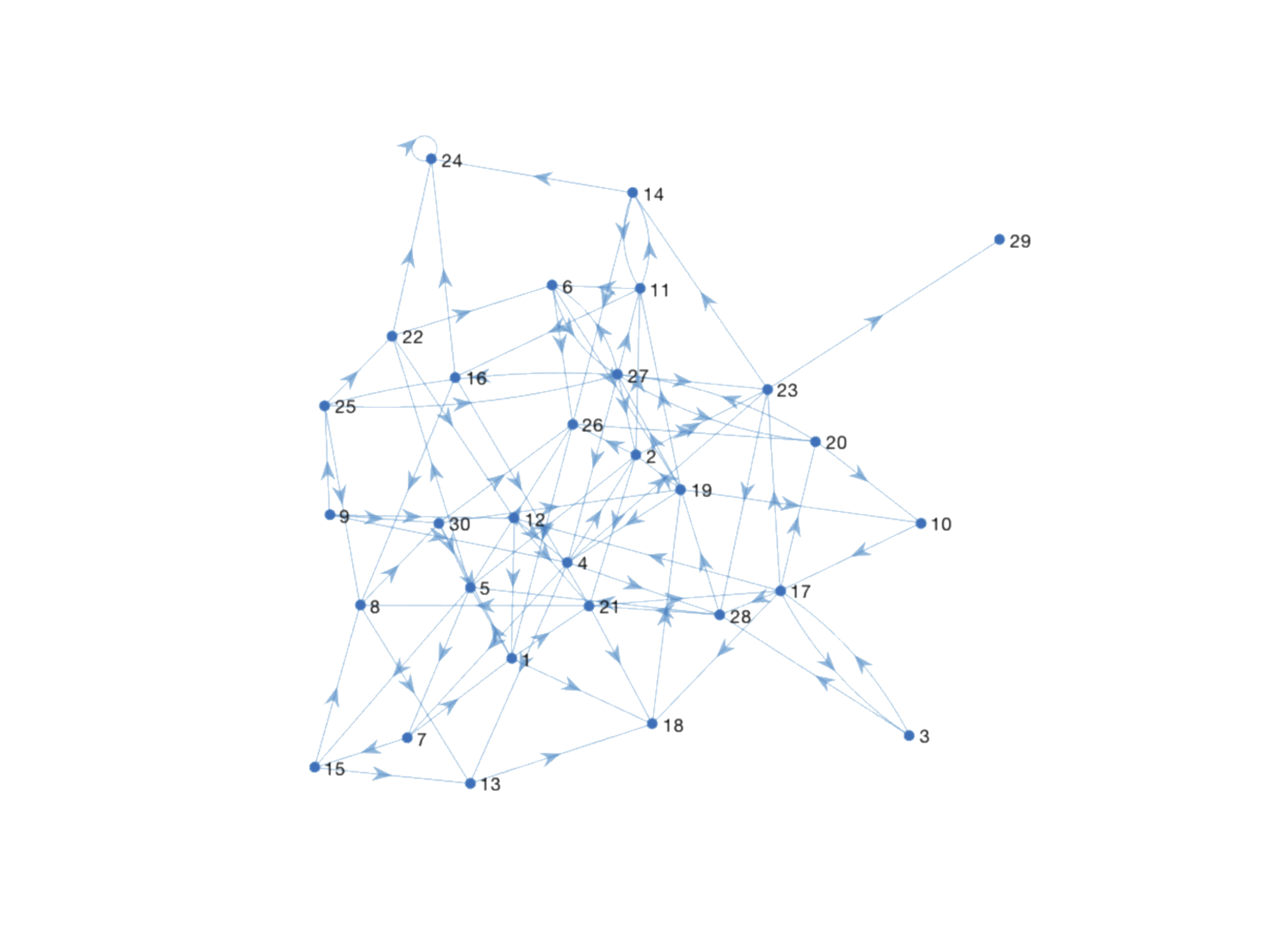}
\end{center}
\begin{center}
\sf Sparse network
\end{center}
\end{minipage}
\medskip
\begin{center}
{ \em{"Most of real social networks are sparse" [Stanford Large Network Dataset Collection]}}
\end{center}
\medskip
\begin{center}
\sf\small{\begin{tabular}{lllll}
Name	& Type	& Nodes &	 Edges	& Network density\\
\hline\\
ego-Facebook&	Undirected&	$\sf4039$	& $\sf88234$	&			$\sf5,408\cdot10^{-3}$\\
ego-Gplus	&Directed	&$\sf107614$&	$\sf13673453$		&			$\sf1,180\cdot10^{-3}$\\
ego-Twitter	&Directed	&$\sf81306$	&$\sf1768149$	&	$\sf2,674\cdot10^{-4}$\\
soc-Epinions1&Directed	&$\sf75879$	&$\sf508837$		&		$\sf8,837\cdot10^{-5}$\\
soc-LiveJournal1	&Directed	&$\sf4847571$&	$\sf68993773$	&			$\sf2,936\cdot10^{-6}$\\
soc-Pokec	 & Directed &	$\sf1632803$	&$\sf30622564$	&			$\sf1,148\cdot10^{-5}$\\

soc-Slashdot0922 &	Directed	&$\sf82168$&	$\sf948464$	&			$\sf1,404\cdot10^{-4}$\\
wiki-Vote &	Directed	&$\sf7115$	&$\sf103689$	 		&		$\sf2,048\cdot10^{-3}$\\
gemsec-deezer	&Undirected	&$\sf143884$	&$\sf846915$	&				$\sf4,090\cdot10^{-5}$\\

gemsec-facebook	&Undirected	&$\sf134833$&	$\sf1380293$	&				$\sf7,592\cdot10^{-5}$\\

\end{tabular}}
\end{center}
\end{tcolorbox}

\begin{multicols}{2}

The distinction of dense and sparse graphs depends on the context. The index with which sparsity is commonly measured in network graphs is edge density\cite{jdmdh:77}.
We will consider the following asymptotic definition of sparsity.
We say that a graph is {\em sparse} if the number of edges is not larger than a quantity that scales linearly in the number of nodes, i.e. $|\E|\leq \alpha n$ with $\alpha \in(0,1)$.
There are other metrics to define sparsity, as the generalization of the Gini Index \cite{Gini} for networks. We refer to \cite{DBLP:journals/isci/GoswamiMD18} for an overview of the sparsity definitions adapted for networks.

Another {important statistical characteristics is} the {in- and out-degree distribution (see the box~\boxref{Centrality measures in weighted graphs})}. \ant{If the in-degree of a node is small compared to the network size}, then the corresponding row in the influence matrix $\Wb$ is \emph{sparse}, \ant{and contains}  few non-zero {entries} (see the box \boxref{Sparse models}).

Many real world networks exhibit power-law degree distributions \cite{Newman:2003}. {Remarkably, such a distribution has been discovered in the early works on sociometry~\cite{MorenoJennings:1938}. The} fraction of nodes with degree $k$ \ant{is distributed as}
\begin{equation}\label{eq:power-law}
\ant{p_{\rm deg}(k)\sim k^{-\gamma}.}
\end{equation}
for some exponent $\gamma>1$ and minimum degree $k_{\min}$.
Networks with power-law distributions are called ``scale-free" because power laws have the same functional form at all scales, i.e.  the power law $p_{\rm deg}(k)$  remains unchanged (other than a multiplicative factor) when rescaling the independent variable $k$, as it satisfies $p_{\rm deg}(\alpha k)=\alpha^{-\gamma}p_{\rm deg}(k)$.

In \cite{ARNABOLDI201626}, the structural properties of Facebook Ego-networks are analyzed. Ego-networks are well studied, as they capture local information about network structure from the perspective of a vertex. The Ego-network of a focal node, called Ego, is defined as the subgraph induced over nodes that are directly connected to it, but excluding the ego itself. Keep in mind that since the ego node is removed from the network, an ego network can be disconnected.
In the box \boxrefb{Degree distribution in Facebook Ego-Networks} it is shown the normalized degree distribution of three Ego-Networks \cite{leskovec2016snap}. As can be seen, some of them are more ``concentrated'' around a mean value, while some others show a power-law decay with smaller exponent $\gamma$.
As discussed in the next section, this concentration property actually plays a crucial role in the inference of trust network from few data.
\end{multicols}
\bigskip

\begin{tcolorbox}[title= \textbf{Degree distribution in Facebook Ego-Networks},colframe=airforcecarmine!10,
colback=airforcecarmine!10,
coltitle=black,
breakable,
fontupper=\sffamily,
fonttitle=\sffamily
]

\begin{minipage}{0.49\columnwidth}
\textbf{Ego-networks} analysis represents a common tool for the investigation of the relationships between individuals and their peers in online social networks \cite{Boldrini:2018}. Moreover, the structural properties of ego networks are shown to be correlated to many aspects of the human social behavior, such as willingness to cooperate and share resources \cite{Arnaboldi:2012}.
\medskip

The Ego-network of a focal node is defined as the subgraph induced over nodes that are directly connected to it, but excluding the ego itself.
\medskip

In this figure the empirical degree distribution of three Facebook Ego-Networks retrieved from the Stanford Network Database \cite{leskovec2016snap} is depicted. It should be noticed that
some of the degree distributions are more concentrated around a mean value, while  others show a power-law decay.
The tails of the distribution are well approximated by \eqref{eq:power-law} with $\gamma\in[1.2,3].$

\end{minipage}
\begin{minipage}{0.49\columnwidth}
\begin{center}
  \includegraphics[trim={0cm 5cm 1cm 5cm},width=0.97\columnwidth]{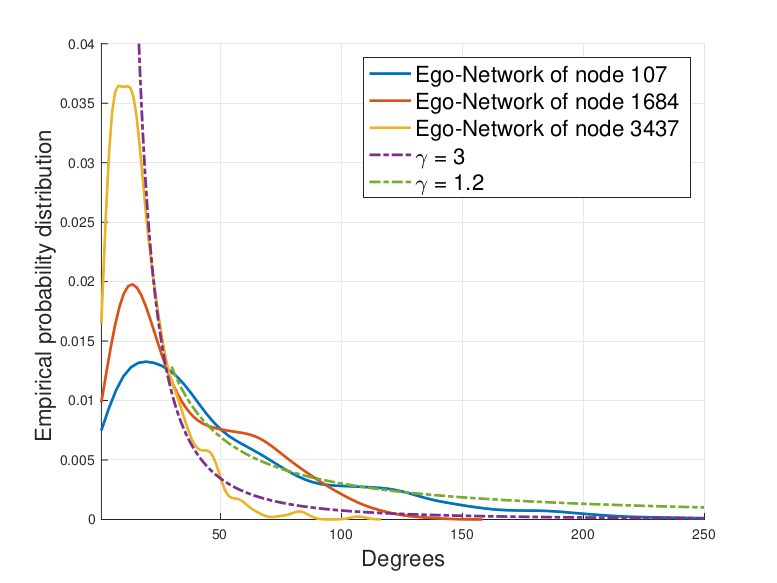}
\end{center}
\end{minipage}
\end{tcolorbox}
\bigskip

\begin{multicols}{2}

Examining the Twitter networks of 14 Destination marketing organizations,
the work~\cite{doi:10.1108/JHTT-09-2016-0057} reveals the presence of few ``leaders'' with high out-degrees, that is, a sparse structure of the influence matrix with few dense columns.

Finally, other networks exhibit the presence of few clusters \cite{ARNABOLDI201626}, i.e. a community of individuals with dense friendship patterns internally and sparse friendships externally.
This inherent tendency to cluster is measured by the {\em average clustering} coefficient \cite{journals/socnet/OpsahlP09}.
These type of networks are described by an influence matrix that can be decomposed as a sum of a low-rank matrix and a sparse matrix.

To better fix some ideas, in the box \boxref{Sparse models} we show different examples that summarize how the notion of sparsity can be exploited for social networks analysis.

\end{multicols}
\newpage
\begin{tcolorbox}[breakable,title= \sf \textbf{\sf \textbf{Sparse models}},colframe=carmine!10,colback=carmine!10,coltitle=black,
]
{\sf\small{Sparse models to represent high-dimensional data have been used in several areas, such as in statistics, signal and image processing, machine learning, coding and control theory \cite{Rish14}.
Intuitively, we say that data are sparse or compressible if they are so highly correlated that only a few degrees of freedom compared to their ambient dimension are significant.
This general definition leads many possible interpretations and alternative measures of sparsity  can be defined according to the data and applications.
\medskip

The simplest definition is the sparsity in the elements. We say that a signal is sparse if the number of non zeros, or significantly different from zero, are few compared to the signal dimension. Moreover, we say that a signal $x\in\R^n$ is $k$-sparse if $\|x\|_0\doteq |\{i\in\{1,\ldots, n\}:\ x_i\neq 0\}|\leq k$ with $k\ll n.$
\bigskip

In social networks analysis the notion of sparsity can be exploited in several ways.
We show here some examples to easily fix some ideas.
\medskip

\textsc{Few friends}. If, from sociological perspective, an agent is influenced from few friends, then the in-degree is low compared to the size of the network. As a consequence, the corresponding adjacency matrix is sparse, i.e. with few non-zero elements. In the following figures a typical sparse adjacency matrix is depicted together with the signal obtained stacking the matrix by columns. It should be noticed that only a small portion of elements is different from zero.
\medskip

\begin{minipage}{0.4\columnwidth}
\begin{center}
  \includegraphics[trim={4.5cm 10cm 4.5cm 8cm},width =0.8\columnwidth]{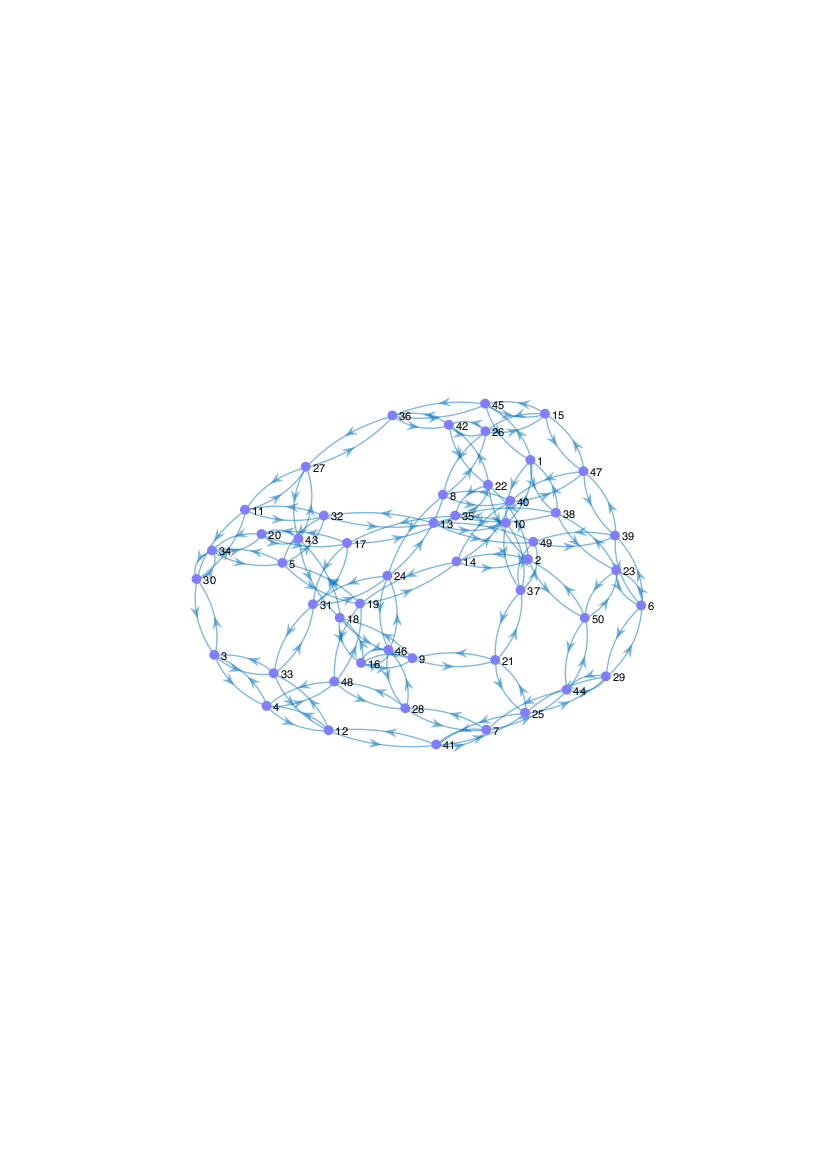}
\end{center}
\begin{center}
\sf\small Few friends
\end{center}
\end{minipage}
\bigskip
\begin{minipage}{0.6\columnwidth}
\begin{center}
  \includegraphics[trim={4.5cm 7cm 4.5cm 7cm},width =0.4\columnwidth]{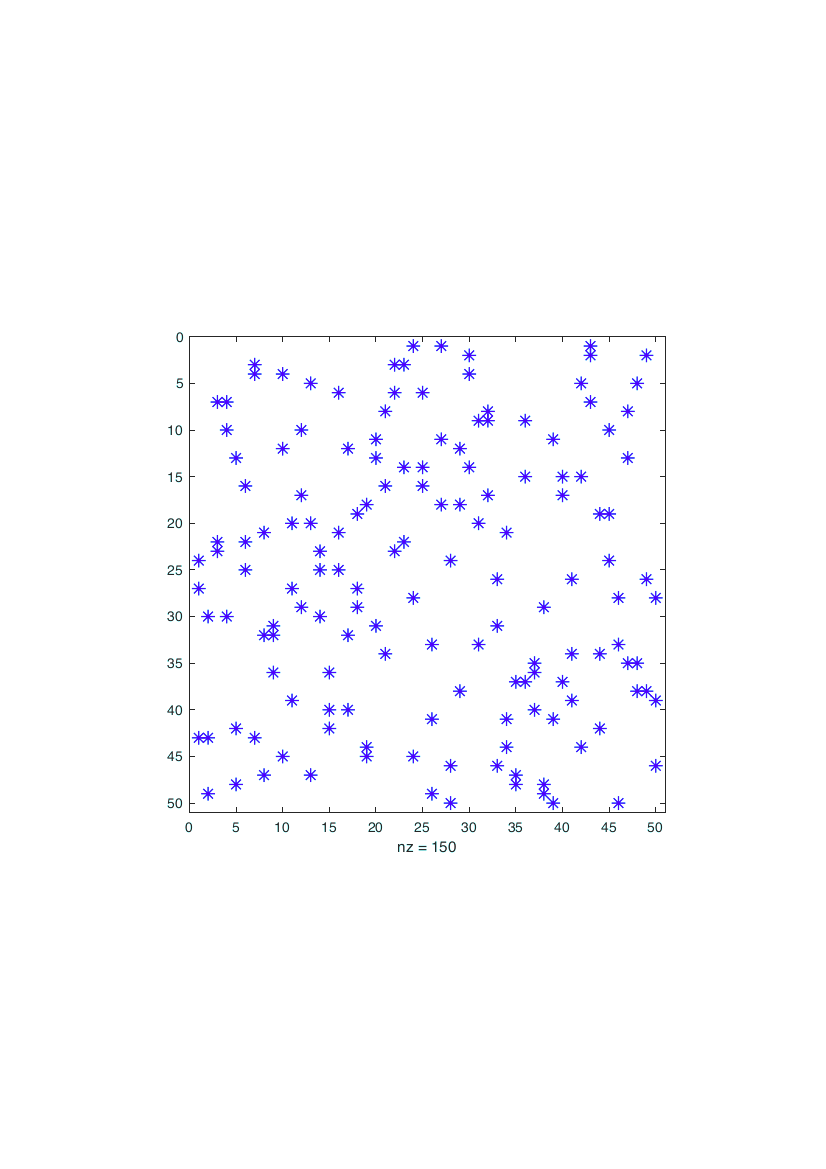}
\end{center}
\begin{center}
\sf\small Sparsity in the elements of adjacency matrix
\end{center}
\end{minipage}

\textsc{Few influencers}. If the networks contains some few leaders, i.e. few individuals influencing many people in the network, then the adjacency matrix will exhibit a sparse structure with few dense columns.
The adjacency matrix of a network with 5 influencers is shown in the following figure on the left. On the right the elements of the matrix, stacked by columns, are shown. It should be noticed that the signal is sparse with few dense patterns. In literature this feature is also known with the name of Block-sparsity.
\medskip

\begin{minipage}{0.4\columnwidth}
\begin{center}
  \includegraphics[trim={4.5cm 10cm 4.5cm 8cm},width =0.8\columnwidth]{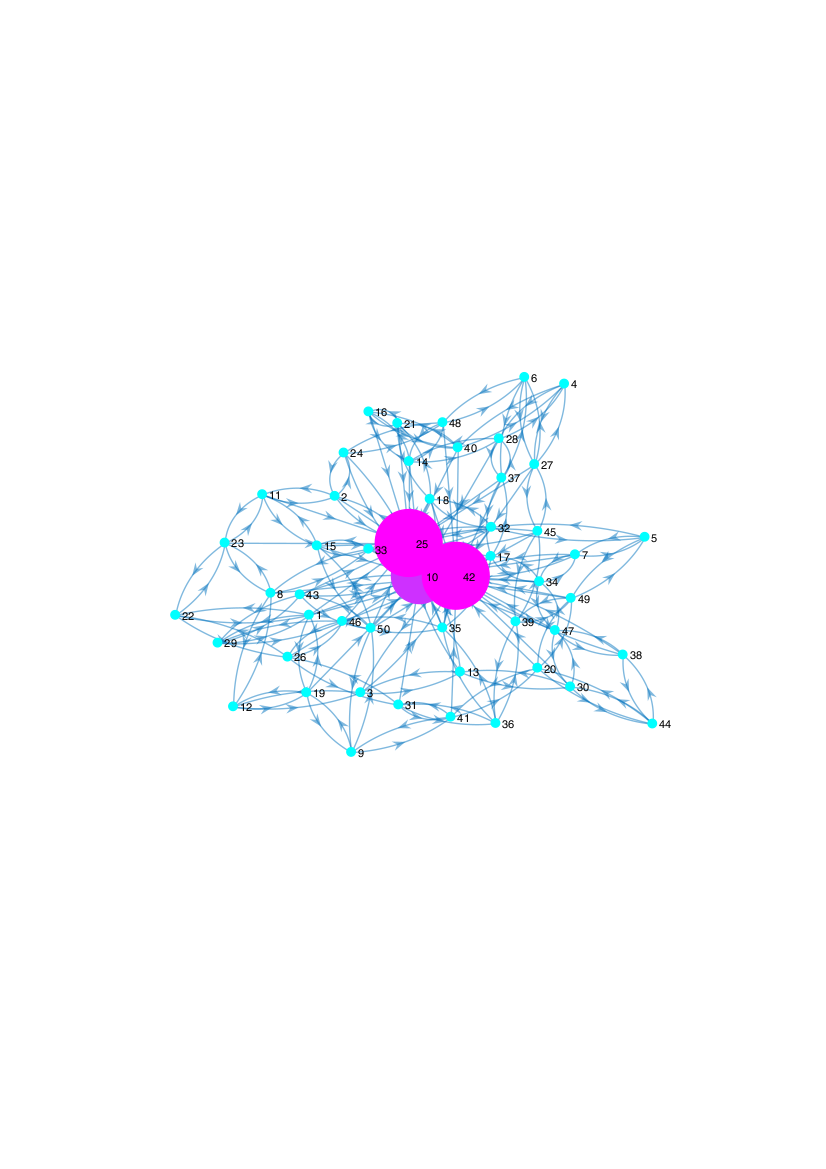}
\end{center}
\begin{center}
\sf\small Few influencers
\end{center}
\end{minipage}
\bigskip
\begin{minipage}{0.6\columnwidth}
\begin{center}
  \includegraphics[trim={4.5cm 7cm 4.5cm 7cm},width =0.4\columnwidth]{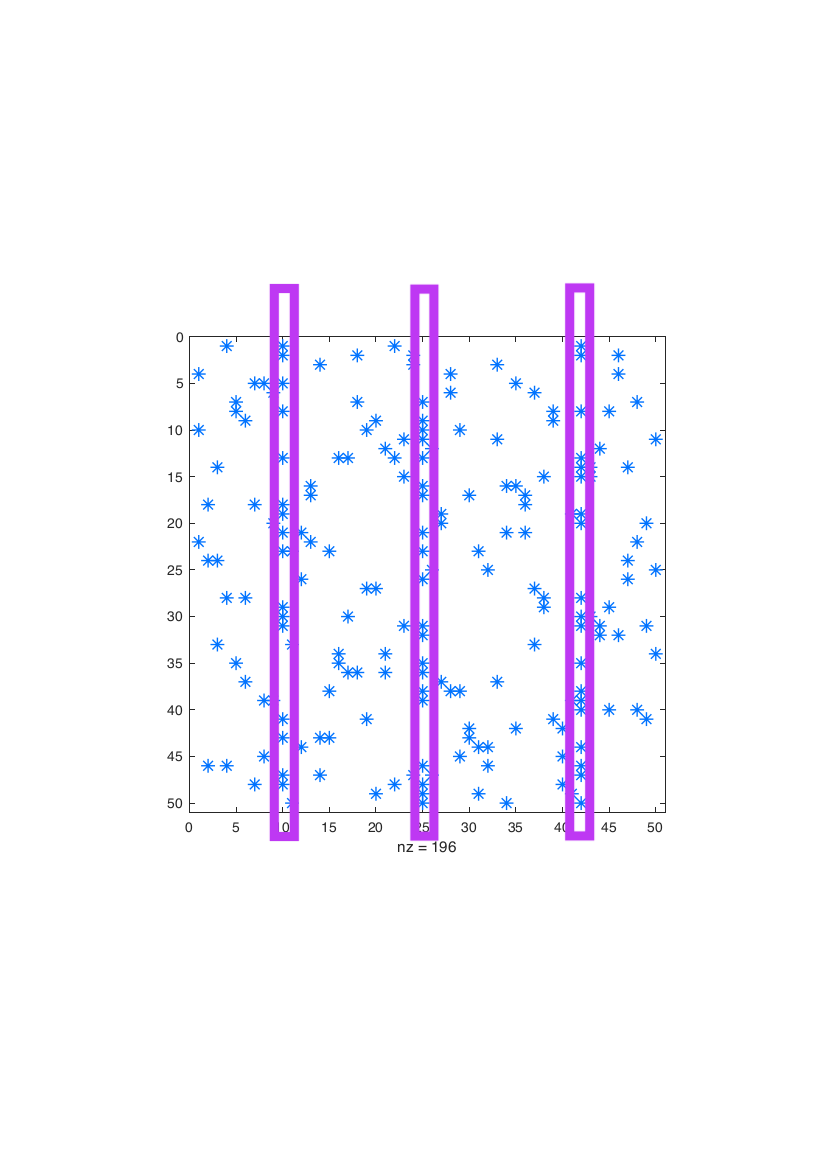}
\end{center}
\begin{center}
\sf\small Block-sparsity in the adjacency matrix
\end{center}
\end{minipage}

\textsc{Few communities}. Finally, other networks show the presence of few communities, i.e. a set of individuals with dense friendship patterns internally and sparse friendships externally. For these type of networks the adjacency matrix that can be decomposed as
$
A=L+S
$, where $L$ is a low-rank matrix and $S$ is a sparse matrix.
In the following figures a typical adjacency matrix of a network with few communities is shown (on the left) and the corresponding eigenvalues in absolute values (on the right). It should be noticed that the eigenvalues are highly compressible and only 4 eigenvalues, even to the number of communities, contain the most energy of the signal.
\medskip

\begin{minipage}{0.4\columnwidth}
\begin{center}
  \includegraphics[trim={4.5cm 10cm 4.5cm 9cm},width=0.99\columnwidth]{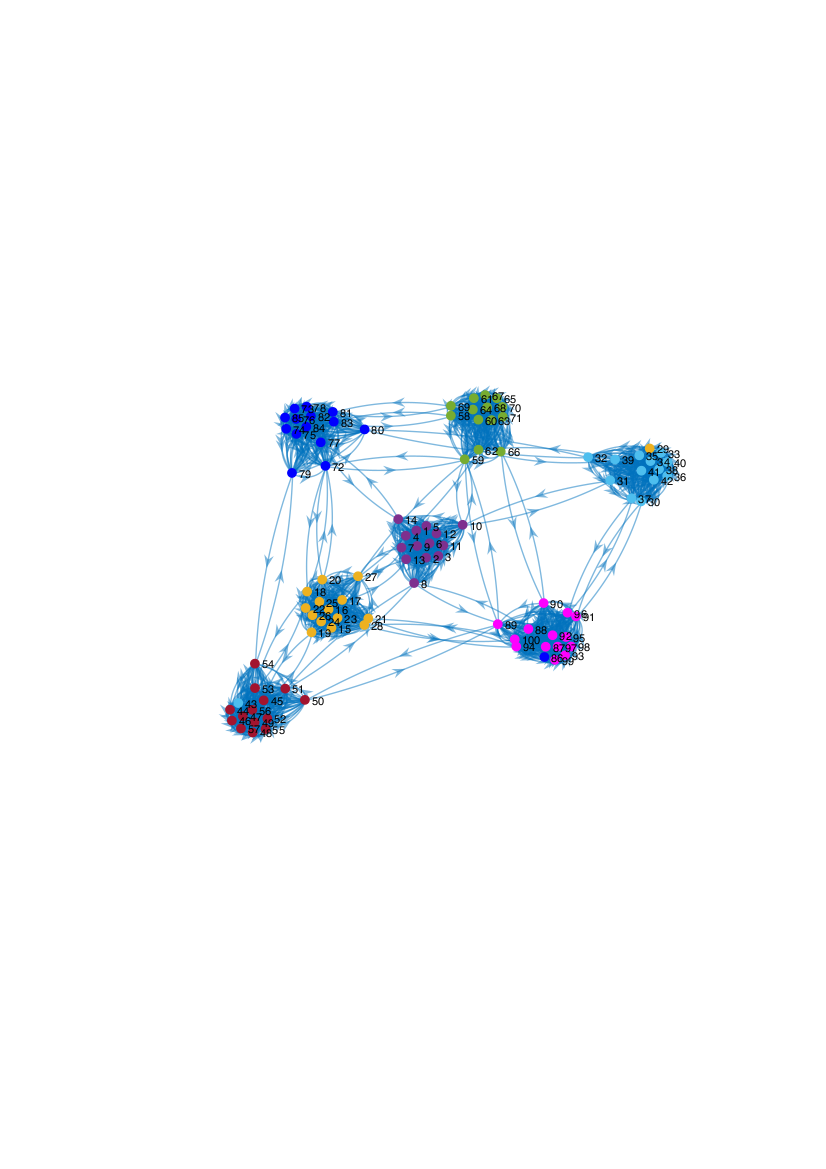}
  \end{center}
\begin{center}
\sf\small Few communities
\end{center}
\end{minipage}
\begin{minipage}{0.6\columnwidth}
\begin{center}
  \includegraphics[trim={4.5cm 7cm 4.5cm 7cm},width=0.43\columnwidth]{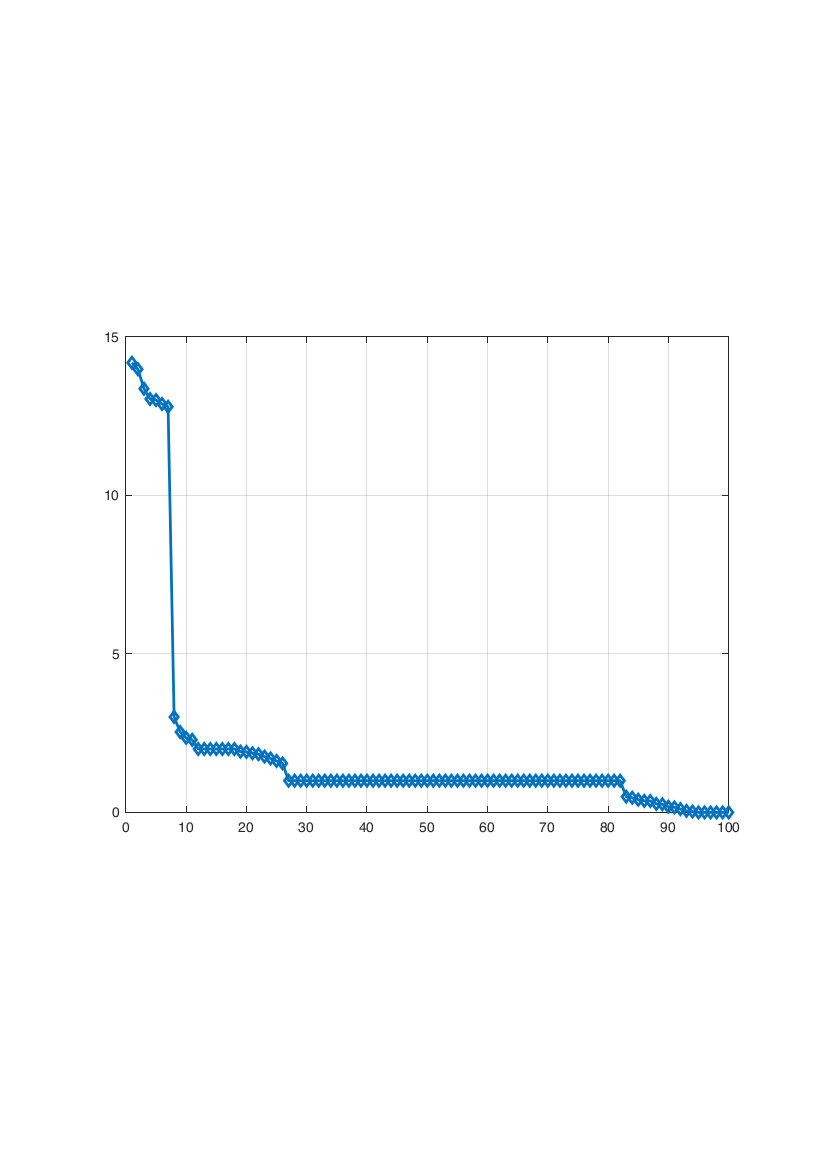}
  \end{center}
\begin{center}
\sf\small Sparsity in the eigenvalues of adjacency matrix
\end{center}
\end{minipage}

%
%
%

}}
        \end{tcolorbox}

\begin{multicols}{2}

\subsection{Sparse models for Multidimensional Networks}\label{sec-hetero}
{\em Multidimensional networks} describe different types of relations among nodes. For example, friendships in a social network may arise for various reasons, e.g. because users are colleagues, 
\ant{teammates in sports or share some other hobbies.}
In this case we consider multiplex networks, where we can distinguish different \ant{layers of
interconnections that correspond}
to \ant{different types of relations}. 
Another example in real life is when a social group discusses several issues in parallel. For example, Twitter is comprised of microblogs and users express opinions on \textit{different topics} and the influence between the users is \emph{topic-dependent} (such networks are sometimes called \emph{heterogeneous}~\cite{Liu2012}). Each layer in multilayer network takes into account the influence \ant{relations arising among people when they discuss a certain subject}.
The analysis of multiplex networks is an active field of research (see \cite{10.5555/3125944} and reference therein).

In fact, if the underlying social network is essentially composed by the same individuals, then we will expect that the social systems share a common feature. The above intuition entails that the networks describing the micro-level mechanisms of social influence with respect to topics are not completely independent. 
It follows that, besides the sparsity model describing the degrees of freedom of each network, the model must be augmented by taking into account the correlations of the networks relative to the different topics.
In this sense, in \cite{Coluccia2015RegularizedCM} two correlated models are introduced.

The first model, \textit{the common component model} ($\Mc_\textrm{cc}$), takes into account the cases where the  networks relative to the different topics $\ell=1,\ldots,n_{\ell}$, only differ for few edges. In this case, all influence matrices share a common \ant{base}, and contain an innovation. \ant{Formally, the} influence matrix $\Wb^{(\ell)}$ describing the interaction network relative to the $\ell$-th topic \ant{is decomposed as}
\begin{equation}\label{eq: M1}
\Wb^{(\ell)}=\bar{\Wb}+\boldsymbol{\delta} \Wb^{(\ell)},
\end{equation}
where the matrices $\bar \Wb$ and $\boldsymbol{\delta} \Wb^{(\ell)}$, representing respectively the common part and the innovation part, are both sparse (see examples in the box \boxrefb{Multidimensional networks}).

The second model, the \textit{common support model} ($\Mc_\textrm{cs}$) instead, describes situations where the topology is equal for all the different topics but the weights are different. This is captured by a model in which all transition matrices share a common support $\Omega \subseteq \{1,\ldots,n\} \times \{1,\ldots, n\}$, i.e.
\begin{equation}\label{eq:M2}
\Wb^{(\ell)}_{ij}\neq0\iff  (i,j)\in\Omega, \forall \ell\in \{1,\ldots, m\}.
\end{equation}
An example of this model can be found in deliberative groups that deal with a sequence of issues, such as department faculties in universities, Boards of Directors in large organizations, to mention just a few. \ant{Empirical findings show}~\cite{nef-pj-fb:14n} that the weights evolve according to a natural social process, known as reflected appraisal \cite{cooley1922human, Fried11_reflected_appraisal}. \ant{The model of this process proposed in~\cite{nef-pj-fb:14n} is squarely based on the Friedkin-Johnsen model of opinion formation and will be considered in Section~\ref{sec-models}.}

\medskip

Summarizing, any efficient technique for social media modeling, analysis and optimization must take into account
the large size of the networks and exploit the notion of sparsity as a structural constraint.

From the previous discussion, we observe that the key ingredient for performing  a social influence analysis is the knowledge of influence matrix $\Wb$.
\ant{In the next sections, we focus on algorithms inferring matrix $\Wb$ and related computational aspects and consider two different approaches. The ``static'' method deals with the} inference of matrix $\Wb$ from samples of some observables~$\{x_j\}_{j\in\V}$ \ant{whereas the ``dynamic'' approach deals with identification of opinion formation models}.
\end{multicols}


\begin{tcolorbox}[breakable,title= \sf \textbf{\sf \textbf{Multidimensional networks}},colframe=airforcecarmine!20,colback=airforcecarmine!20,coltitle=black,
]
{\sf\small{

Multidimensional (\ant{multiplex, multi-layer}) networks allow to distinguish among different kinds of \ant{links between the nodes and naturally arise in social sciences \cite{10.5555/3125944}, economics and finance~\cite{Perillo18}, transportation~\cite{WU202058}, and biology\cite{Halu100370}.}
\medskip

There are multiple ways to define a multidimensional network. For simplicity, we only consider networks denoted by a triple $\G = (\V, \E, \mathcal{L})$, where $\V$ is a set of nodes; $\mathcal{L}$ is the set of layers; $\E=\bigcup_{d\in\mathcal{L} }\E_d$ is the set of edges; and $\E_d$ the set of edges at layer $d$.
Temporal networks are a special type of multiplex networks with explicit dimensions and can be represented as a sequence of graphs, where we see a single dimension as a separate layer.
Therefore, in this new framework it is mandatory: (a) generalizing the centrality measures defined for classical mono-dimensional networks; and (b)
studying the correlations 
among dimensions in order to capture hidden relationships among different layers.

\vspace{0.5cm}
\begin{multicols}{2}
\begin{minipage}{0.95\columnwidth}
\begin{center}
  \includegraphics[trim={1cm 0cm 1cm 0cm},width=0.99\columnwidth]{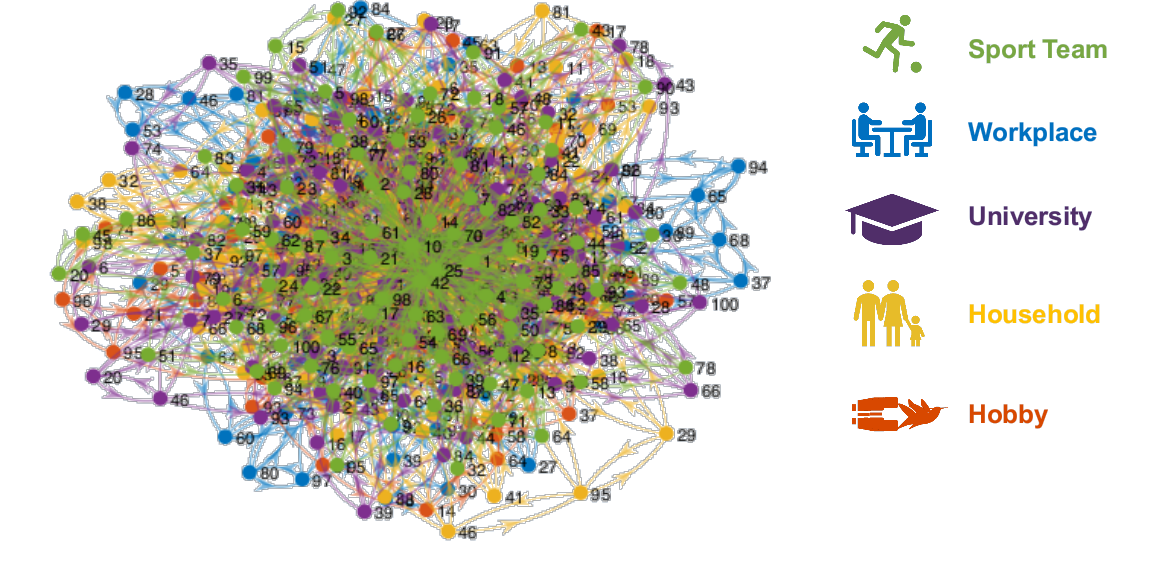}
  \end{center}
\begin{center}

\end{center}
\end{minipage}

\begin{minipage}{0.95\columnwidth}
\bigskip
This picture represents the relations among different individuals. Different colors correspond to different origin of the friendship, e.g. friendships arisen in sport teams, at workplace, at university, etc.
We can consider the single type of networks as separate graphs or the multiplex networks given by the union of all graphs. In this case each layer has been generated independently as an Erd\H{o}s-R\'enyi graph.
 \end{minipage}
 \vspace{0.5cm}

\begin{minipage}{0.95\columnwidth}
\begin{center}
  \includegraphics[trim={0cm 0cm 0cm 0cm},width=0.9\columnwidth]{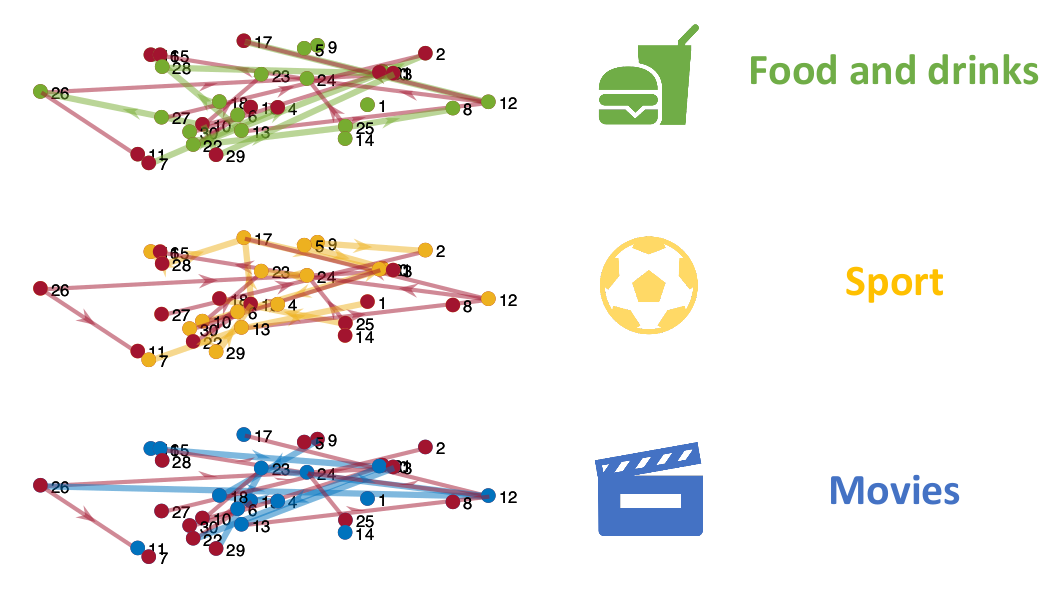}
  \end{center}
  \vspace{0.05cm}
\begin{center}
\end{center}
In this example we consider a 3-dimensional networks. The network is constituted by different layers and represent the influence network of a community based on the topic under discussion, e.g. Food and drinks, Sport and Movies. Each layer is constituted by a sparse common component (the subgraph in red color) and a sparse innovation component (the subgraph in orange, purple, light blue color, respectively).
\end{minipage}
\end{multicols}

\begin{minipage}{0.95\columnwidth}
\bigskip
In \cite{Berlingerio13} a measure is defined in order to quantify how much two dimensions are similar.
These measures can be seen as an extension of the classical Jaccard correlation coefficient in order to cope with more than two sets. Let us consider an $\mathcal{L}$-dimensional network.
Let $\mathcal{D} \subseteq \mathcal{L}$ be a set dimensions of a network $\G = (\V, \E, \mathcal{L})$. The Pair $\mathcal{D}$-Correlation is defined as
$$
\rho_{\mathcal{D}}=\frac{|\bigcap_{d\in\mathcal{D}}\mathcal{E}_d|}{|\bigcup_{d\in\mathcal{D}}\mathcal{E}_d|}.
$$
In the example
$\rho_{\mathcal{{\{\sf{M},\sf{S}\}}}}=0.3636$,  $\rho_{\mathcal{{\{\sf{M},\sf{F\&D}\}}}}=0.3871$ and $\rho_{\mathcal{{\{\sf{S},\sf{FD}\}}}}=0.3333$.

 \end{minipage}

}}
        \end{tcolorbox}


\begin{multicols}{2}

\section{Learning graphs from data} \label{sec:graphID}

The influence  network estimation problem discussed in this paper represents a special instance of the general problem of reconstructing the graph topology from  data measured on the nodes. This problem, known under the name of \textit{graph learning} or \textit{network inference}, has seen an increasing interest in the past years. We refer the reader to the excellent survey \cite{Frossard-learning}, on which this section is largely based.

In the literature, we can distinguish several approaches for the influence network estimation. These mainly  depend on  the assumptions on the networks under observation and on the available data. We categorize the methods into three classes: i) Statistical models, ii) Learning models for Social Similarity and Influence, and iii) Model-based approach.
Most methods deal with undirected graphs and non-dynamical (static) variables, and the extensions to directed and/or dynamically varying topologies usually turn out to be rather complex. For this reason, in this section we will mostly focus on the simpler case of static undirected graphs.

\subsection{Statistical models}

Statistical models presume the availability 
of $N$ measurements (usually scalar) at each node $i\in\V$
\[
x_i(1),\ldots,x_i(N), \quad i\in\V.
\]
The main idea behind statistical models is to interpret the observed data as independent realizations of random variables $\{x_i\}, i\in\V$ whose  joint probability distribution is determined in some way by the topology of the graph $\mathcal{G}$. Hence, a connection between two nodes translates into a
statistical correlation between the signals at those nodes.

In particular, one introduces the so-called \textit{probabilistic graphical models} \cite{jordan2004,Koller:2009:PGM:1795555}, in which data are interpreted as multiple outcomes of random experiments, and a graphical model is introduced to capture the conditional dependence between random variables.
In the simple case of undirected graphs and continuous variables, the most popular models proposed in the literature are \textit{Markov random fields} (MRF) \cite{Rue:2005:GMR:1051482}
and \textit{Gaussian graphical models} \cite{Wainwright:2008:GME:1498840.1498841}. Given a graph $\G=\{\V,\E\}$, Markov random fields are postulated by requiring that the random variables at the different nodes satisfy a series of local Markov properties: of particular interest is the so-called \textit{pairwise Markov property}, which states that two variables are conditionally independent given the all other variables if and only if they are not connected by an edge. That is, we can write
 \[
x_i \bigCI x_j \mid \{x_{w}\}_{w\in\V\setminus\{i,j\}}\iff (i,j)\notin\E
\]
where the notation $\bigCI$ denotes statistical independence, that is it holds that
\[
p(x_i |\{x_j\}_{j\in\V\setminus i})
=p(x_i |\{x_j\}_{j\in\N_i}).
\]

It can be shown \cite{Wainwright:2008:GME:1498840.1498841} that this property is guaranteed by the exponential family of distributions
of the form
\[
p(\xb|\boldsymbol{\Theta}) = \frac{1}{Z(\boldsymbol{\Theta})} \exp \left(\sum_{v\in\V} \theta_{ii}x_j^2 + \sum_{(i,j)\in\E} \theta_{ij}x_i x_j
\right)
\]
where $ \xb\doteq[x_1,\ldots,x_n]^\top$ is a collection of the random variables across nodes,
$\boldsymbol{\Theta}=[\theta_{ij}]$ is a parameter matrix, and
 $Z(\boldsymbol{\Theta})$ is a normalization constant.
 Conditional independence between $x_i$ and $x_j$ translates into $\theta_{ij}=0$. In other words, the parameter matrix  $\boldsymbol{\Theta}$ is adapted to the graph.
This class goes under the name of \textit{exponential random graphs}, or $p^*$ models. In the literature, estimation schemes for such graphs, based on Monte Carlo maximum likelihood estimation have been proposed, see e.g.~\cite{Robins2006}.
As observed in  \cite{Giannakis2018}, these methods naturally extend 
the classical statistical approach based on estimating the \textit{partial} (Pearson)  correlation coefficient starting
from the observations $\xb$.

A commonly adopted assumption in exponential random graphs stipulates that the observations are realizations of the multivariate Gaussian distribution
\[
p(\xb|\boldsymbol{\Theta}) = \frac{\det(\boldsymbol{\Theta})^{1/2}}{(2\pi)^{N/2}} \exp \left(-\frac{1}{2} \xb^\top \boldsymbol{\Theta} \xb\right),
\]
where $\boldsymbol{\Theta}$ represents the so-called precision matrix, i.e.\ the inverse of the covariance matrix, i.e.~$\boldsymbol{\Theta}=\boldsymbol{\Sigma}^{-1}$. This leads to the family of Gaussian graphical models. It can be observed that in this case the existence of a non-zero entry in the precision matrix, immediately implies a partial correlation between the corresponding random variables. The goal then becomes to estimate the precision matrix from the observed data $\{x_j\}_{j\in\V}$.
To this end, in the literature several procedures have been proposed for computing the Maximum-Likelihood (ML) estimator via a log-determinant program. In this class of algorithms, the so-called  Graphical Lasso (G-Lasso) \cite{mazumder2012} method has become extremely popular, see 
the box \boxrefb{Gaussian graphical models and G-Lasso}.

We observe that, although the convergence of G-Lasso is guaranteed under suitable conditions, this method has some drawbacks. First, the whole procedure only works in the case of undirected networks. Second, in many contexts, as for instance  in the opinion formation processes discussed in this article, data are the result of a dynamic process. This situation is not well captured by the Graphical Lasso framework, since for these problems data can be highly correlated leading to a dense precision matrix. Finally, the sample covariance matrix may fail to have full rank due to the lack 
of observed data, giving rise to numerical problems in the identification of the network.

It is worth emphasizing that the estimation of the precision matrix via Graphical models does not allow a direct interpretation of social influence but is able only to reflect pairwise correlation between opinions in the social system. The estimation of social influence, however, is primarily aimed at
predicting a direct causal effect of this influence.

\end{multicols}
\newpage
\begin{tcolorbox}[breakable,title= \sf \textbf{\sf \textbf{Gaussian graphical models and G-Lasso}},colframe=airforcecarmine!20,colback=airforcecarmine!20,coltitle=black,
]
{{\sf \small
Graphical models are graphs capturing the relationships between many variables,  providing a compact representation of joint probability distributions. In these models the nodes correspond to random variables and edges represent statistical dependencies between pairs of them.

\medskip

\indent
In GGM, the variables at each node are Normally distributed, $\mathbf{x}\sim\mathcal{N}(\bold{0},\boldsymbol{\Sigma}_\xb)$, and for any $i$ and $j\in\V$ a zero in the $i,j$ entry of the precision matrix
means conditional independence (given all other variables):
$$x_i \bigCI x_j \mid \{x_{w}\}_{w\in\V\setminus\{i,j\}}\iff \theta_{ij}=0\quad \boldsymbol\Theta=\boldsymbol{\boldsymbol{\Sigma}_\xb}^{-1}.$$
}}
\bigskip
         \begin{center}\begin{tikzpicture} [scale=.4,every node/.style={circle,fill=white!20,inner sep=6pt}]
  \node (n1) at (-3,8) {};
  \node (n2) at (-5,6)   {};
  \node (n3) at (2,5) {};
  \node (n4) at (0,4) {};
  \node (n5) at (-7,9)   {};
  \node (n6) at (-8,2)   {};
  \node (n7) at (-6,4)   {};
  \node (n8) at (-3,1)   {};
  \node (n9) at (-1,6)   {};

    \tikzset{every node/.style={}}
  \node (n10) at (15,6)   {\footnotesize{$\Theta=\left[\begin{array}{ccccccccc}
  0&0&0&0&\star&0&0&0&\star\\
  0&0&0&\star&\star&0&\star&0&0\\
  0&0&0&\star&0&0&0&0&\star\\
  0&\star&\star&0&0&\star&0&0&0\\
  \star&\star&0&0&0&0&0&0&0\\
  0&0&0&\star&0&0&\star&\star&0\\
  0&\star&0&0&0&\star&0&\star&0\\
  0&0&0&0&0&\star&\star&0&0\\
  \star&0&\star&0&0&0&0&0&0\\
  \end{array}\right]
  $}};
 \node (n22) at (16,1)   {$ (i,j)\notin\mathcal{E}\Longrightarrow \theta_{ij}=0
$};

    \foreach \from/\to in {n2/n5,n1/n5,n2/n7,n2/n4,n3/n4, n9/n3, n6/n8, n6/n4, n8/n7, n7/n6, n9/n1}
    \path (\from) edge[-,bend right=3] (\to);
     \tikzset{mystyle/.style={-}}

  \node (n11) at (-3,8) {\footnotesize$x_1$};
  \node (n12) at (-5,6)   {\footnotesize$x_2$};
  \node (n13) at (2,5)  {\footnotesize$x_3$};
   \node (n14) at (0,4)  {\footnotesize$x_4$};
  \node (n15) at (-7,9)    {\footnotesize$x_5$};
  \node (n16) at (-8,2)   {\footnotesize$x_6$};
  \node (n17) at (-6,4)   {\footnotesize$x_7$};
  \node (n18) at (-3,1)    {\footnotesize$x_8$};
  \node (n19) at (-1,6)  {\footnotesize$x_9$};

\end{tikzpicture}
\end{center}

\par\medskip

{{\sf \small Consider $N$ observations $\{\xb(1),\xb(2),\ldots, \xb(N)\}$ from multivariate Gaussian distribution,
we are interested in estimating the precision matrix $\Theta=\boldsymbol{\boldsymbol{\Sigma}}^{-1}$.
The classical Maximum Likelihood (ML) estimator is obtained by solving the following optimization problem
\[
\hat{\boldsymbol{\Theta}}_{\text{ML}}=  \max_{\Theta\succeq0}\mathrm{log}\ \mathrm{det}(\boldsymbol{\Theta})-\mathrm{tr(\Sb\boldsymbol{\Theta})}\qquad \text{with}\quad
 \Sb=\frac{1}{n}\sum_{k=1}^n\xb(k)\xb(k)^{\top}.
\]
Classical theory guarantees that in the high-dimensional regime $\hat{\boldsymbol{\Theta}}_{\text{ML}}$ converges to the truth as sample size $N\rightarrow\infty $.\\

In practice, we are often in the regime where sample size $N$ is small compared to the dimension $n$. Therefore, $\Sb$ is not full rank and ML estimation problem does not admit a unique solution.
The main approach in these cases is to assume that many pairs of variables are conditionally independent, i.e. many links are missing in the graphical model or, equivalently, $\boldsymbol{\Theta}$ \textit{is sparse}.
The key idea in Graphical-lasso \cite{friedman_sparse_2008} is to apply lasso by treating each node as a response variable and solving the following convex program
\[
\hat{\boldsymbol{\Theta}}_{\text{G-lasso}}=  \max_{\boldsymbol{\Theta}\succeq0}\mathrm{log}\ \mathrm{det}(\boldsymbol{\Theta})-\langle \Sb,\boldsymbol{\Theta}\rangle-\rho\|\boldsymbol{\Theta}\|_1
\]
where $\rho$ tunes the number of zero entries in $\boldsymbol{\Theta}$.
}}
\end{tcolorbox}

\begin{multicols}{2}
\subsection{Graph Signal Processing}

Recent years have witnessed a growing interest of the signal processing community in analysis of signals that are supported on the vertex set of weighted graphs, leading to the field of Graph Signal Processing (GSP), \cite{journals/spm/ShumanNFOV13}. By generalizing classical signal processing concepts and tools, GSP enables the processing and analysis of signals that lie on structured but irregular domains.
In particular, GSP allows to re-define concepts as such as Fourier transform, filtering and frequency response for data residing on graphs.

Note that the signals in the graph are not time-dependent: they instead vary spatially, and their spatial dynamics are governed by the underlying graph. A brief overview of GSP is provided in the box \boxrefb{Graph signal processing}.

\end{multicols}

\begin{tcolorbox}[title= \sf \textbf{\sf \textbf{Graph signal processing}},colframe=airforcecarmine!20,colback=airforcecarmine!20,coltitle=black,
]
{\sf \small
The rapidly growing field of {graph signal processing} (GSP) provides tools to represent signals that are supported on the vertices of a graph.
A \textit{graph signal} is defined as a mapping $\xb:\V\to \mathbb{R}^n$, from the vertices of the graph to the real numbers. It can be represented as a vector $\xb\in\mathbb{R}^n$,
where $x_i$ stores the value of the signal on  the $i$-th vertex.\\

A simple way to understand the basics of  GSP is to consider how the classical concept of shift operator is extended to graph signals.
First observe that a periodic discrete-time signal can be represented by a circular directed unweighted graph, in which the $k$-th node represents the value of the signal $x$ at the discrete-time instant $k$. The next figure represent a signal $\xb$ (left) and its shifted version
$\xb^+$ (right), which follow the classical relationship $\xb^+={\textsc{Shift}}(\xb)\doteq\Sb\xb$ defined as follows
\begin{eqnarray*}
x^+_i &=& x_{i-1}, \quad i= 2,\ldots,N\\
x^+_1 &=& x_{N},
\end{eqnarray*}
where the last equation follows from the \textit{circular shift} assumption.
Note that, in this case, the shift operator $\Sb$ coincides with the adjacency matrix $\Adj$ of the directed graph.

  \begin{center}\begin{tikzpicture} [scale=.4,every node/.style={circle,fill=white!20,inner sep=6pt}]

  \node (c1) at (0,4) {};
  \node (c2) at (-3,7)   {};
  \node (c3) at (-7,8) {};
  \node (c4) at (-10,7) {};
  \node (c5) at (-13,4)   {};
 \node (c6) at (-11,1)   {};
  \node (c7) at (-8,0)   {};
  \node (c8) at (-5,0)   {};
  \node (c9) at (-2,1)   {};
  \tikzset{every node/.style={}}
  \node (c11) at (0,4) {\footnotesize$x_1$};
  \node (c12) at (-3,7)   {\footnotesize$x_2$};
  \node (c13) at (-7,8)  {\footnotesize$x_3$};
   \node (c14) at (-10,7)  {\footnotesize$x_4$};
  \node (c15) at (-13,4)    {\footnotesize$x_5$};
  \node (c16) at (-11,1)   {\footnotesize$x_6$};
  \node (c17) at (-8,0)   {\footnotesize$x_7$};
  \node (c18) at (-5,0)    {\footnotesize$x_8$};
  \node (c19) at (.-2,1)  {\footnotesize$x_9$};
   \foreach \from/\to in {c1/c2,c2/c3,c3/c4,c4/c5,c5/c6,c6/c7,c7/c8,c8/c9,c9/c1}
    \path (\from) edge[<-,bend right=3] (\to);
     \tikzset{mystyle/.style={-}}

 \tikzset{every node/.style={circle,fill=carmine!90,inner sep=2pt}}
 \node (sc1) at (0,6) {};
  \node (sc2) at (-3,10)   {};
  \node (sc3) at (-7,12) {};
  \node (sc4) at (-10,10) {};
 \node (sc5) at (-13,7)   {};
  \node (sc7) at (-8,-1.5)   {};
  \node (sc8) at (-5,-3)   {};
  \node (sc9) at (-2,-1)   {};
\foreach \from/\to in {c1/sc1,c2/sc2,c3/sc3,c4/sc4,c5/sc5,c7/sc7,c8/sc8,c9/sc9}
    \path (\from) edge[draw=carmine!90] (\to);
     \tikzset{mystyle/.style={-}}


 \tikzset{every node/.style={circle,fill=white!20,inner sep=6pt}}
  \node (rc1) at (17,4) {};
  \node (rc2) at (14,7)   {};
  \node (rc3) at (10,8) {};
  \node (rc4) at (7,7) {};
  \node (rc5) at (4,4)   {};
  \node (rc6) at (6,1)   {};
  \node (rc7) at (9,0)   {};
  \node (rc8) at (12,0)   {};
  \node (rc9) at (15,1)   {};
  \tikzset{every node/.style={}}
  \node (rc11) at (17,4) {\footnotesize$x_1^+$};
  \node (rc12) at (14,7)   {\footnotesize$x_2^+$};
  \node (rc13) at (10,8)  {\footnotesize$x_3^+$};
   \node (rc14) at (7,7)  {\footnotesize$x_4^+$};
  \node (rc15) at (4,4)    {\footnotesize$x_5^+$};
  \node (rc16) at (6,1)   {\footnotesize$x_6^+$};
  \node (rc17) at (9,0)   {\footnotesize$x_7^+$};
  \node (rc18) at (12,0)    {\footnotesize$x_8^+$};
  \node (rc19) at (15,1)  {\footnotesize$x_9^+$};
   \foreach \from/\to in {rc1/rc2,rc2/rc3,rc3/rc4,rc4/rc5,rc5/rc6,rc6/rc7,rc7/rc8,rc8/rc9,rc9/rc1}
    \path (\from) edge[<-,bend right=3] (\to);
     \tikzset{mystyle/.style={-}}


 \tikzset{every node/.style={circle,fill=carmine!90,inner sep=2pt}}
 \node (src1) at (17,8) {};
  \node (src2) at (14,12)   {};
  \node (src3) at (10,10) {};
  \node (src4) at (7,10) {};
 \node (src6) at (6,-0.5)   {};
  \node (src7) at (9,-3)   {};
  \node (src8) at (12,-1)   {};
  \node (src9) at (15,3)   {};
\foreach \from/\to in {rc1/src1,rc2/src2,rc3/src3,rc4/src4,rc6/src6,rc7/src7,rc8/src8,rc9/src9}
    \path (\from) edge[draw=carmine!90] (\to);
     \tikzset{mystyle/.style={-}}

\end{tikzpicture}
\end{center}
\centerline{a) A periodic signal on a directed graph $\xb=x_1,x_2;\ldots,x_8,x_9$ and b) its shifted version.}

\vskip .3cm
\[
\xb^+=\Sb \xb, \qquad
\Sb=\left[\begin{array}{ccccccccc}
  0&1&0&0&0&0&0&0&0\\
  0&0&1&0&0&0&0&0&0\\
  0&0&0&1&0&0&0&0&0\\
  0&0&0&0&1&0&0&0&0\\
  0&0&0&0&0&1&0&0&0\\
  0&0&0&0&0&0&1&0&0\\
  0&0&0&0&0&0&0&1&0\\
   0&0&0&0&0&0&0&0&1\\
  1&0&0&0&0&0&0&0&0\\
  \end{array}\right]
\]

This parallelism between shift operators and graphs may be extended to general graphs. In particular, in GSP, given a graph $\G$, a \textbf{graph shift operator} is defined as a  matrix $\Sb\in\mathbb{R}^{n,n}$ \textit{adapted to the graph}, and the shift operation is given by $\Sb\xb$. Different choices of $\Sb$ define different shifts.

For undirected graphs, the most typical choice of graph shift operator is the Laplacian $\Lb$: for any graph signal $\xb$, one defines the new signal $\xb^+=\Sb\xb=\Lb\xb$,
whose element $x^+_u$ is given by
\[
x^+_u = [\Lb\xb]_u = \sum_{v\in\N_u} w_{ij}(x_i-x_j).
\]
From the above formulation, it can be easily observed that the Laplacian acts as a difference operator on graph signals.

         \begin{center}\begin{tikzpicture} [scale=.4,every node/.style={circle,fill=white!20,inner sep=6pt}]
  \node (n1) at (10,8) {};
  \node (n2) at (8,6)   {};
  \node (n3) at (15,5) {};
  \node (n4) at (13,4) {};
  \node (n5) at (6,9)   {};
  \node (n6) at (5,2)   {};
  \node (n7) at (7,4)   {};
  \node (n8) at (10,1)   {};
  \node (n9) at (11,6)   {};

     \tikzset{every node/.style={}}

    \foreach \from/\to in {n2/n5,n1/n5,n2/n7,n2/n4,n3/n4, n9/n3, n6/n8, n6/n4, n8/n7, n7/n6, n9/n1}
    \path[->] (\from) edge[-,bend right=3] (\to);
     \tikzset{mystyle/.style={-}}

  \node (n11) at (10,8) {\footnotesize$x_1$};
  \node (n12) at (8,6)   {\footnotesize$x_2$};
  \node (n13) at (15,5)  {\footnotesize$x_3$};
   \node (n14) at (13,4)  {\footnotesize$x_4$};
  \node (n15) at (6,9)    {\footnotesize$x_5$};
  \node (n16) at (5,2)   {\footnotesize$x_6$};
  \node (n17) at (7,4)   {\footnotesize$x_7$};
  \node (n18) at (10,1)    {\footnotesize$x_8$};
  \node (n19) at (11,6)  {\footnotesize$x_9$};

 \tikzset{every node/.style={circle,fill=carmine!90,inner sep=2pt}}
 \node (sn1) at (10,11) {};
  \node (sn2) at (8,7.5)   {};
  \node (sn3) at (15,3.5) {};
  \node (sn4) at (13,2.5) {};
 \node (sn5) at (6,11)   {};
  \node (sn6) at (5,0)   {};
  \node (sn7) at (7,5.5)   {};
  \node (sn8) at (10,-2)   {};
  \node (sn9) at (11,8.5)   {};

\foreach \from/\to in {n1/sn1,n2/sn2,n3/sn3,n4/sn4,n5/sn5,n6/sn6,n7/sn7,n8/sn8,n9/sn9}
    \path (\from) edge[draw=carmine!90] (\to);
     \tikzset{mystyle/.style={-}}

         \tikzset{every node/.style={}}
  \node (n100) at (29,6)   {\footnotesize{$
\Sb=\Lb=\left[\begin{array}{ccccccccc}
  2&0&0&0&-\tfrac{1}{2}&0&0&0&-\tfrac{1}{2}\\
  0&3&0&-\tfrac{1}{3}&-\tfrac{1}{3}&0&-\tfrac{1}{3}&0&0\\
  0&0&2&-\tfrac{1}{2}&0&0&0&0&-\tfrac{1}{2}\\
  0&-\tfrac{1}{3}&-\tfrac{1}{3}&3&0&-\tfrac{1}{3}&0&0&0\\
  -\tfrac{1}{2}&-\tfrac{1}{2}&0&0&2&0&0&0&0\\
  0&0&0&-\tfrac{1}{3}&0&3&-\tfrac{1}{3}&-\tfrac{1}{3}&0\\
  0&-\tfrac{1}{3}&0&0&0&-\tfrac{1}{3}&3&-\tfrac{1}{3}&0\\
  0&0&0&0&0&-\tfrac{1}{2}&-\tfrac{1}{2}&2&0\\
  -\tfrac{1}{2}&0&-\tfrac{1}{2}&0&0&0&0&0&2\\
  \end{array}\right]
$}};

\end{tikzpicture}
\end{center}
\centerline{b) A signal defined on a generic \emph{undirected} graph, and the corresponding graph shift operator defined in terms of the Laplacian}

From the definition of graph shift operator the extension of the concept of Fourier transform to graph signal descends almost immediately \cite{6808520}. Indeed, for the choice $\Sb=\Lb$, if one considers the eigenvalue decomposition of $\Lb=\Lb^{\top}$
\[
\Lb =\Ub  \boldsymbol{\Lambda} \Ub^\top, \quad \boldsymbol{\Lambda}=\mathrm{diag}(\lambda_1,\ldots,\lambda_n), \quad \Ub=[\ub_1\cdots \ub_n]
\]
where $\Vb$ is the eigenvector matrix, i.e.\ the matrix containing the eigenvectors of $\Lb$ as columns (which are orthonormal, being $\Lb$ symmetric) and
$\lambda_i$-s are the eigenvalues, which are real and ordered, with $0<\lambda_2\le \cdots \le \lambda_n$.
Then, the  \textbf{Graph Fourier Transform} (GFT) associated to the Laplacian may then be defined as
\[
\tilde{x}_k \doteq  \ub_k^\top \xb = \sum_{j=1}^n x_j [\ub_{k}]_j.
\]

It should be noted that the Laplacian-based GFT only works for undirected graphs. Extensions to directed graphs are
\ant{nontrivial, since} the GFT definition does not cover situations where $\Lb$ has complex eigenvalues or is not diagonalizable.
}
\end{tcolorbox}

\begin{multicols}{2}

While the main directions of research in GSP focus on the development of methods for analyzing signals defined over given \textit{known}
graphs, the inverse problem has also been considered concerned with
\emph{learning} the graph topology from measurements of the signals on the graph.
The existing mathematical results adopt
specific assumptions on the characteristics of 
the graph Fourier transform.

In particular, the most common approach for GSP-based graph topology reconstruction is based on the assumption that the underlying graph signal is \textit{smooth} on the graph. That is, the links in the graph should be chosen in such a way that  signals on neighboring nodes are close to each other.
As a measure of  smoothness of the signal $\xb$ on the graph $\G$ the so-called Laplacian quadratic form is usually adopted, see e.g.~\cite{Frossard-smooth} and references therein
\begin{equation}
\label{QL}
\xb^\top\Lb\xb = \frac{1}{2} \sum_{i,j} w_{ij} (x_i -x_j)^2
\end{equation}
Several approaches have been proposed in the literature for learning a graph (or, in this case, its Laplacian matrix $\Lb$) such that the Laplacian quadratic cost \eqref{QL} is small, that is the signal variations on the resulting graph is small. The reader is referred to \cite{Frossard-learning} for a detailed overview of this approach
whose central step is to solve the
an optimization problem
\[
\min_{\Lb,\yb} \|\xb-\yb\|_2^2 + \alpha \yb^\top\Lb\yb.
\]
The first term enforces data fidelity, the second one enforces smoothness of the signal.
This approach is extended in subsequent works, see e.g. \cite{Frossard-smooth,Chepuri2017}, by adding additional constraints on the Laplacian $\Lb$, allowing to fix the volume of the graph and to impose specific structures on the graph.

Other  GSP-base approaches for deriving topological characteristics of the graph are based on graph signal measurement
and assume that the graph signals are generated by applying a graph-filtering operation to a latent  signal.
In particular, the  graph signal $\xb$ is assumed to be generated by a \textit{diffusion process} of the form
\[
\xb=\sum_{k=0}^K \alpha_k \Sb^k \ub
\]
where $\Sb$ is a given graph operator (again, usually  the Laplacian matrix $\Lb$) capturing the graph connectivity.
The ensuing algorithm are  hence well-suited for learning graph topologies when the observations are the result of a diffusion process on a graph. The existing methods for reconstruction stem from the observation that, when the ``input" signal  $\ub$ is uncorrelated (white noise), and the graph is undirected, then the eigenvalues of $\Sb$ coincide with the eigenvalues of the covariance matrix $\boldsymbol{\Sigma}_\xb$ of $\xb$. This, in turn, may be approximated via the sample covariance.

Finally, we mention approaches using
spectral graph dictionaries 
for efficient signal representation, see e.g.\ \cite{Frossard-dictionaries} and references therein. In this case,  a graph signal diffusion model is envisioned,
representing data as
(sparse) combinations of overlapping local patterns that reside on the graph.

\subsection{Model based learning of directed and dynamical graphs}

As has been discussed, the large majority of the graph learning approaches available in the literature deal with \textit{undirected} and \textit{stationary} graphs, whereas their extensions to directed graphs meet
serious difficulties. In the case of probabilistic graphical models, for instance, directed graphical models also called Bayesian Networks or Belief Networks (BNs), need the introduction of  a more complicated notion of independence, which takes into account
the asymmetry of interconnections.
In GSP-based techniques, directionality of the graph destroys the symmetry of its operator $\Sb$,
complicating thus
the mere definition of Graph Fourier transform. 

On the other hand, in many contexts, as in the case of social interactions reconstruction which represents the main focus of this work,  learning directed graphs is more desirable, especially for those cases   where the edges  directions translates in  a causal dependencies between the variables that the vertices represent.
In this case, model-based approaches appear more natural. The main assumption is that data are result of a dynamical process, and the problem is cast as an inverse optimization problem exploiting prior information on the model. This research line is related to sparse vector autoregressive (SVAR) estimation \cite{Mei2017,Bolstad2011,Giannakis2018}, inverse optimization from partial samples \cite{Bertsimas2015} and models from opinion dynamics \cite{Wai2016ActiveSO,Ravazzi2018}.
In the next section, we overview the main models introduced in the literature.

\section{Social influence in opinion dynamics} \label{sec:FJ}
\label{sec-models}

The approaches to social influence we have discussed up to now represent a social network as either a weighted graph or a probabilistic graphical model. An alternative approach, leading to {the so-called \textit{social influence network theory}} (SINT)~\cite{friedkin_1998,FriedkinJohnsenBook}, considers a social network as a \emph{dynamical system}. The relevant mathematical models describe diffusion of some information over the network, which can manifest itself as evolution of individual opinions, attitudes of beliefs: the individuals interact (during face-to-face meetings or via social media) and display their opinions on some issues to each other. Based on the opinions displayed to him/her, each individual updates their own opinion on an issue. The within-individual mechanisms of opinion assimilation are related to psychological studies on information integration~\cite{Anderson:81} and cognitive dissonance~\cite{FestingerBook}. Their mathematical models are currently limited to simple opinion update rules, such as iterative averaging. In such simplified models, social influence is naturally represented by \emph{influence weights} an individual assigns to her own and others' opinions. We consider models stemming from the French-Harary-DeGroot's model of iterative opinion pooling.

\end{multicols}
\begin{tcolorbox}[title= \sf \textbf{Original French's model~\cite{French:1956}},colframe=carmine!10,colback=carmine!10,coltitle=black,]
\sf \small
\begin{itemize}
\item a group of $n$ individuals are associated with nodes of a directed graph;
\item individual $i$ holds an \emph{opinion} $x_i$, assumed to be a scalar real value;
\item individual $j$ discloses his/her opinion to individual $i$ if the graph has a directed arc $(j,i)$;
\item individuals know their own opinions, and thus each node in the graph has a self-arc;
\item at each period $k=0,1,\ldots$, an individual updates their opinion to the \emph{mean values} of all opinions displayed to them.
\end{itemize}

\begin{center}
 \includegraphics[width =0.6\columnwidth]{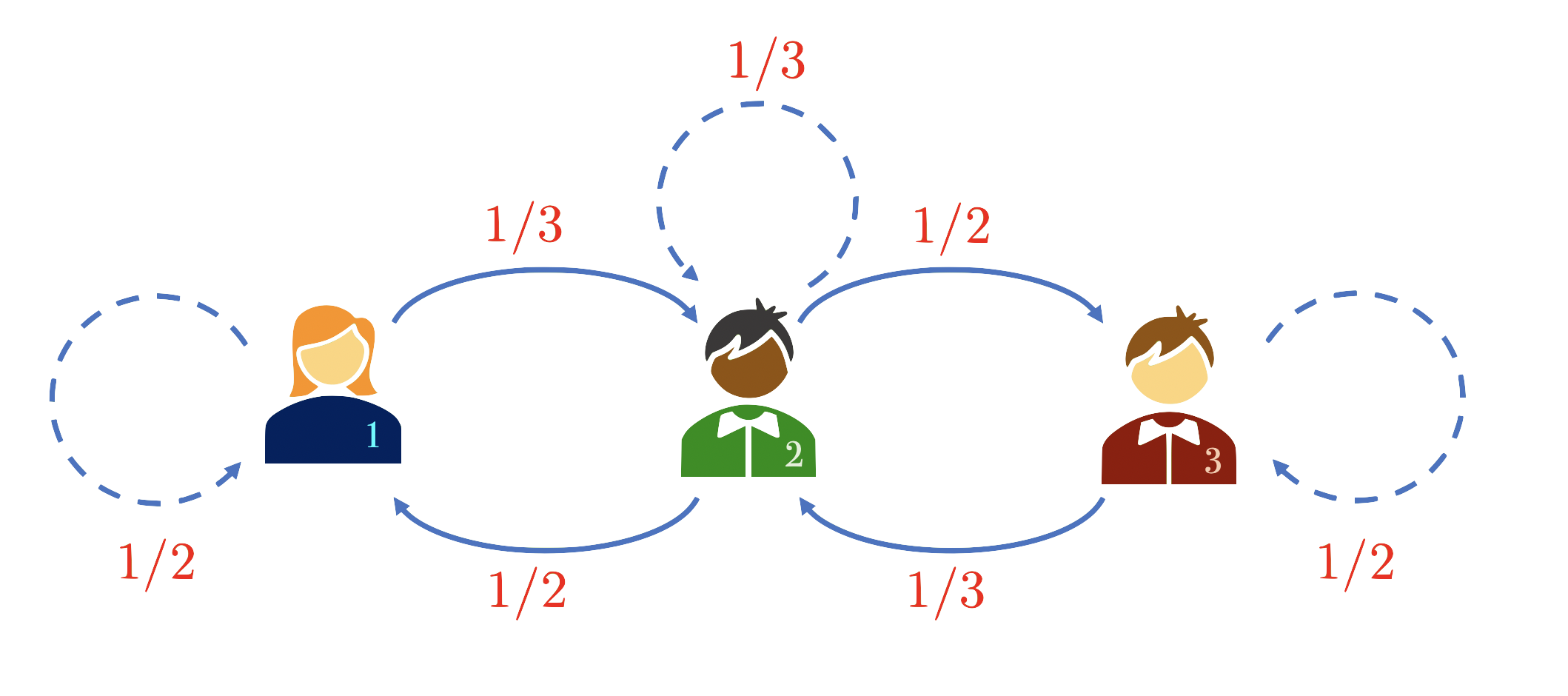}
\end{center}
For instance, the graph shown in the picture induces the opinion formation process
\[
\begin{bmatrix}
x_1(k+1)\\
x_2(k+1)\\
x_3(k+1)
\end{bmatrix}=
\begin{pmatrix}
1/2 & 1/2 & 0\\
1/3 & 1/3 & 1/3\\
0 & 1/2 & 1/2
\end{pmatrix}
\begin{bmatrix}
x_1(k)\\
x_2(k)\\
x_3(k)
\end{bmatrix}
\]

The most typical behavior of the model is eventual consensus (unanimity) of the opinions. Consensus (for an arbitrary initial condition) is achieved if and only if~\cite{Harary:1959,JacksonBook2008,ProTempo:2017-1} the graph has a node from which all other nodes are reachable, that is, there is an agent who influences, directly or indirectly, all other agents.
\end{tcolorbox}

 \begin{multicols}{2}
 \subsection{The French-Harary-DeGroot model}

One of the simplest models of opinion formation has been proposed by French in his seminal work on social power~\cite{French:1956} and later examined by the renowned graph theorist F. Harary~\cite{Harary:1959,HararyBook:1965}, and discussed in the box \boxref{Original French's model}. The most known, however is its generalized version proposed by DeGroot~\cite{Degroot_74} (and, independently, by Lehrer~\cite{Lehrer:1976,LehrerWagnerBook})
\begin{equation}\label{eq.degroot}
\mathbf{x}_i(k+1)=\sum_{j=1}^nw_{ij}\mathbf{x}_j(k),\quad i=1,\ldots,n.
\end{equation}
Here $\mathbf{x}_i(k)$ stands for the opinion of agent $i$ at the $k$th stage of the opinion evolution, and $\Wb=[w_{ij}]$ is a row-stochastic matrix (a non-negative matrix whose rows sum to $1$).

It is remarkable that the work~\cite{Degroot_74}, unlike the pioneer works~\cite{French:1956,Harary:1959}, introduced \emph{multidimensional} opinions. Such opinions can represent individual's positions on several issues, for instance, optimal distribution of resources between several entities~\cite{friedkin2019mathematical} or subjective probability distribution of outcomes in some random experiment~\cite{Scaglione:2013,Degroot_74}.
Unless otherwise stated, we assume the opinions to be \emph{row} vectors
\[
\mathbf{x}_i(k)=[x_{i}^{(1)}(k),\ldots,x_{i}^{(m)}(k)].
\]
It is convenient to stack theses rows one on top of another obtaining thus a $n\times m$ \emph{matrix} of opinions
\begin{equation}\label{eq.opinion-matrix}
\mathbf{X}(k)=
\begin{bmatrix}
\mathbf{x}_1(k)\\
\vdots\\
\mathbf{x}_n(k)
\end{bmatrix}=[\mathbf{x}^{(1)}(k),\ldots,\mathbf{x}^{(m)}(k)]\in\mathbb{R}^{n\times m}.
\end{equation}
The $\ell$-th column of this matrix $\mathbf{x}^{(\ell)}(k)=(x_1^{(\ell)},\ldots,x_n^{(\ell)})^{\top}$ contains the actors' positions on issue $\ell=1,\ldots,m$.

DeGroot's model is then rewritten in the matrix form
\begin{equation}\label{eq.degroot-m}
\mathbf{X}(k+1)=\mathbf{W}\mathbf{X}(k),\quad k=0,1,\ldots.
\end{equation}

According to the DeGroot model, at each stage of the opinion iteration individuals update their opinions to convex combinations of all opinions disclosed to them. This update is simultaneous. The weights $w_{ij}$ of this convex combination serve as natural measures of mutual \emph{influences} among individuals~\cite{Friedkin:1986,Friedkin:2015}.
Social influence can be thought of as a finite resource individuals distribute between themselves and their peers (this is modeled as a distribution of chips in the Friedkin-Johnsen experiment, see p.~\pageref{exp.chips})

The weight $w_{ij}\ge 0$ assigned by agent $i$ to another agent~$j$ measures importance of $j$'s opinion for $i$. If $w_{ij}=1$ (maximal value), agent $i$ fully relies on agent $j$'s opinion on the issue and is insensitive to the opinions of the others
\[
w_{ij}=1\Longleftrightarrow \mathbf{x}_i(k+1)=\mathbf{x}_j(k).
\]
An individual assigning the maximal weight $w_{ii}=1$ to self is often called \emph{stubborn} (radical), being completely closed to social influence and keeping their opinion unchanged
\begin{equation}\label{eq.stubborn}
\mathbf{x}_i(k+1)=\mathbf{x}_i(k)=\ldots=\mathbf{x}_i(0).
\end{equation}
If $w_{ij}=0$, the opinion of agent $j$ is either not disclosed to agent $i$ or does not taken into account by him/her.
Mathematically, agent $j$'s opinion at step $k$ does not influence the opinion of agent $i$ at the consecutive step $k+1$, however, it can \emph{indirectly} influence $i$'s opinions at the subsequent steps $k+2,k+3,\ldots$ through the opinions of other individuals (a chain of influence $j\xrightarrow{} j'\xrightarrow{} j''\xrightarrow{}\ldots \xrightarrow{}i$).
\end{multicols}
\begin{tcolorbox}[title= \sf \textbf{DeGroot's model as a dynamics over a graph},colframe=carmine!10,colback=carmine!10,coltitle=black,]
\sf \small
Social network $\leftrightarrow$ Weighted graph
$\mathcal{G}=(\mathcal{V},\mathcal{E},\Wb)$  \begin{itemize}
\item agents $\leftrightarrow$   $v\in \mathcal{V}$
\item interactions $\leftrightarrow$  $\mathcal{E}\subseteq\mathcal{V}\times \mathcal{V}$
\item influences $\leftrightarrow$  $\Wb\in\mathbb{R}^{\mathcal{V}\times \mathcal{V}}$
\item $w_{ij}=0$ if $(i,j)\notin\mathcal{E}$
\item opinions on issue $\ell$ $\leftrightarrow$
 $x^{(\ell)}_v(k)\in \mathbb{R}$
\item the \emph{row} vectors of multidimensional opinions $\mathbf{x}_i(k)=(x_i^1(k),\ldots,x_i^m(k))$ obey~\eqref{eq.degroot};
\item the vectors of positions on each issue $\mathbf{x}^{(\ell)}(k)=(x_1^{(\ell)}(k),\ldots,x_n^{(\ell)}(k))^{\top}$ evolve as 
$\mathbf{x}^{(\ell)}(k+1)=\Wb\mathbf{x}^{(\ell)}(k)$;
\item the matrix of opinions~\eqref{eq.opinion-matrix} evolves in accordance with~\eqref{eq.degroot-m}.
\end{itemize}
\medskip
 \begin{center}
\includegraphics[width=0.75\columnwidth]{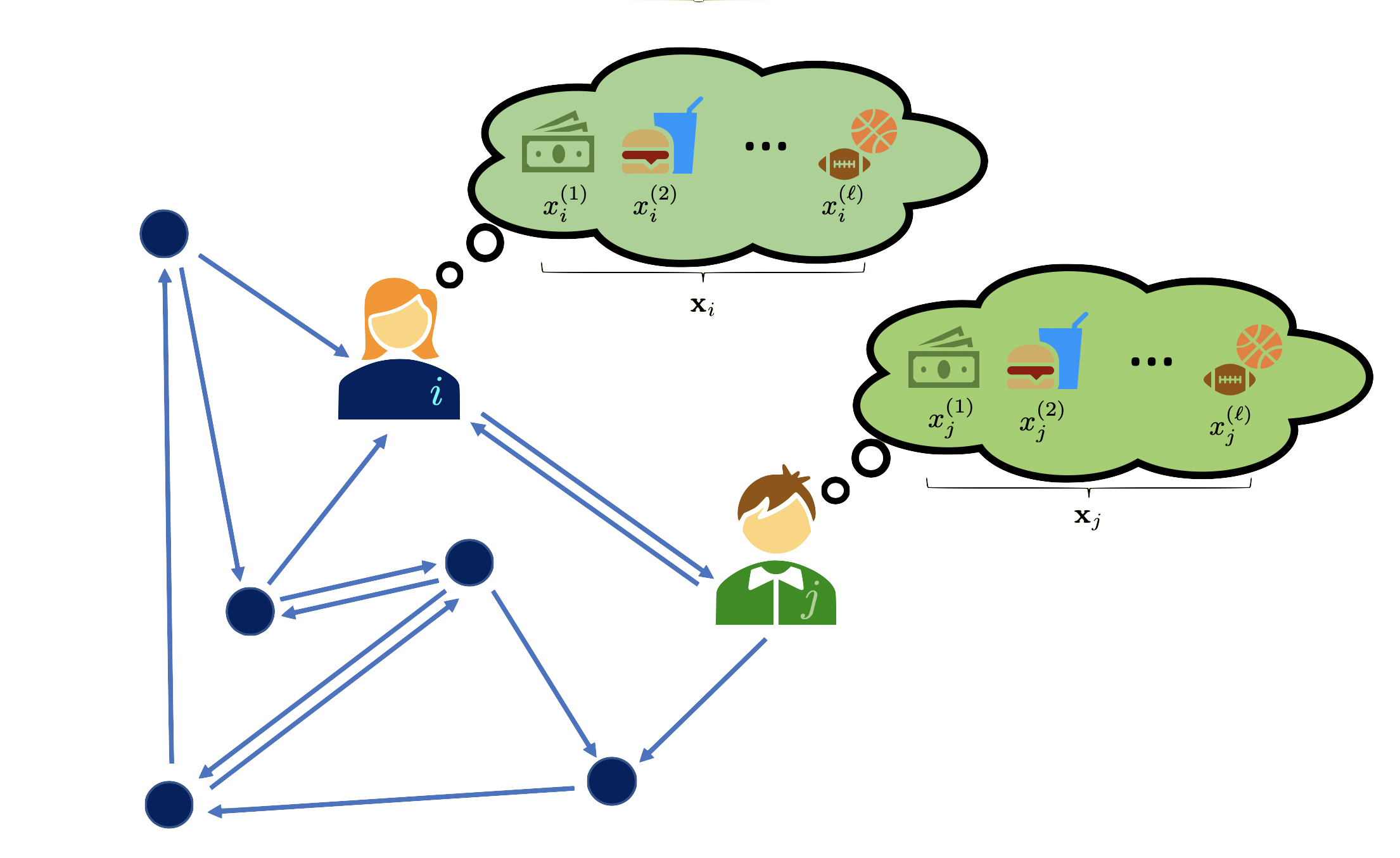}
\end{center}
\end{tcolorbox}

\begin{multicols}{2}
\subsection{From consensus to disagreement}

Along with French's model, the DeGroot model~\eqref{eq.degroot} typically also predicts consensus of opinions. This holds, for instance, when $\mathbf{W}$ is a primitive (irreducible and aperiodic)~\cite{GantmacherVol2,HornJohnsonBook1991} matrix, that is, $\mathbf{W}^m$ has strictly positive entries for sufficiently large exponent $m$~\cite{Degroot_74}. Another standard criterion guarantees consensus if all diagonal entries $w_{ii}$ are positive and the graph corresponding to $\mathbf{W}$ has a globally reachable node (that is, an individual influence all others directly or indirectly) \cite{RenBeardBook,Bullo-Book}. A necessary and sufficient graph-theoretic condition for consensus can be found in~\cite{JacksonBook2008,ProTempo:2017-1}. The aforementioned criteria can be extended, under some additional restrictions to the DeGroot model over a dynamic graph (the matrix of influence weights $\mathbf{W}$ is time-varying)~\cite{RenBeardBook,CaoMorse:08Part1,ProTempo:2018}.

Since real social groups often fail to reach consensus, realistic models of opinion formation should be capable to explain not only ``regular'' consensus behavior, but also various ``disordered'' behaviors featured by disagreement of opinions. To find such a model is a problem that has been studied since 1960s~\cite{Abelson:1964,Abelson:1967} and is known as \textit{Abelson's diversity puzzle} or the problem of \textit{community cleavage}~\cite{Friedkin:2015}.

Most of the models portraying community cleavage, in particular, convergence of the opinions to several clusters, replace the DeGroot equation by nonlinear dynamics, portraying various effects of information assimilation and integration within individual and communication between individuals~\cite{HunterDanesCohenBook_1984,DandekarPNAS2013,Hegselmann.Krause2002OpinionDynamicsand,
DeffuantWeisbuch:2000,Macy:2003,Urbig:2003,Flache:2011,MasFlacheHelbing:2010,MasFlache:2013,MotschTadmor:2013,CeragioliFrasca:18}.
The most studied are \emph{bounded confidence} models~\cite{Hegselmann.Krause2002OpinionDynamicsand,DeffuantWeisbuch:2000,MotschTadmor:2013,ProTempo:2018,Scaglione:2013} capturing the effect of homophily in social groups and assuming that individuals tend to assimilate opinions of like-minded individuals and meet dissimilar opinions with discretion or even ignore them. Identifiability properties of nonlinear models are, however, almost unexplored. The models' behaviors are very sensitive to the structures of nonlinear couplings (e.g. the lengths of confidence intervals in bounded confidence models), noises and numerical errors~\cite{hegselmann2015optimal}.
Hence, in spite of some recent progress in identification of nonlinear networks~\cite{JinYuan:2017,Mangan:2016,KaiseKutz:2018},  identification of social influence from nonlinear opinion dynamics to a great extent remains a challenge.

In linear models of opinion formation, the disagreement of opinions is typically explained by two factors: antagonistic interactions between individuals and their stubborness (reluctance to change the initial opinion). Models of the first type revise the basic assumption on convex combination mechanism of opinion evolution and allowing not only attraction of the agents' opinions, but also their \emph{repulsion}~\cite{Altafini:2012,Altafini:2013,CaoJiao2018,XiaCaoJohansson:16,LiuChenBasar:2017,XiaCao:11,ShiBarasJohansson:15,ShiBarasJohansson:16}. The presence of negative influence is typically explained by ``boomerang'', reactance and anticonformity effects~\cite{Abelson:1967,CaoJiao2018}, that is, the \emph{resistance of some individuals to social influence}.

The theory of signed (or ``coopetitive'') dynamical networks developed in the aforementioned literature is extremely important due to various applications in economics, physics and biology~\cite{ShiAltafiniBaras:2019}. However, its applicability to social influence systems is still disputable for two reasons. First, evidence of ubiquity of negative influence has not been secured experimentally. Since the first definition of boomerang effects \cite{HovlandBook}, the empirical literature has concentrated on the special conditions under which these effects might arise in dyadic interpersonal interactions. The existence of negative ties in large-scale social networks has been questioned in some recent works~\cite{TakacsFlacheMas:16}. The second reason is that clustering of opinions in presence of antagonistic interactions is typically proven under some restrictive assumptions on the network such as \emph{structural balance} of positive and negative ties~\cite{Altafini:2013,ProMatvCao:2014,ProMatvCao:2016,LiuChenBasar:2017,ShiBarasJohansson:15}.

At the same time, the usual DeGroot model is able to explain disagreement of the opinions, assuming the existence of several stubborn individuals that are closed to social influence and keep their opinions unchanged (equivalently, their self-weights are maximal $w_{ii}=1$)~\cite{Acemoglu:2013:OFD:2448396.2448397}. Further development of the DeGroot model with stubborn individuals has naturally lead to the Friedkin-Johnsen (FJ) model, considered in the next subsection. Unlike many other models proposed in physical and engineering literature, the FJ model has been experimentally assessed on small- and medium-size groups~\cite{Friedkin:Johnsen:2011,Friedkin:Johnsen:1999,Friedkin:2012,friedkin2019mathematical}.
An essential part of these experiments is the empirical procedure of matrix $\Wb$ reconstruction, see the box \boxref{Friedkin-Johnsen experiment}.

\subsection{The Friedkin-Johnsen model}

Whereas the DeGroot model allows stubborn individuals that are completely closed to social influence, the FJ model allows also ``partial'' stubborness, measured by a \emph{susceptibility} coefficient $\lambda_i\in [0,1]$. An agent with minimal susceptibility is the stubborn individual retaining its initial opinion~\eqref{eq.stubborn}, whereas the agent with maximal susceptibility assimilates the others' opinions in accordance with the conventional DeGroot mechanism~\eqref{eq.degroot}. In general, the individual opinion at each iteration is influenced by both the others' opinions and their initial opinion
\begin{equation}\label{eq.fj}
\mathbf{x}_i(k+1)=\lambda_{i}\sum_{j=1}^nw_{ij}\mathbf{x}_j(k)+(1-\lambda_i)\mathbf{x}_i(0).
\end{equation}
The matrix $\mathbf{W}$ is stochastic and has the same meaning as in DeGroot's model, namely, $w_{ij}$ stands for the influence weight individual $i$ accords to individual $j$. Without loss of generality, it can be assumed that $\lambda_i=0$ for the agents with the maximal self-weights $w_{ii}=1$ as both conditions imply the full stubborness in the sense of~\eqref{eq.stubborn} (see the box \boxref{Simple properties of the FJ model}). As discussed in~\cite{Friedkin:Johnsen:1999}, individuals' anchorage at their initial opinions can be explained by an ongoing effect of some exogenous factors that had influenced the social group in the past. An initial opinion can be also considered as an individual's \emph{prejudice}~\cite{FrascaTempo:2013,Parsegov2017TAC,proskurnikov2017opinion} that influence their opinion on each subsequent steps.

\medskip

\begin{tcolorbox}[breakable, title= \sf \textbf{Simple properties of the FJ model},colframe=carmine!10,colback=carmine!10,coltitle=black,
]
\sf \small
Using induction on $k=0,1,\ldots$, a number of properties of the FJ model can be proven.
\begin{enumerate}
\item \emph{(self-weight and stubborness)} An agent with maximal self-weight $w_{ii}=1$ is stubborn independent of the susceptibility value, that is, $\mathbf{x}_i(k)=\mathbf{x}_i(0)$. For this reason, it is convenient to assume that $\lambda_i=0$ whenever $w_{ii}=1$.
\item \emph{(consensus preservation)} If the initial opinions are in consensus $\mathbf{x}_1(0)=\ldots=\mathbf{x}_n(0)=\xb_0^*$, this consensus is not deteriorated
    \[
    \mathbf{x}_1(k)=\ldots=\mathbf{x}_n(k)=\xb_0^*\quad\forall k;
    \]
\item \emph{(containment property)} More generally, at each stage of the opinion iteration the opinions are contained by the \emph{convex hull} of their initial values, that is, $\mathbf{x}_i(k)\in\mathfrak{X}_0$, where
    \[
    \mathfrak{X}_0=\left\{\sum_{i=1}^na_i\mathbf{x}_i(0):a_i\ge 0,\sum_{i=1}^na_i=1\right\}.
    \]
\end{enumerate}
Whereas the containment property is very intuitive in the case of scalar opinions (where the set $\mathfrak{X}_0$ is nothing else than the interval $[\min_ix_i(0),\max_ix_i(0)]$), its validity in higher dimensions is a non-trivial property of a social influence network, predicted by the Friedkin-Johnsen theory. Even for three-dimensional opinions, it is difficult to visualize the convex hull $\mathfrak{X}_0$ (being a convex polyhedron) without special software. Nevertheless, experiments on rational decision making on resource allocation~\cite{friedkin2019mathematical}, illustrate that multidimensional decisions of individuals typically stay in the convex polyhedron $\mathfrak{X}_0$.
\end{tcolorbox}

Similar to DeGroot's model, the opinions $\mathbf{x}_i(k)$ may be scalar or multidimensional. Stacking them one on top of another in order to get the opinion matrix $\mathbf{X}(k)$, the FJ system~\eqref{eq.fj} can be rewritten in the matrix form
\begin{equation}\label{eq.fj-m}
\mathbf{X}(k+1)=\mathbf{\Lambda}\Wb\mathbf{X}(k)+(\mathbf{I}_n-\mathbf{\Lambda})\mathbf{X}(0).
\end{equation}
Here $\mathbf{\Lambda}=\diag(\lambda_{1},\ldots,\lambda_n)$ stands for the diagonal matrix composed of the susceptibility coefficients. DeGroot's model arises as a special case of~\eqref{eq.fj-m} with $\mathbf{\Lambda}=\mathbf{I}_n$.
\medskip
\end{multicols}

\begin{tcolorbox}[breakable,title= \sf \textbf{\sf \textbf{Schur stability criteria}},colframe=airforcecarmine!20,colback=airforcecarmine!20,coltitle=black,
]
\sf \small
Consider the graph of social influence $\G[\mathbf{W}]$, associated with the matrix $\mathbf{W}$ and let $\mathcal{S}\subseteq\{1,\ldots,n\}$ stand for the set of individuals that are fully or partially stubborn (anchored at their initial opinions)
\[
\mathcal{S}=\{i:\lambda_i<1\}.
\]
As discussed in~\cite{Friedkin:Johnsen:1999}, an individual's attachment to their initial opinion may be explained as a direct ongoing effect of some previous experience or other external factors that had influenced the group in the past. The FJ model is Schur stable if and only if the opinions of the remaining individuals (with $\lambda_i=1$) also remain influenced by these factors via paths of influence, that is, any node $l\in\V\setminus\mathcal{S}$ is connected by a walk to a node from $\mathcal{S}$.
\vskip0.5cm
\textbf{Theorem.}~\cite{FrascaTempo:2013,Parsegov2017TAC} \sf The matrix $\mathbf{\Lambda}\mathbf{W}$ is Schur stable if and only if every node of $\G[\mathbf{W}]$ either belongs to $\mathcal S$ or is connected to a node from $\mathcal S$ by a walk. This holds, e.g., if $\mathcal{S}\ne\emptyset$ and $G[\mathbf{W}]$ is a strongly connected graph.
\vskip0.5cm

\begin{minipage}{0.49\columnwidth}
 \begin{center}
\includegraphics[width=0.49\columnwidth]{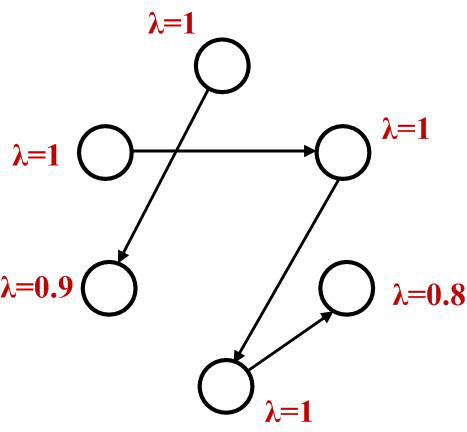}
\end{center}
\end{minipage}
\begin{minipage}{0.49\columnwidth}
\begin{center}
\includegraphics[width=0.49\columnwidth]{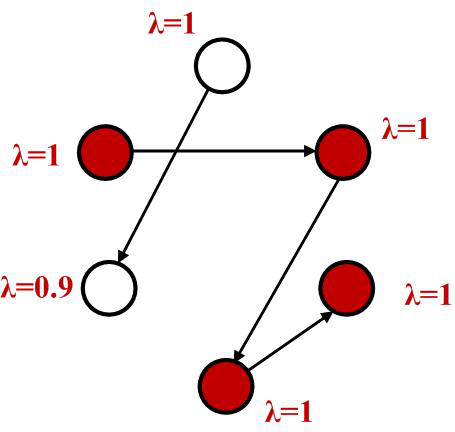}
\end{center}
\end{minipage}
\bigskip

The social group shown on the left figure corresponds to a Schur stable matrix $\mathbf{\Lambda}\mathbf{W}$ (each node with $\lambda=1$ is connected to one of the nodes with $\lambda<1$). {For} the group shown on the right figure, {the matrix $\mathbf{\Lambda}\mathbf{W}$} is not Schur stable: the group of red nodes is not connected to
{the unique node with $\lambda<1$.}
\end{tcolorbox}

\begin{multicols}{2}
Due to presence of fully or partially stubborn agents, the FJ dynamics usually does not lead to consensus of opinions (except for special situations, where e.g. the FJ system reduces to DeGroot's model). However, in generic situations the opinions converge. The most interesting case where such a convergence can be established is the case where the matrix $\mathbf{\Lambda}\mathbf{W}$ is \emph{Schur stable}, that is, all its eigenvalues $\mu_1,\ldots,\mu_n$ belong to the open unit disk $|\mu_j|<1$. A graph-theoretical criterion of Schur stability is available~\cite{FrascaTempo:2013,Parsegov2017TAC,ProTempo:2017-1}, summarized in the box \boxrefb{Schur stability criteria}.

If $\mathbf{\Lambda}\mathbf{W}$ is a Schur stable matrix, then the matrix of opinions converges~\cite{FrascaTempo:2013,Parsegov2017TAC}
\begin{equation}
\label{Xinfinity}
\begin{gathered}
\mathbf{X}(\infty)=\lim_{k\to\infty}\mathbf{X}(k)=\mathbf{V}\mathbf{X}(0),\\ \mathbf{V}=(\mathbf{I}_n-\mathbf{\Lambda}\mathbf{W})^{-1}(\mathbf{I}_n-\mathbf{\Lambda}).
\end{gathered}
\end{equation}
The matrix $\mathbf{V}=[v_{ij}]$ appears to be \emph{row-stochastic}~\cite{Friedkin:2015,ProTempo:2017-1} and is referred to as the \emph{control matrix} as it determines the ability of individuals to control the final opinion of others (see the box \boxref{Control matrix and Friedkin's centrality}).
\begin{tcolorbox}[breakable,title= \sf \textbf{Control matrix and influence centrality},colframe=carmine!10,colback=carmine!10,coltitle=black,
]
\sf\small When agents' opinions converge,
the final opinion of agent $i$ can be represented in the following form
\[
\mathbf{x}_i(\infty)=\sum_{j=1}^nv_{ij}\mathbf{x}_j(0).
\]
In this sense, the entry $v_{ij}$ serves {as} a measure of \emph{social power}~\cite{Friedkin:1986}) of individual $j$ over individual $i$, that is, $j$'s ability to influence $i$'s terminal opinion. The \emph{average} power of individual $j$ over the group
\[
c_j=\frac{1}{n}\sum_{i=1}^nv_{ij}
\]
serves as a natural \emph{measure of centrality} for the nodes of the social network. Choosing different matrices $\mathbf{\Lambda}$, a whole family of centrality measures is obtained for the weighted graph $G[\mathbf{W}]$ that have been first introduced by Friedkin~\cite{Friedkin:1991} for the case where $\mathbf{\Lambda}=\alpha \mathbf{I}_n$ with a scalar $\alpha\in (0,1)$. In this situation,
\[
\mathbf{V}=(1-\alpha)(I-\alpha\mathbf{W})^{-1},
\]
and the vector of influence centralities $\mathbf{c}=(c_1,\ldots,c_n)^{\top}$ can be found as
\[
\mathbf{c}=\frac{1}{n}\mathbf{V}^{\top}\1_n=(1-\alpha)\left(I-\alpha\mathbf{W}^{\top}\right)^{-1}\1_n.
\]
{For a specially chosen matrix (see e.g.~~\cite{HI-RT:14}) $\mathbf{W}$ and $\alpha=1-m$, where $m\in (0,1)$,} the latter vector coincides with the PageRank {centrality measure}, which had appeared in~\cite{Friedkin:1991} seven years earlier than the seminal work by Brin and Page~\cite{SB-LP:98}. Relations between the influence centrality and PageRank are discussed in more detail in~\cite{FriedkinJohnsen:2014,ProTempo:2017-1,ProTempoCao16-1}.
\end{tcolorbox}
\subsection{Dynamics of reflected appraisal}

The concept of influence centrality (see the box \boxref{Control matrix and Friedkin's centrality})
serves a base for the dynamical models describing the evolution of influence matrix $\mathbf{W}$ and known as dynamics of \emph{reflected appraisals}. As argued in~\cite{nef-pj-fb:14n}, in deliberative groups (such as e.g. standing policy bodies and committees, boards of directors, juries and panels of judges) {\em an individual's influence centrality on an issue alters his or her expectation of future group-specific influence on issues}. In other words, the influence matrix may evolve as the social group discusses a sequence of different issues, see the box
\boxrefb{Reflected appraisal model}.

\subsection{Extensions of the FJ model}

The seminal FJ model can be extended in many directions, among which we consider only three. The first extension is concerned with the dynamics of multidimensional opinions, which stand for the agents' positions on several logically related issue. Such an opinion may be considered as a special case of a \emph{belief system}, defined as ``a configuration of ideas and attitudes in which the elements are bound together by some form of constraint or functional interdependence''~\cite{Converse:1964}. Contradictions and other inconsistencies between beliefs, attitudes and ideas may trigger tensions and discomfort (``cognitive dissonance'') that can be resolved by a within-individual process.
This process, studied in cognitive dissonance and cognitive consistency theory, is thought to be an automatic process of the  brain, enabling humans to develop coherent systems of beliefs~\cite{FestingerBook,GawronskiBook}. Modeling the dynamics of opinions on interrelated issues is a challenging problem, and only a few models have been proposed in the literature, and most of them are featured by nonlinear dynamics~\cite{Noorazar:2018,Xiong:2017,Noorazar:2020}. The box \boxrefb{A model of a belief system's dynamics} is devoted to a simple linear model proposed in~\cite{Parsegov2015CDC,Parsegov2017TAC,FriedkinPro2016}.
In general, the presence of the logical relations between the issues can affect the recoverability of the influence network from partial observations~\cite{RavazziTempoDabbene:2018}.

Another extension of the FJ model revises the restrictive assumption on simultaneous communication.
As stated in \cite{Friedkin:Johnsen:1999}, {\em{the interpersonal influences do not occur in the simultaneous way}} and the assumption on synchronous rounds of interactions is too simplistic. In other words, individuals in real social groups are featured by asynchronous {\em ad-hoc} interactions. More realistic models, assuming that only a couple of individuals can interact at each step, have been introduced in \cite{FrascaTempo:2013,FrascaIshiiTempo:2015,Parsegov2017TAC}.
Such multi-agent communication protocols are known as \emph{gossiping}~\cite{Boyd:06}. The model is summarized in the box  \boxref{Asynchronous gossip-based FJ model}. In~\cite{Parsegov2017TAC}, a gossip-based version of the extended FJ model~\eqref{eq.fj-mult-m} is considered.
Another potential ``culprit'' of randomness is a noise, representing the effects of individuals' free will and unpredictability of their decisions (one model with noise is discussed in the box \boxref{(Dynamics on multiplex networks)}.
\end{multicols}

\begin{tcolorbox}[breakable,title= \sf \textbf{\sf \textbf{Reflected appraisal model}},colframe=carmine!10,colback=carmine!10,coltitle=black,
]
{\sf\small{In psychology the theory of Reflected appraisal states that people's perception is influenced by the evaluation of others (see \cite{Fried11_reflected_appraisal} and reference therein). In \cite{nef-pj-fb:14n}, the evolution of power across a series of issues over time is explained as the result of direct and indirect interpersonal influences on group members.
More formally, the phenomenon is described by the following dynamical system
\begin{equation}\begin{split}\label{eq:dynamical_system}
\Wb^{(s)}={\bf{I}-\bf{\Lambda}}^{(s)}+{\bf{\Lambda}}^{(s)}{\bf{C}}\\ c(s+1)^{\top}=\frac{\mathds{1}^{\top}}{n}({\bf{I}}-{\bf{\Lambda}}^{(s)}\Wb^{(s)})^{-1}({\bf{I}-\bf{\Lambda}}^{(s)}),\\
{\bf{\Lambda}}^{(s)}=I-\diag(\mathbf{c}^{(s)})==I-\diag(\Wb^{(s)}).
\end{split}
\end{equation}
where $c(s)$ is influence centrality vector during the discussion on issue $s$ and ${\bf{C}}$ is a constant matrix \ant{with zero diagonal entries}. \ant{From control-theoretic viewpoint, this
mechanism can be interpreted as a nonlinear feedback. The social power $c_i(s)$ individual has acquired in the discussion on issue $s-1$ influences his/her self-weight (and thus also the weights assigned to the others) during the discussion on issue $s$. Notice that the structure of the graph remains unchanged, being encoded in the matrix $\mathbf{C}$, the mechanism alters only the influence weights.}

\begin{multicols}{2}
\begin{minipage}{0.95\columnwidth}
\begin{center}
  \includegraphics[trim={1cm 6cm 3cm 8cm},width=0.9\columnwidth]{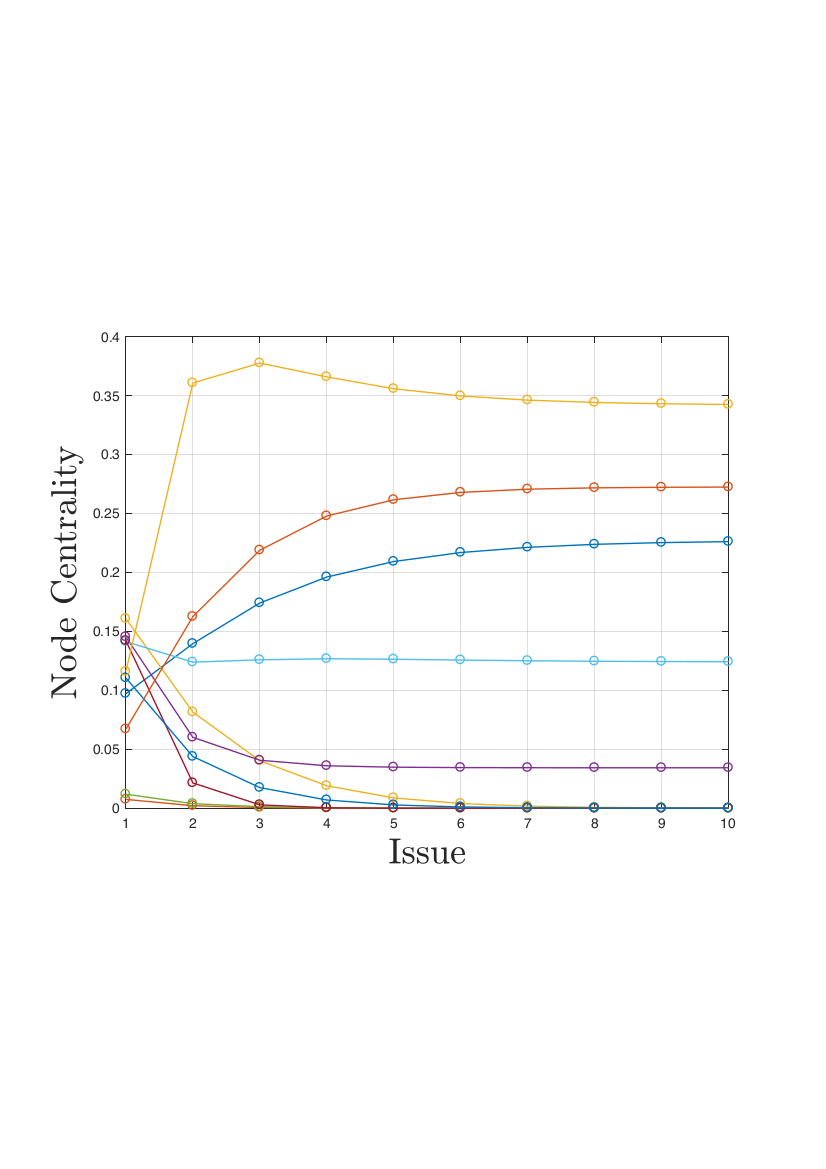}
  \end{center}
\end{minipage}

\begin{minipage}{0.95\columnwidth}\vspace{1cm}
In these figures the evolution of the influence network and the social power is depicted as a function of issue sequence, respectively. It should be noted that the topology of the networks is the same at each layer but the strength of influence changes across issue sequence. This is captured by a model in which all transition matrices share a common support $\Omega \subseteq \{1,\ldots,n\} \times \{1,\ldots, n\}$, i.e.
\begin{equation}\label{eq:M2}
\Wb^{(s)}_{ij}\neq0\qquad\forall  (i,j)\in\Omega, \forall \ell\in \{1,\ldots, m\}.
\end{equation}
\end{minipage}
\end{multicols}

\medskip

\begin{minipage}{1\columnwidth}
\begin{center}
  \includegraphics[trim={1cm 0cm 1cm 0cm},width=0.99\columnwidth]{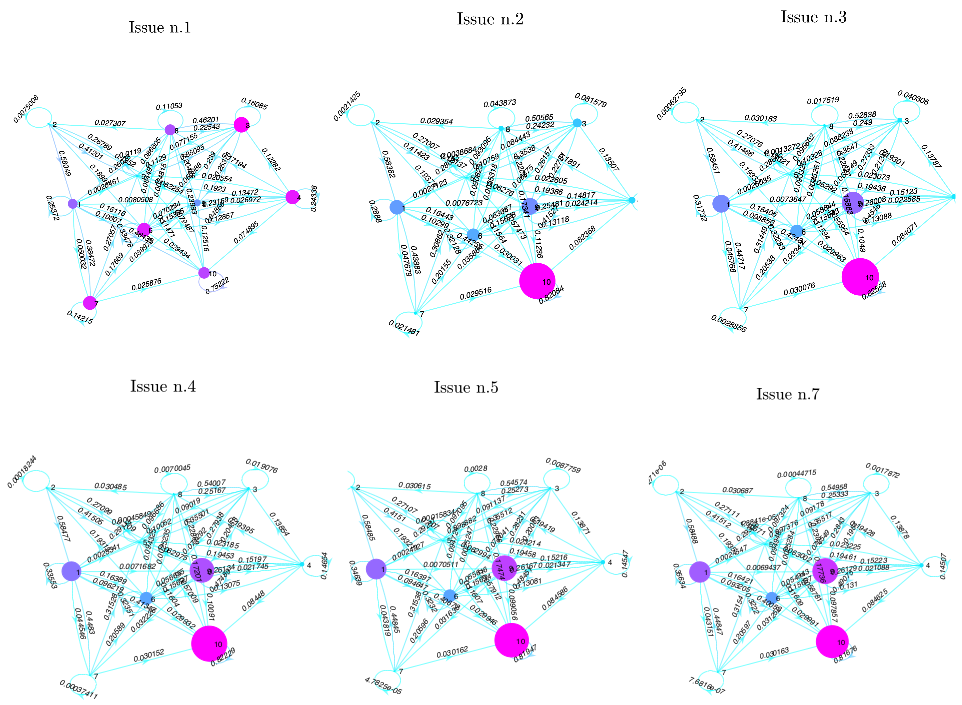}
  \end{center}
\end{minipage}

}}

   \end{tcolorbox}

\color{black}

\begin{tcolorbox}[title= \sf \textbf{A model of a belief system's dynamics},colframe=airforcecarmine!20,colback=airforcecarmine!20,coltitle=black,
]
\sf\small
Adjusting their position on one of the interdependent issues, an individual might have to adjust the positions on several related issues simultaneously in order to maintain the belief system's consistency. Such an adjustment can be thought of as an operator $\mathbf{x_i}\mapsto \mathcal{C}_i(\mathbf{x_i})$, which preserves the vector's dimension.
Whereas the actual mathematical representation of introspective tension-resolving processes is unknown, it was conjectured in~\cite{FriedkinPro2016} that in some situations the operator $\mathcal{C}_i$ may be linear, and is represented by a matrix $\mathbf{C}_i$, so that $\mathbf{x}_i\mapsto \mathbf{x}_i\mathbf{C}_i^{\top}$ (recall that, according to our conventions, the multidimensional opinion is represented by a row ${m}$-dimensional vector, so that $\mathbf{C}_i\in\mathbb{R}^{{m \times m}}$). Assuming that the tension resolving process follows the integration of opinions from the neighbors, the FJ model~\eqref{eq.fj} is replaced by the dynamics
\begin{equation}\label{eq.fj-mult}
\mathbf{x_i}(k+1)=\lambda_i\left(\sum_{j=1}^nw_{ij}\mathbf{x_j}(k)\right)\mathbf{C_i}^{\top}+(1-\lambda_i)\mathbf{x}_i(0).
\end{equation}

It has been shown (see the Supplementary Material to~\cite{FriedkinPro2016}) that if the matrix $\mathbf{\Lambda}\mathbf{W}$ is Schur stable and all matrices $\mathbf{C}_i$ are row-stochastic or, more generally, $\mathbf{C}_i=(c_{lm}^{(i)})$ where
$\sum_m|c_{lm}^{(i)}|\le 1$ for each $l$ and $i$, then the linear operator
\[
\mathbf{X}\mapsto
\mathbf{\Lambda}\mathbf{W}
\begin{pmatrix}
\mathbf{x}_1\mathbf{C_1}^{\top}\\
\hline\\[-3mm]
\mathbf{x}_2\mathbf{C_2}^{\top}\\
\hline\vdots\\
\hline\\[-3mm]
\mathbf{x}_n\mathbf{C_n}^{\top}
\end{pmatrix}
\]
is Schur stable, in particular, the opinion matrix $\mathbf{X}(k)$ determined by~\eqref{eq.fj-mult} converges as $k\to\infty$.

The model~\eqref{eq.fj-mult} becomes more elegant in the case of homogeneous agents $\mathbf{C}_1=\ldots=\mathbf{C}_n=\mathbf{C}$. In this situation, the equations~\eqref{eq.fj-mult} may be rewritten in the matrix form very similar to~\eqref{eq.fj-mult-m}
\begin{equation}\label{eq.fj-mult-m}
\mathbf{X}(k+1)=\mathbf{\Lambda}\mathbf{W}\mathbf{X}(k)\mathbf{C}^{\top}+(\mathbf{I}_n-\mathbf{\Lambda})\mathbf{X}(0).
\end{equation}
The FJ model is a special case of~\eqref{eq.fj-mult-m}, corresponding to $\mathbf{C}=\mathbf{I}_n$ (if the issues are not logically related, it is natural to assume that the different dimensions of the opinion evolve independently). In~\cite{Parsegov2017TAC}, $\mathbf{C}$ is referred to as the \emph{MiDS} (multi-issue dependency structure) matrix.

An example of the system~\eqref{eq.fj-mult-m} with three-dimensional opinions has been considered in~\cite{FriedkinPro2016}. It was conjectured that the speech of Colin Powell, the highly
respected U.S. Secretary of State, in the UN Security Council presented a logic structure on three truth statements:
\begin{enumerate}[i)]
\item Saddam Hussein has a stockpile of weapons of mass destruction;
\item Hussein's weapons of mass destruction are real and present dangers
to the region and the world;
\item An invasion of Iraq would be a just war.
\end{enumerate}
It was a logic structure high certainty of belief on statement i) implies high certainty
of belief on statements ii) and iii). On the other hand, if statement i) is false then statements ii) and iii) are
also false. This corresponds to the MiDS matrix
\[
\mathbf{C}=
\begin{pmatrix}
1 & 0 & 0\\
1 & 0 & 0\\
1 & 0 & 0
\end{pmatrix}.
\]
A numerical example considered in~\cite{FriedkinPro2016} shows that if the population has initially a high certainty on statement i), the then the belief system dynamics over a random graph generates a consensus
that a preemptive invasion is a just war. At the same time, if statement i) is considered to be false, then the the population's certainty belief on all three statements is dramatically lowered. Hence, the model can explain the fluctuation of the public opinion on the Iraq invasion.
\end{tcolorbox}

\begin{tcolorbox}[title= \sf \textbf{Asynchronous gossip-based FJ model},colframe=carmine!10,colback=carmine!10,coltitle=black,
]

\sf\small
The FJ model \cite{FriedkinJohnsen:1990} can be extended to the case where the interactions follow a model more consistent with the ``usual'' social network interactions where only a few agents interact at a time. In this case, the opinions evolve as follows \cite{FrascaTempo:2013}:
\ant{
\begin{itemize}
\item each agent $i\in \V $ starts from an initial belief $x_i(0)\in \real$;
\item at each period  $k\in\integernonnegative$, a subset of \emph{active} nodes $\V_k$ is randomly selected from a uniform distribution over $\V$;
\item the opinions of inactive agents remain unchanged, where each active agent $i \in \V_k$ interacts with a randomly chosen neighbor $j$ and updates its belief according to
    a rule that resembles the FJ mechanism, which results in the equations
    \begin{eqnarray}\label{eq:gossip-friedkin}
        x_i(k+1)&=&\lambda_{i}\big((1-w _{ij})x_i(k)+w _{ij}x_j(k)\big)+(1-\lambda_{i})x_i(0)
\qquad\qquad \forall i\in\V_k\nonumber \\
x_\ell(k+1)&=&x_\ell(k)\qquad \forall \ell\in \V\setminus\V_k,
\end{eqnarray}
\end{itemize}
}
By  denoting the set of neighbors of node $i\in\V$ by  $\mathcal{N}_i \doteq\{j\in\V : (i,j)\in\E\}$, introducing the out-degree $d_i \doteq|\mathcal{N}_i|$, 
the dynamics \eqref{eq:gossip-friedkin} can be formally rewritten in the following form: given $\V_k$ and letting $\theta(k)\doteq\{\theta_i\}_{i\in\V_k}$, we have
\begin{equation}\label{eq:system_ergodic}
\mathbf{x}(k+1)=\mathbf{\Gamma}(k)\mathbf{x}(k)+\mathbf{B}(k)\xb(0),
\end{equation}
where \ant{the coefficients are defined as}
$$
\begin{gathered}
\mathbf{\Gamma}(k)\doteq\left(\mathbf{I}_n \!-\!\!\sum_{i\in\V_k}e_ie_i^{\top} (\mathbf{I}_n - \mathbf{\Lambda})\!\right)\!\!\! \left(\mathbf{I}_n \!\!+\!\! \sum_{i\in\V_k}\mathbf{W}_{i\theta_i} (\mathbf{e}_i\mathbf{e}_{\theta_i}^{\top}  -\mathbf{e}_i\mathbf{e}_i^{\top} ) \!\right),\qquad
\Bb(k)\doteq \sum_{i\in\V_k}e_ie_i^{\top}(\mathbf{I}_n - \mathbf{\Lambda})
\end{gathered}
$$
\ant{and $\theta_i$ is a uniformly distributed random element of $\mathcal{N}_i$, that is,
$\theta_i=j\in\mathcal{N}_i$ with probability $1/d_i$.}

It can be shown that the sequence $\{\xb(k)\}_{k\in\integernonnegative}$ is a Markov process~\cite{grimmett2001probability}, \ant{which} 
fails to converge in a deterministic sense, and shows persistent oscillations.

\begin{center}
  \includegraphics[trim={0cm 6cm 0cm 6cm},width=.49\columnwidth]{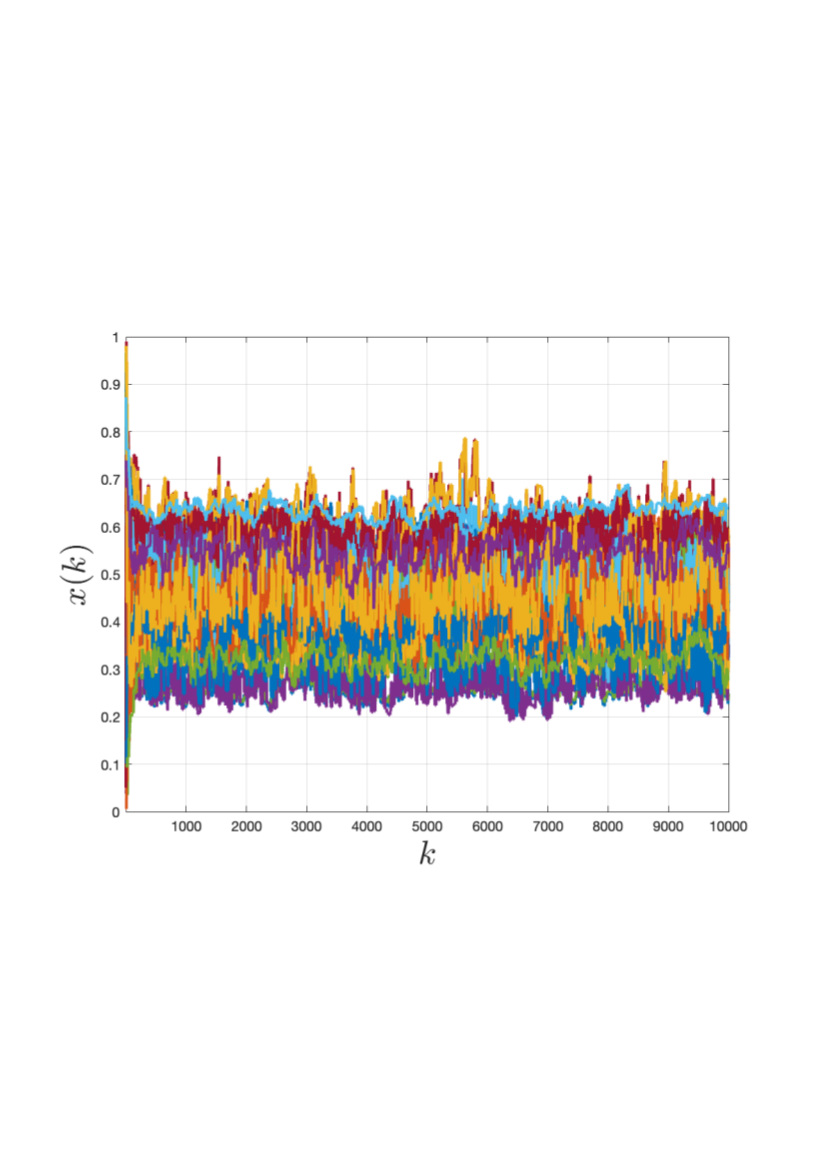}
  \includegraphics[trim={0cm 6cm 0cm 6cm},width=.49\columnwidth]{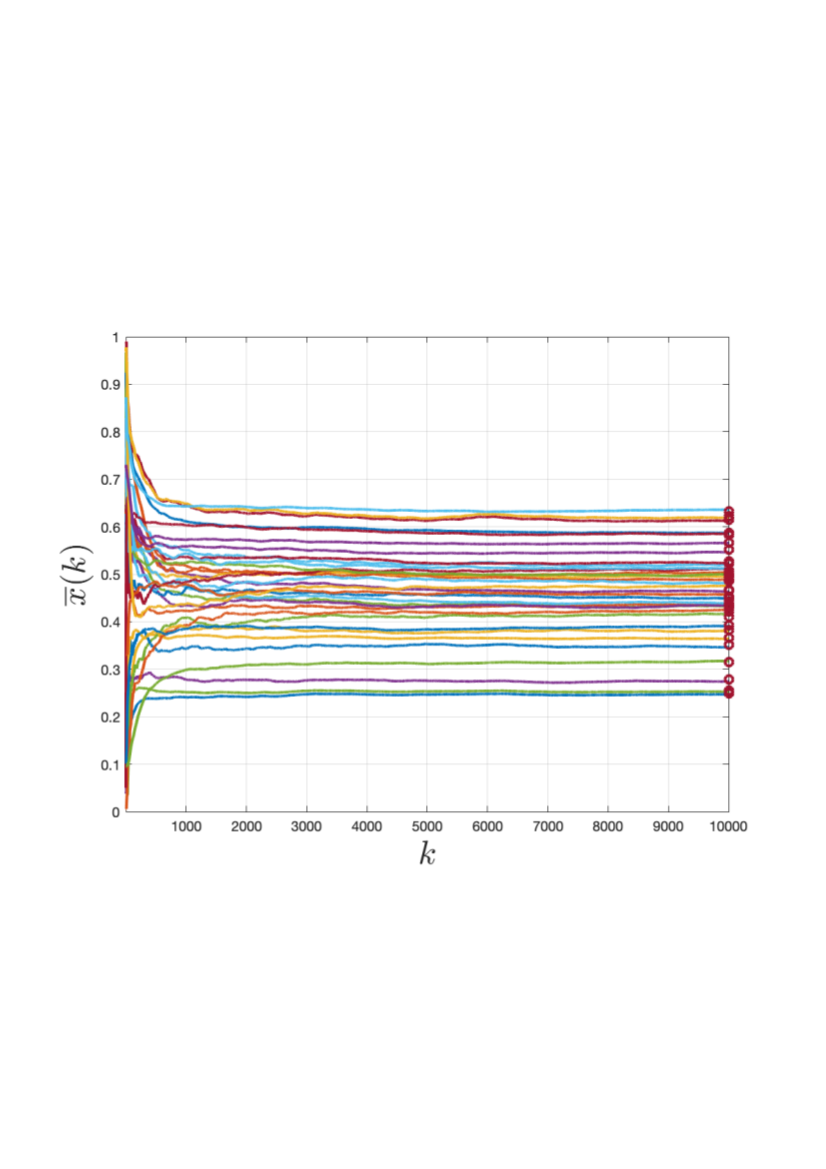}
\end{center}

However, if the matrix $\mathbf{\Lambda}\mathbf{W}$ is Schur stable (see the box~\boxrefb{Schur stability criteria}), the convergence of the \emph{expectations} and the ergodicity of the oscillations can be ensured. Namely, it was shown in~\cite{FrascaTempo:2015} that the opinions' expected values obey the equation
	$$
	\mathbb{E}[\mathbf{x}(k+1)]=\overline{\mathbf{\Gamma}} \mathbb{E}[\mathbf{x}(k)]+\overline{\mathbf{b}}
	$$
	where
	\begin{eqnarray*}
		\overline{\mathbf{\Gamma}}\doteq\mathbb{E}[\mathbf{\Gamma}(k)]=\left(1-\beta\right)\mathbf{I}_n+\beta\mathbf{\Lambda}(\mathbf{I}_n- \mathbf{D}^{-1}(\mathbf{I}-\mathbf{W})),		\qquad
		\overline{\mathbf{b}}\doteq\beta(\mathbf{I}_n-\mathbf{\Lambda})\xb(0),
	\end{eqnarray*}
	$\beta=|\V_k|/|\V|,$ and $D$ is the degree matrix of the network, a diagonal matrix whose diagonal entry is equal to the degree $d_i = |\mathcal{N}_i|$.
	Moreover, the sequence $\mathbb{E}[\mathbf{x}(k)]$ converges  to
	\[
	\mathbb{E}[\mathbf{x}(\infty)]=(\mathbf{I}_n-\overline{\mathbf{\Gamma}})^{-1}\overline{\mathbf{b}}.
	\]
	
	The opinion sequence has a few more interesting ergodicity properties that can be exploited in estimation algorithms. Namely, i) $\mathbf{x}(k)$ converges in distribution to a random variable $\mathbf{x}_{\infty}$ and the distribution is the unique invariant distribution for \eqref{eq:gossip-friedkin}; ii) the process is ergodic, iii) the limit random variable satisfies $\mathbb{E}[\mathbf{x}_{\infty}]=(\mathbf{I}_n-\overline{\mathbf{\Gamma}})^{-1}\overline{\mathbf{b}}$
and \ant{iv) the Ces\'aro averages converge almost surely (and in the sense of $p$-th moment for each $p\geq 1$):}
\[
\mathbf{\bar x}(k)=\frac{1}{k+1}\sum_{\ell=0}^k\mathbf{x}(\ell)\xrightarrow[k\to\infty]{}
\mathbf{x}_{\infty}
\]
\vskip0.1cm
\ant{To identify influences in a social networks, the opinions cross-correlation matrix are useful that are} defined as follows
$$
\boldsymbol{\boldsymbol{\Sigma}}^{[\ell]}(k)\doteq \mathbb{E}\left[\mathbf{x}(k)\mathbf{x}(k+\ell)^{\top}\right].
$$
As shown in \cite{RAVAZZI2020}, these correlation matrices satisfy
	\begin{equation}\label{eq:Sigmak2}
	\mathbf{\Sigma}^{[\ell+1]}(k)=\mathbf{\Sigma}^{[\ell]}(k)\overline{\mathbf{\Gamma}}^{\top}+\mathbb{E}[\mathbf{x}(k)]\overline{\mathbf{b}}^{\top}.
	\end{equation}
	Moreover $\mathbf{\Sigma}^{[\ell]}(k)$  converges \ant{(as $k\to\infty$) to the limit} $\mathbf{\Sigma}^{[\ell]}({\infty})$ for all $\ell \in\integernonnegative$, satisfying
	\begin{equation}\label{eq:YW2}
	\mathbf{\Sigma}^{[\ell+1]}(\infty)=\mathbf{\Sigma}^{[\ell]}(\infty)\overline{\mathbf{\Gamma}}^{\top}+\mathbb{E}[\mathbf{x}(\infty)]\overline{\mathbf{b}}^{\top}.
	\end{equation}

Notice that the relation in \eqref{eq:YW2} is a sort of Yule-Walker equation \cite{Yonina-1}
used for estimation in autoregressive processes.

\end{tcolorbox}

\begin{tcolorbox}[title= \sf \textbf{\color{black}{Dynamics on multiplex networks}
}, colframe=carmine!10,
colback=carmine!10,
coltitle=black,
]\sf \small
The FJ model previously described can be extended to the cases where the social network discusses on several issues and the influence network matrix is different depending on the topic under discussion. However, being the underlying social network essentially the same, we expect that the social systems will share some common feature. In this sense, $\mathcal{M}_{cc}$ \ant{(the common component model)} takes into account the cases where the networks differ  in few components. $\mathcal{M}_{cs}$ \ant{(the common support model)}, instead, describes situations where the topology is equal for all systems but the weights are different (see Section \ref{sec-hetero} and the box~\boxrefb{Multidimensional networks}).

\begin{multicols}{2}
More precisely, we can consider the following set of dynamical equations
\begin{align*}
&\mathbf{x}^{(s)}(k+1)=\boldsymbol{\Lambda}^{(s)} \mathbf{W}^{(s)}\mathbf{x}(k)+(\mathbf{I}-\boldsymbol{\Lambda}^{(s)})\mathbf{u}^{(s)}+\boldsymbol{\eta}^{(s)}, \\
 &\mathbf{x}^{s}(0)=\mathbf{u}^{(s)}
\end{align*}
where the values $\mathbf{x}^{(s)}$ represent the agents' opinions on a specific subject $s$ and $\boldsymbol{\eta}^{(s)}(t)\sim\mathcal{N}(\boldsymbol{0},\mathbf{Q}_{\eta})$ is an additive noise.
The Markov process exhibit persistent fluctuations, due to the random uncertainty in the dynamical system.
Also in this case in can be shown that the expected opinions and cross correlations matrices converge to a final pattern of values.


\includegraphics[width=0.95\columnwidth]{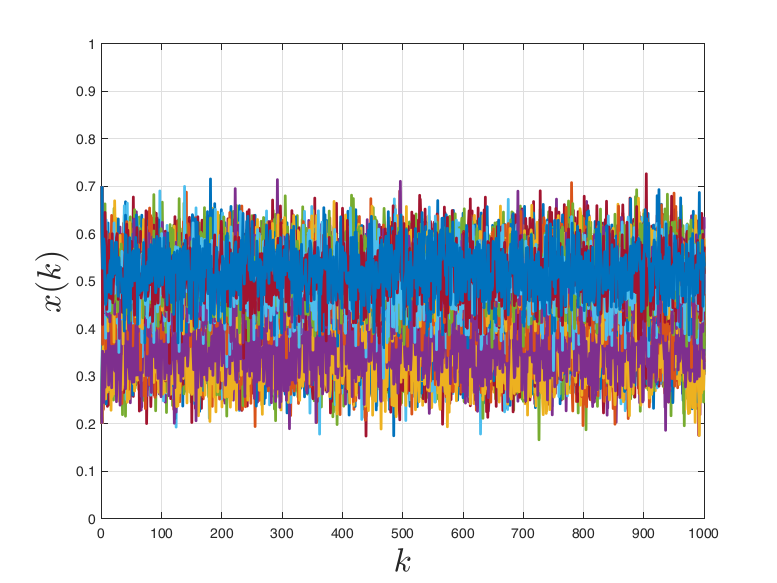}

\end{multicols}
\medskip

\begin{multicols}{2}
\begin{center}
{Model $\mathcal{M}_{cc}$: Influence matrices differ in few components}
\smallskip

\includegraphics[width=0.65\columnwidth, trim={6cm 6cm 6cm 6cm},clip]{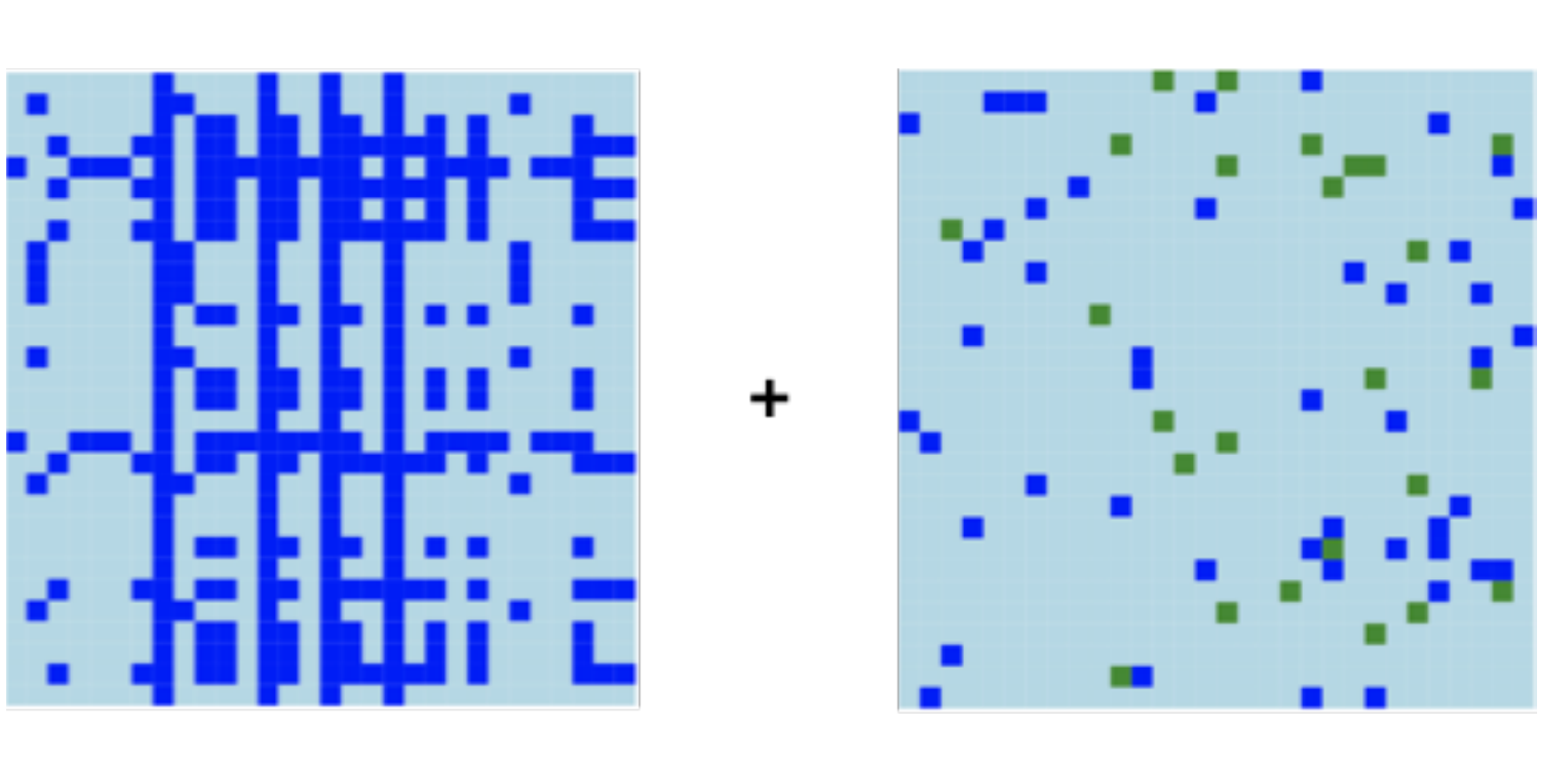}
\end{center}
\newpage

\begin{center}
{Model $\mathcal{M}_{cs}$: Influence matrices have common topology}
\bigskip

\includegraphics[width=0.9\columnwidth]{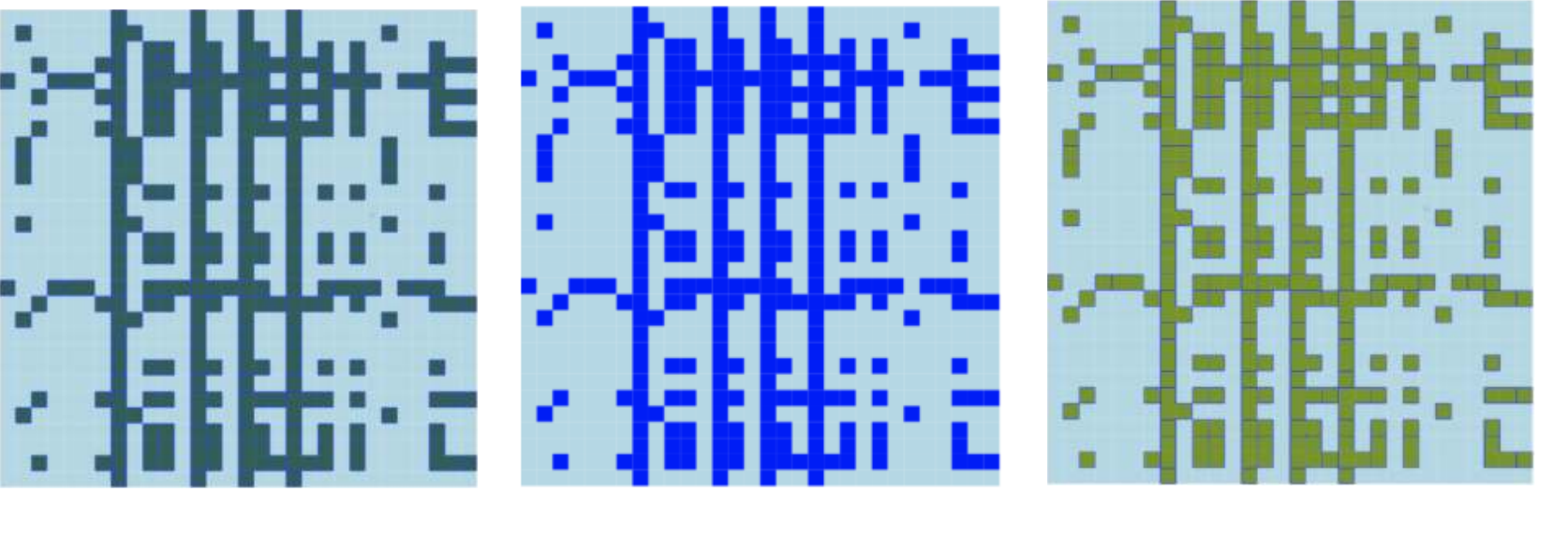}
\end{center}
\end{multicols}

However, if the \ant{matrices $\boldsymbol{\Lambda}^{(s)}\mathbf{W}^{(s)}$ are} Schur stable (see the box~\boxref{Schur stability criteria}), the convergence of the \emph{expectations} and the ergodicity of the oscillations can be ensured. The sequence $\mathbb{E}[\mathbf{x}(k)]$ converges  to
	\[
	\mathbb{E}[\mathbf{x}^{(s)}(\infty)]=(\mathbf{I}_n-\boldsymbol{\Lambda}^{(s)} \mathbf{W}^{(s)})^{-1}(\mathbf{I}_n-\boldsymbol{\Lambda}^{(s)} )\mathbf{u}^{(s)}.
	\]
	and the opinions' cross-correlation matrices satisfy the following relations (see \cite{8796302})
		\begin{equation}\label{eq:Sigmak3}
	\mathbf{\Sigma}_{(s)}^{[\ell+1]}(k)=\mathbf{\Sigma}_{(s)}^{[\ell]}(k)(\overline{\mathbf{\Gamma}}^{(s)})^{\top}+\mathbb{E}[\mathbf{x}^{(s)}(k)](\overline{\mathbf{b}}^{(s)})^{\top}.
	\end{equation}
	with $\overline{\mathbf{\Gamma}}^{(s)}=\boldsymbol{\Lambda}^{(s)} \mathbf{W}^{(s)}$ and $\overline{\mathbf{b}}=(\mathbf{I}_n-\boldsymbol{\Lambda}^{(s)})\mathbf{u}$ and
	\begin{align}\begin{split}\label{eq:Sigmak4}
	\mathbf{\Sigma}_{(s)}^{[0]}(\infty)&=\boldsymbol{\Lambda}^{(s)} \mathbf{W}^{(s)}\mathbf{\Sigma}_{(s)}^{[0]}(\infty)(\mathbf{W}^{(s)})^{\top}\boldsymbol{\Lambda}^{(s)}+\boldsymbol{\Lambda}^{(s)} \mathbf{W}^{(s)}	\mathbb{E}[\mathbf{x}(\infty)]\mathbf{u}^{\top}(\mathbf{I}_n-\boldsymbol{\Lambda}^{(s)} )\\
	&+(\mathbf{I}_n-\boldsymbol{\Lambda}^{(s)}) \mathbf{u}\mathbb{E}[\mathbf{x}(\infty)]^{\top}(\mathbf{W}^{(s)})^{\top}\boldsymbol{\Lambda}^{(s)}	+(\mathbf{I}_n-\boldsymbol{\Lambda}^{(s)} )\mathbf{u}\mathbf{u}^{\top}(\mathbf{I}_n-\boldsymbol{\Lambda}^{(s)} )+\mathbf{Q}_{\eta}.
\end{split}
\end{align}

%

\end{tcolorbox}

\begin{multicols}{2}

\section{Inference over networks}

The models presented in the last sections have proven to be powerful tools for the analysis of interactions in social networks. However, in general, the structure of the network is not available. Hence, the following question arises: \textit{Given measurements of the evolution of the opinions, and a model of the opinion evolution, how can one estimate the interaction graph and the strength of the connections?}

With this in mind, in the second part  of this paper, we describe recent approaches arising in the literature to infer the social influences in a given a group of individuals, whose opinions on $m$ independent issues are supposed to evolve according to a pre-specified model.
Here, we will limit our study to the Friedkin-Johnsen model. The approaches described can be  adapted to the \ant{DeGroot model and some other models of opinion formation. Two strategies to estimate the interactions in the network are considered} that are referred to as {\em persistent measurement} and {\em sporadic measurement} identification procedures.

In the experiments of the first kind (persistent measurement) the opinions are observed during $T$ rounds of conversation, and the influence matrix is estimated as the matrix best fitting the dynamics for $0 \leq k < T$. In such cases, the available results on parsimonious systems identification \ant{can be used to determine the unknown parameters}~\cite{c45774ed5f4e4222999fe72642cd6adb,Giannakis2018}.

To exemplify how such an approach can be used in the context of network inference estimation, assume that the opinions evolve according to the multidimensional Friedkin-Johnsen model \eqref{eq.fj-m}, which we recall here for readability
\begin{equation}\label{eq.fj-m2}
\mathbf{X}(k+1)=\mathbf{\Lambda}\Wb\mathbf{X}(k)+(\mathbf{I}_n-\mathbf{\Lambda})\mathbf{X}(0).
\end{equation}
Assume that
measurements of $\mathbf{X}(k)$ are available for $k=0,1,2,\ldots,T$ (This assumption can be relaxed by assuming that a sufficient number of measurement pairs
$(\mathbf{X}(k+1),\mathbf{X}(k))$ are available.).
To simplify the development to follow, we first rewrite this model in a standard system identification form, as follows
\[
\mathbf{X}(k+1)=\mathbf{A}\mathbf{X}(k)+\mathbf{B}\mathbf{X}(0),
\]
with $\Ab\doteq \mathbf{\Lambda}\Wb$, $\Bb=\mathrm{diag}(b_1,b_2,\ldots,b_n)\doteq \mathbf{I}_n-\mathbf{\Lambda}$.

Now, denote by $\mathrm{off}(\Ab)\doteq\Ab- \mathrm{diag}(\Ab)$ the matrix composed by the off-diagonal elements of $\Ab$ and having a  zero diagonal. Then, given measurement error $\epsilon$, the problem of estimating the sparsest interaction graph that is compatible with the measurements collected can be formulated as the following optimization problem
\begin{align*}
\min_{\Ab,\Bb} \  & \|\mathrm{off}(\Ab)\|_0 \\
\text{subject to } & \| \mathbf{X}(k+1) - \mathbf{A}\mathbf{X}(k) -\mathbf{B}\mathbf{X}(0)\|_\infty \leq \epsilon \\
& \hskip 1.4in k=0,1,2\ldots, T-1 \\
& \Bb=\text{diag}(b_1,b_2,\ldots,b_n) \\
& \sum_{j=1}^n \mathbf{A}_{i,j} = 1-b_i \text{ for } i=1,2,\ldots n \\
& \mathbf{A}_{i,j} \geq 0 \text{ and } 0 \leq b_i \leq 1 \text{ for } i,j=1,2,\ldots,n
\end{align*}
where~$\|\mathbf{A}\|_0$ denotes the number of non-zero entries of the matrix~$\mathbf{A}$.

The optimization problem above is a non-convex combinatorial problem due to the presence of the zero-norm cost. To approximate the solution, a commonly used convex relaxation is to relax this norm to the $\ell_1$-norm  (see box on Compressed Sensing for additional details)

\begin{align*}
\min_{\Ab,\Bb} \  & \|\mathrm{off}(\Ab)\|_1\\
\text{subject to } & \| \mathbf{X}(k+1) - \mathbf{A}\mathbf{X}(k) -\mathbf{B}\mathbf{X}(0)\|_\infty \leq \epsilon \\
& \hskip 1.4in k=0,1,2\ldots, T-1 \\
& \Bb=\text{diag}(b_1,b_2,\ldots,b_n) \\
& \sum_{j=1}^n A_{i,j} = 1-b_i \text{ for } i=1,2,\ldots n \\
& A_{i,j} \geq 0 \text{ and } 0 \leq b_i \leq 1 \text{ for } i,j=1,2,\ldots,n
\end{align*}
\ant{Since $\|\mathrm{off}(\Ab)\|_1\doteq \sum_{i=1}^n\sum_{j\ne i}  A_{ij}$}, the latter problem  can be decomposed into $n$ independent problems.
This is especially useful in very large networks.

The drawback of these methods is that they require knowing the discrete-time indices for the observations made and to store a sufficiently long subsequence of opinions $\mathbf{x}(k),\ \mathbf{x}(k+1),..,\ \mathbf{x}(k+M-1)$. This knowledge may be difficult to obtain in general and the collection may involve a large amount of data.  The loss of data from one of the agents in general requires to restart the experiment.
Moreover, the system could be updated with an unknown interaction rate and the interaction timing between agents can be \ant{unobservable in practice}~\cite{citeulike:1449789, Wang2015103}. {These considerations \ant{make} the persistent measurement approach \ant{inapplicable} in many practical situations as also discussed in~\cite{Wai2016ActiveSO}.}

To circumvent {these} issues, in this paper we describe two approaches that fall in the second class of methods; i.e., they only use sporadic data and, therefore, complete history of agents' opinions is not required and the interactions are not limited to any prescribed number of rounds. In the first one, and similar to the experiments from \cite{Wai2016ActiveSO}, the agents interact until their opinions stabilize and the identification problem considers only the initial and the final opinions. In the second one, it is only assumed that one has access to random measurements of the agents' opinions and statistics of the measurement process are used to estimate the structure of the social network that generated the measurements.




\section{The influence estimation problem: infinite horizon approach}

As a first approach to the problem of estimating the structure of social network from infrequent data, we consider the Friedkin-Johnsen model in \eqref{eq.fj-m2}, and assume that we have knowledge of
$n$ the prejudices $\mathbf{X}(0)$ and the final opinions $\mathbf{X}({\infty})=\lim_{k \rightarrow \infty}\mathbf{X}(k)$. Then, our goal is to estimate $\Wb$ from this data only, under the assumption of network sparsity. To this end, some model identifiability considerations need to be made. This is done in the next subsection

\subsection{Model identifiability}

We first notice that, due to the \textit{consensus preservation} property discussed in the box \boxref{Simple properties of the FJ model}, whenever the initial opinions are at consensus, then also the final opinions are at consensus. It is noticed  that in this case the problem is not well posed, since any stochastic matrix $\Wb$ will be consistent with the data.
 Motivated by this consideration, from now on we assume that for all $\ell=1,\ldots,m$ there exists $i,j\in\mathcal{V}$ such that $x^{(\ell)}_i(0)\neq x^{(\ell)}_j(0)$.

Similarly, we know that when all agents are completely susceptible, i.e.~$\mathbf{\Lambda}= \mathbf{I}_n$), \ant{the FJ model reduces to DeGroot's model, typically leading to consensus of opinions.} Clearly, the problem is not well posed also in this case, since there are infinitely many matrices  leading  the dynamics in \eqref{eq.fj-m2} to the same value of consensus. This fact is illustrated in the following example, borrowed from \cite{Ravazzi2018}.
\begin{example}
Let $\mathbf{\Lambda}=\mathbf{I}_n$, and let $\mathbf{W}\in\R^{n\times n}$ be any doubly stochastic matrix that is \ant{irreducible (the graph $\G[\Wb]$ is strongly connected)} and aperiodic.
Then, Perron Frobenius theorem \cite{Perron} guarantees that
$$
\mathbf{X}(\infty)
=\mathds{1}\mathds{1}^{\top}\mathbf{X}(0)/n=\mathds{1}\mathds{1}^{\top}\mathbf{X}(0)/n.$$
%
Similarly, if $\mathbf{\Lambda}=0$, then $\mathbf{X}(\infty)=\mathbf{X}(0)$ and all 
stochastic matrices $\Wb$ are consistent with the data.
\end{example}

We remind that an agent $i$ with susceptibility $\lambda_i=0$ is totally stubborn, i.e.\  is not influenced by any other agent. Hence, to avoid ambiguities, in the rest of the paper we suppose that $\lambda_i\neq 0$ for all $i\in\V$, $\mathbf{\Lambda}\neq \mathbf{I}_n$, and for any node $i\in \V$ there exists a path from
$i$ to a node $j$ such that  $\lambda_{j}<1$ (each agent is influenced by at least one partially stubborn agent). With these assumptions, for any initial profile the opinion dynamics leads asymptotically to an equilibrium point that can be computed from the weights, the obstinacy levels, and the initial opinions.

It follows that  recovering $\Wb$ amounts at solving the following system of equations
\begin{equation}\label{system-id}
 \begin{cases}(\mathbf{I}_n-\mathbf{\Lambda} \mathbf{W})\mathbf{x}^{(\ell)}({\infty})=(\mathbf{I}_n-\mathbf{\Lambda})\mathbf{x}^{(\ell)}(0),\\ \mathbf{W}\mathds{1}=\mathds{1},\\ \mathbf{W}\geq0, \mathbf{\Lambda}\geq0
 \end{cases}
\end{equation}
However, as shown in \cite{Ravazzi2018}, this system contains an implicit ambiguity: if  $(\mathbf{\Lambda},\mathbf{W})$ is a solution of \eqref{system-id} then it is possible to construct a different solution
pair $(\mathbf{\Lambda}',\mathbf{W}')$ as follows:
\begin{gather*}\label{eq:problem}
\mathbf{\Lambda}'=\mathbf{I}_n-\mathbf{D}(\mathbf{I}_n-\mathbf{\Lambda})\\
\mathrm{off}(\mathbf{\Lambda}'\mathbf{W}')=\mathbf{D}\ \mathrm{off}(\mathbf{\Lambda} \mathbf{W})\\
\mathrm{diag}(\mathbf{\Lambda}'\mathbf{W}')=\mathds{1}-\mathbf{D}((\mathbf{I}_n-\mathbf{\Lambda})\mathds{1}+\mathrm{off}(\mathbf{\Lambda} \mathbf{W})\mathds{1}),
\end{gather*}
for any non-negative diagonal matrix $\mathbf{D}$ with $[\mathbf{D}]_{ii}\in[0,1]$.
The ambiguity described, which was already pointed out in  \cite{Wai2016ActiveSO} in the setting of De Groot models with stubborn agents,
 is due to the fact that the information about the rate of social interactions is missing, and
it can not be removed without making the additional assumption that
the susceptibilities  $\mathbf{\Lambda}$ are known.
In this case, for $m\geq n$, if the system in \eqref{system-id} is full rank then the problem in \eqref{system-id} admits a unique solution, and may be easily solved, e.g using linear programming or any solver for convex optimization \cite{cvx}.

Following these considerations, from now on we assume that $\mathbf{\Lambda}$ is known, and we focus the more interesting case when $m<\!\!\!<n.$
It should be remarked that, if also the matrix $\mathbf{\Lambda}$ is part of
the learning, we must define an invariant quantity among the ambiguous solutions, for instance by defining equivalence classes and resolve the ambiguity by imposing constraints on $\mathrm{diag}(\mathbf{W})$. This is further discussed at the end of Section \ref{sec:l1}.

\subsection{Sparse identification problem}
Motivated by the discussion in Section \ref{subsec:Structure_OSN}, we base our identification approach on  the observation that a social network is typically sparse, in the sense that the interactions among the agents are few when compared to the network dimension. For given $\mathbf{\Lambda}$, $\mathbf{X}(0)$, and $\mathbf{X}(\infty)$, this leads us to estimate the social influence networks by solving a \emph{sparsity problem}.
Formally, determining the sparsest network that is compatible with the available information can be expressed as the following  $\ell_0$-minimization problem\cite{citeulike:2688127}
\begin{gather}\label{eq:id_problemo}
\min_{\mathbf{W}\in\R^{n\times n}}\|\mathbf{W}\|_0,\qquad
\text{s.t. }
 \begin{cases}\mathbf{\Phi}\mathbf{W}^{\top}=\mathbf{\Psi}^{\top},\\ \mathbf{W}\mathds{1}=\mathds{1},\\ \mathbf{W}\geq0.
  \end{cases}
  \end{gather}
where $\|\|\mathbf{W}\|_0\|$ counts the number of nonzeros of the matrix $\mathbf{W}$, $\mathbf{\Phi}\doteq \mathbf{X}({\infty})^{\top}$, $\mathbf{\Psi}\doteq\mathbf{\Lambda}^{-1}[\mathbf{X}({\infty})-(\mathbf{I}_n-\mathbf{\Lambda})\mathbf{X}(0)]$.

It should be noticed that this problem is separable into $n$ subproblems, since each row of $\Wb=[\wb_1^{\top},\ldots,\wb_n^{\top}]^{\top}$ can be learned independently from the others. More precisely,
\begin{align}\begin{split}\label{eq:id_problem}
&\min_{w_j\in\R^{n}}\|\wb_j\|_0,\qquad
\text{s.t. }
 \begin{cases}\mathbf{\Phi}\wb_j=\boldsymbol{\psi}_j,\\\mathds{1}^{\top}\wb_ j=1,\\ \wb_j\geq0.
  \end{cases}\end{split}
\end{align}
where $\boldsymbol{\psi}_j$ equals to $j$-th row of $\mathbf{\Psi}$ for every $j\in[n]$.

As discussed in the box \boxref{Schur stability criteria}, the reachability of each node from a partially stubborn node is an assumption that the true network must satisfy to guarantee the stability of the affine dynamics in \eqref{eq.fj-m2} and the existence of the final opinion profile. However, as it should be noticed in the optimization problem \eqref{eq:id_problem}, this constraint is not imposed in the recovery problem.


\end{multicols}

\begin{tcolorbox}[title= \sf \textbf{Compressed sensing},colframe=airforcecarmine!20,colback=airforcecarmine!20,coltitle=black,
]
\sf\small

          The optimization problems in \eqref{eq:id_problem} are a particular case of the so-called sparse recovery problem starting from compressed measurements \cite{citeulike:2688127}, a problem also known as Compressed Sensing (CS). More precisely, sparse recovery problems are of the form
          \begin{align*}
          \min_{\mathbf{z} \in \R^n} & \|\mathbf{z}\|_0 \\
          \text{s.t. } & \mathbf{\Phi} \mathbf{z} = \boldsymbol{\psi}
          \end{align*}
          where $\mathbf{\Phi} \in \R^{m \times n}$ with $m <n$, and
          $\|\mathbf{z}\|_0$ defines the  $\ell_0$ quasi-norm, which corresponds to the   \cl{\sout{as the}} number of nonzero elements of $\mathbf{z}$.
          It should be noted that the linear system of equations in the optimization problem above is underdetermined and admits infinitely many solutions. However, a sufficient condition for determining a solution can be derived exploiting the sparsity of the desired solution and using the notion of {\em{spark}} of a matrix \cite{eldar2012compressed}.
          \vskip .1in

          \noindent
          \textbf{Spark of a matrix.}
          	The spark of a given matrix $\mathbf{\Phi}$, denoted with $\mathrm{spark}(\mathbf{\Phi})$, is the smallest number of columns from $\mathbf{\Phi}$ that are linearly dependent.
      	  \vskip .1in
          When dealing with sparse vectors, the spark concept provides a complete characterization of when sparse recovery is possible. The interested reader can refer to \cite[Theorem 1.1]{eldar2012compressed} for a proof.
          \vskip .1in
          \noindent
          \textbf{Proposition 1.}\label{prop:spark}
          	For any vector $\boldsymbol{\psi}$, there exists at most
          	one vector $\mathbf{z}$ such that $\boldsymbol{\psi}=\mathbf{\Phi}\mathbf{z}$ if and only if \[\mathrm{spark}(\mathbf{\Phi})>2\|\mathbf{z}\|_0.\]
          \vskip .1in

          Computing the spark of a matrix involves checking the dependence of combinations of columns. Testing the condition in Proposition 1 is computationally expensive for practical purposes, as it requires a combinatorial search.
          Moreover the CS problem described above is known to be, in general,  NP-hard. For this reason  $\ell_1$-based relaxations are often used to approximate the solution of CS problems. More precisely, this relaxation has the form
           \begin{align*}
          \min_{\mathbf{z} \in \R^n} & \|\mathbf{z}\|_1 \\
          \text{s.t. } & \mathbf{\Phi} \mathbf{z} = \boldsymbol{\psi}
          \end{align*}

          Much of the theory concerning explicit performance bounds for the relaxation described above  relies on the concept of
          Restricted Isometry Property (RIP). RIP characterizes matrices which are nearly orthonormal, at least when acting on sparse vectors \cite{citeulike:4109966}.

          \vskip .1in
          \noindent
          \textbf{Restricted Isometry Property.}
          	Let $\mathbf{\Phi}\in\R^{m\times n}$. Suppose that there exists a constant
          	$\delta _{s}\in (0,1)$ such that
          	$$(1-\delta _{s})
          	\|\zb\|_{{2}}^{2}\leq \|\mathbf{\Phi}\zb\|_{{2}}^{2}\leq (1+\delta _{s})\|\zb\|_{{2}}^{2}.
          	$$
          	for all $\zb\in\mathcal{Z}_s=\{\zb\in\R^{n}:\|\zb\|_0\leq s\}$.
          	Then, the matrix $\mathbf{\Phi}$ is said to satisfy the $s$-restricted isometry property with restricted isometry constant $\delta_s$.

          \vskip .1in

          Denote with $\mathbf{\Phi}_S$ the matrix with columns indexed by $S\subseteq[n]$.
          It can be shown (see \cite{citeulike:4109966}) that, if a given matrix satisfies the RIP of order $2s$ with a constant $\delta _{2s}\in(0,1/(\sqrt{2}+1))$, then one can uniquely recover a $s$-sparse vector using the $\ell_1$ relaxation described above.
          It is straightforward to see that
          $$(1-\delta _{s})\leq
          \mu_{\min} (\mathbf{\Phi}_S^{\top}\mathbf{\Phi}_S)\leq\mu_{\max}(\mathbf{\Phi}_S^{\top}\mathbf{\Phi}_S)\leq (1+\delta _{s}).
          $$
          where  $\mu_{\min}$ and $\mu_{\max}$ denote the smallest and largest eigenvalue. Consequently, it must hold
          $$1\approx\frac{\mu_{\max}(\mathbf{\Phi}_S^{\top}\mathbf{\Phi}_S)}{\mu_{\min} (\mathbf{\Phi}_S^{\top}\mathbf{\Phi}_S)}\leq 2,
          $$
          for all $S\subseteq[n]$ with $|S|\leq s$.

            \vskip .1in

          It is well known \cite{citeulike:4109966} that sub-Gaussian random matrices with i.i.d. entries satisfy the RIP of order $2s$ with constant $\delta_{2s}$ with probability close to 1  if $$m \geq \frac{cs}{\delta_{2s}^{2}} \log \left(\frac{n}{\delta_{2s}s}\right)$$ where $c>0$ is a positive constant.
          It is worth noticing that, in the classical framework of CS the sensing matrix is generally chosen by the user and is independent of the signal to be recovered.

          \par\medskip
        \end{tcolorbox}

\begin{multicols}{2}

\clearpage
\newpage
\subsection{Recovery via convex optimization}\label{sec:l1}
The convex relaxation of \eqref{eq:id_problem} \ant{(where the constraint $\wb_j\geq0$ is removed)}
\begin{align}\begin{split}\label{eq:l1}
&\min_{\wb_j\in\R^{n}}\|\wb_j\|_1,\qquad
\text{s.t. }
 \begin{cases}\mathbf{\Phi}\wb_j= \boldsymbol{\psi}_j,\\\mathds{1}^{\top}\wb_ j=1,
  \end{cases}
  \end{split}
\end{align}
can be formulated as a linear program and has been extensively studied; see box \boxrefb{Compressed Sensing}.  A large amount of algorithms have been proposed aiming at  solving it efficiently especially for the case where the dimension of the vector to be sparsified is high \cite{eldar2012compressed}.

It is well known that under certain conditions on the matrix $\mathbf{\Phi}$, the number of measurements $m$,  and the sparsity of $\wb_{j}$, both \eqref{eq:id_problem} and \eqref{eq:l1} have the same unique solution \cite{citeulike:2688127}.
However, in the case of influence estimation in SDNs considered in this paper and contrary to other problems in compressed sensing, the sensing matrix  $\mathbf{\Phi}$ cannot be designed to satisfy the recovery properties mentioned above \ant{because it depends on the model's parameters}
$$\mathbf{\Phi}=\mathbf{X}(\infty)^{\top}=\mathbf{X}(0)^{\top}(\mathbf{I}_n-\mathbf{\Lambda})(\mathbf{I}_n-\mathbf{\Lambda} \mathbf{W})^{-1\top}.
$$
In the case where the initial opinions are independent and identically distributed Gaussian random variables  and the agents are all ``very stubborn'' ($\mathbf{\Lambda}$ has small diagonal entries) one could in principle consider $\mathbf{\Phi}\approx \mathbf{X}(0)^{\top}$ which would  satisfy the RIP recovery condition with high probability. However, this is a very special case that does not cover most of the SDN recovery problems of interest.
If $\mathbf{\Lambda}$ is not close to zero, then $\mathbf{\Phi}$ is a random variable whose entries are coupled and available results on compressed sensing do not apply. In the remainder of this section, we review the results in \cite{Ravazzi2018} where recovery conditions specific to SDNs are derived.

From now on, in this section, we  assume that
the initial opinions $\xb^{(\ell)}(0)$ on topic $\ell$ are independent and identically distributed random variables having a Gaussian distribution with zero mean and unit variance.
The hypothesis on the Gaussian distribution of the initial condition is a common assumption in opinion dynamics literature \cite{MirtabatabaeiJiaBullo2014}.
It can be explained also by the fact that
 initial opinions can be argued to be pre-averaged opinions or several criteria which can be treated as independent random variables. Therefore, the distribution of initial opinions has a Gaussian-like shape due to the central limit theorem \cite{Lorenz09arxiv}.

Moreover, the assumption on the zero mean and identity covariance matrix is not a restrictive one. Given any   Gaussian distribution of the initial opinions, one can always perform a linear transformation and obtain an equivalent problem that satisfies the assumptions. If $x^{(\ell)}(0)$ have nonzero  expected value, then we can consider $x^{(\ell)}(0)-\overline{x}^{(\ell)}(0)\mathds{1}$ and $z^{(\ell)}(\infty)=\mathbf{V}(x^{(\ell)}(0)-\overline{x}^{(\ell)}(0)\mathds{1})$ where $\mathbf{V}$ is the total effects matrix. Since the total effects matrix is  stochastic
then $z^{(\ell)}(\infty)=x^{(\ell)}(\infty)-\overline{x}(0)^{(\ell)}\mathds{1}$.

At this point, we should note that for SDNs influence estimation problem formulated in this section, the probability of violating the RIP condition can be can ve very close to one  even for very simple graphs.
More precisely, in \cite{Ravazzi2018} it is proven that for Gaussian initial conditions
$x(0)^\ell\sim\mathcal{N}(0,I)$ for all $\ell\in[m]$ and for an arbitrary set $S\subseteq [n]$ of size $|S|=s$, for the defined $\widehat{\boldsymbol{\boldsymbol{\Sigma}}}_{SS}=\mathbf{\Phi}_S^{\top}\mathbf{\Phi}_S/m$, with probability greater than $1-2\e^{-m/32}$ we have
$$
\frac{\mu_{\max}(\widehat{\boldsymbol{\boldsymbol{\Sigma}}}_{SS})}{\mu_{\min}(\widehat{\boldsymbol{\boldsymbol{\Sigma}}}_{SS})} \geq\frac{1}{3}\frac{\mu_{\max}(\boldsymbol{\boldsymbol{\Sigma}}_{SS})}{\mu_{\min}(\boldsymbol{\boldsymbol{\Sigma}}_{SS})}
$$
where $\boldsymbol{\boldsymbol{\Sigma}}=(I-\mathbf{\Lambda} \mathbf{W})^{-1}(I-\mathbf{\Lambda})^2(I-\mathbf{\Lambda} \mathbf{W})^{-\top}$.


For this reason, more powerful tools are need for the analysis of the performance of the problem at hand. More precisely, we need the so-called ``nullspace property''~\cite{citeulike:2688127} which provides a necessary and sufficient condition for recovery. This \ant{property is summarized in} in the box \boxref{Necessary and Sufficient Conditions for Recovery}.
\medskip

\begin{tcolorbox}[title= \sf \textbf{Necessary and sufficient conditions for recovery},colframe=carmine!10,colback=carmine!10,coltitle=black,
	]
	\sf \small
To be able to derive more general conditions for sparse recovery we need the concept of Null Space Property (see~\cite{citeulike:2688127})

\vskip .1in
\noindent
\textbf{Null Space Property (NSP).} The matrix $\mathbf{\Phi}\in\R^{m\times n}$ satisfies the NSP of order $s$ if, given
$$
\mathcal{C}(\ell)=\{\eta\in\R^n:\|\eta_{S^{c}}\|_1\leq \|\eta_S\|_1\},
$$
$$
\mathcal{C}(\ell)\cap\mathsf{Ker}(\mathbf{\Phi})=\{0\}.
$$
for all index set $S$ with $|S|\leq s$ where $\text{Ker}$ denotes the kernel of a matrix.
\vskip .1in

With this definition at hand, Theorem 1 in \cite{citeulike:2688127} provides additional results on when one can recover sparse solutions from systems of linear equations. More precisely, consider matrix $\mathbf{\Phi}\in\R^{m\times n}$. Then, the optimization problem
$$\{\min{\|\mathbf{z}\|_1: \mathbf{\Phi}\mathbf{z} = \boldsymbol{\psi}}\}$$
uniquely recovers all $s$-sparse vectors $\mathbf{z}^{\star}$
from measurements $$\boldsymbol{\psi} = \mathbf{\Phi}\mathbf{z}^{\star}$$ if and only if
$\mathbf{\Phi}$ satisfies the nullspace property with order $2s$.
To further analyze when can sparse solutions can be recovered, lets introduce the concept of Restricted Eigenvalue Criterion.

\begin{definition}[Restricted Eigenvalue Condition (REC)]
We say that a matrix $\mathbf{\Phi}$ satisfies the REC of order $s$ if there exists a $\delta_s>0$ such that
$$
\frac{1}{m}\|\mathbf{\Phi}\mathbf{z}\|_2^2\geq\delta_s^2\|\mathbf{z}\|_2^2
$$
for all $\mathbf{z}\in\mathcal{C}(\ell)$, uniformly for all index sets $S\subseteq[n]$ with $|S|\leq s$.
\end{definition}
It is straightforward to see that REC  is equivalent to NSP.
For random matrices $\mathbf{\Phi}$ with i.i.d. entries drawn from particular distributions or for unitary matrices, it can be shown that, if enough measurements are available, then REC condition is satisfied with the prescribed $\delta_s$ with probability close to 1 \cite{Candes:2005:DLP:2263433.2271950}.
\end{tcolorbox}

These  necessary and sufficient conditions on sparse recovery enable one to study when is it  possible to recover sparse models for the SDN at hand. In \cite{Ravazzi2018} it is shown that if the initial condition
$\mathbf{x}^{(\ell)}(0)\sim\mathcal{N}(0,\mathbf{I}_n)$
and if the number of considered topics satisfies 
\begin{equation}\label{eq:condition_m}m\geq 4c \frac{(1+\lambda_{\max})^2(1-\lambda_{\min})^2}{(1-\lambda_{\max})^4} d_{\max}\log{n}\end{equation}
then the solution to \eqref{eq:l1} is unique and coincides with that of \eqref{eq:id_problem} with probability at least $1-c'\e^{-c''m}$, where $c,c'$ and $c''$ are positive constants, $d_{\max}=\max_{v\in\V}|\mathcal{N}(v)|$, $\lambda_{\max}=\max_{j}\lambda_{j}$, and $\lambda_{\min}=\min_{j}\lambda_{j}$.

If we look closely at  condition \eqref{eq:condition_m}, one can see that, to be able to recover a sparse influence model, the sensitivity  to other opinions cannot be high.
More precisely, if $\lambda_{\max}\rightarrow 1$ then the number of measurements needed for recovery diverges to infinity.
This is reasonable, since the final opinions are a function of preconceived opinions and the network sensing performance should depend on the strength of the influencing power of the prejudices.

Moreover, as conjectured in \cite{Wai2016ActiveSO}, another important issue that affects the reconstruction performance is the degree distribution in the social network. More precisely, for a fixed total number of edges, it is easier to recover a network with a concentrated degree distribution (e.g., the Watts-Strogatz network \cite{watts1998cds}) while a network with power law degree distribution (e.g., the Barabasi-Albert network \cite{Albert:2001:SMC:933363}) is more difficult to recover.

To finalize the discussion in this section, recall  that if $\mathbf{\Lambda} $ is not known then the identification problem is not well-posed. The ambiguity is due to the missing information about the rate of social interactions. This ambiguity can not be removed without making additional assumptions. However, we can determine an invariant quantity among the ambiguous solutions by defining equivalence classes and resolve the ambiguity by imposing constraints on $\mathrm{diag}(\mathbf{W})$. More precisely, in \cite{Ravazzi2018}, it is shown that the problem of learning sensitivity matrix $\mathbf{\Lambda}$
 can be cast as in \eqref{eq:l1} with $\mathbf{\Phi}\doteq [\mathbf{X}(\infty)^{\top}, \mathbf{x}_{j}(0)-\mathbf{x}_{j}(\infty)]$ and $\mathbf{B}=\mathbf{X}(0)$, where $\mathbf{x}_j(\infty), \mathbf{x}_j(0)$ are the column vector corresponding to $j$-th row of $\mathbf{X}(\infty)$ and $\mathbf{X}(0)$, respectively, with the additional constraint that $w_{jj}=0.$

%
%
%
%

\begin{figure*}[t]
\includegraphics[trim={0cm 0cm 0 0cm},clip, width=2\columnwidth]{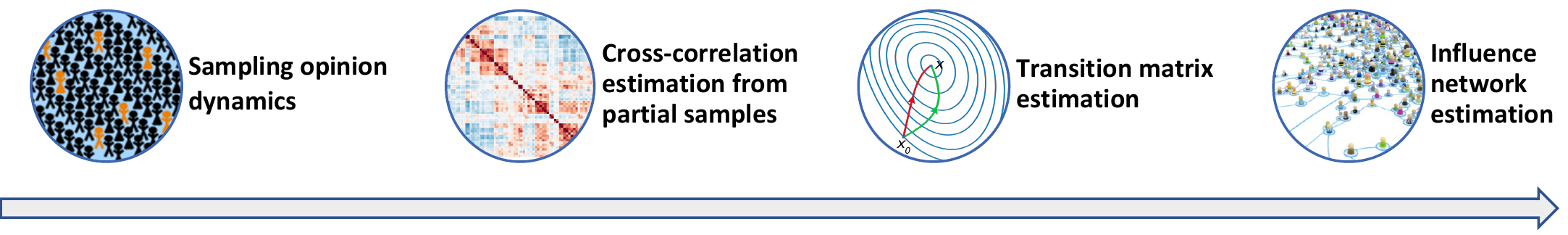}\caption{Main stream of the methodology.}\label{fig:Flusso}.
\end{figure*}

The main stream of the methodology is summarized in Figure \ref{fig:Flusso}

\section{Influence estimation from random opinion measurements} \label{sec:SIENNA}

In this section, we review an alternative approach to the social network estimation problem that exploits the availability of ``intermittent'' measurements of the opinions  to identify the dynamics of the evolution of the opinions and, as a consequence, the influence matrix \cite{Yonina-1,Yonina-2}. Such an approach is especially useful in the case where not all opinions are updated at the same time and random sampling of the opinions might be a less onerous way of estimating the behavior of the network. 

Hence, in this section, we concentrate on the asynchronous gossip-based FJ model and assume that random measurements of
opinions are available. For simplicity, as in \cite{RAVAZZI2020}, we consider the case when a single  topic is discussed, but the reasoning can be easily extended to cases involving multiple topics.

\subsection{Observation models}
As mentioned above, lets consider the gossip opinion dynamics in \eqref{eq:gossip-friedkin} where the influence matrix $\mathbf{W}$ is unknown. We assume that, at each time $k$ we not not have complete knowledge of the opinion vector~$\mathbf{x}(k)$, only partial information is available. More precisely, we assume the following random model for the observations
\begin{equation}\label{eq:obs}
\mathbf{z}(k)=\mathbf{P}(k)\mathbf{x}(k)
\end{equation}
where the diagonal matrix $\mathbf{P}(k)$ is a random measurement matrix defined by
$$
\mathbf{P}(k)=\mathrm{diag}(\mathbf{p}(k))
$$
and $\mathbf{p}(k)\in\{0,1\}^{n}$ is a random selection vector with known distribution representing which opinions are measured at time $k$. Different  probability distributions of the matrix $\mathbf{P}(k)$ lead to very different observation models. For example,
%
if 
\[
\mathbf{p}(k)=\begin{cases}\1&\text{w.p.}\ \rho\\
0&\text{otherwise}
\end{cases}
\]
then we have the so-called \emph{intermittent observation model } where  at $k\in\integernonnegative$ all observations are available with probability $\rho$ or no observations at all are observed.
This model allows to capture the typical situation in which the actual rates at which the interactions occur is not perfectly known,
and thus sampling time is different from interaction time.

Moreover, if at each time $k\in\integernonnegative$ the selection vector is $p_i(k)\sim \mathrm{Ber}(\rho_i)$ for all $i\in\V$ then we have the so-called \emph{independent random sampling model} \cite{Yonina-1,Yonina-2} where the opinions are observed independently with probability $\rho_i\in[0,1]$. In the case where the observations are made with equal probability $\rho_i=\rho$ for all $i\in\V$ this model is referred to as  {\em independent and homogeneous sampling}. If $\rho= 1$ we have full observations, and if $\rho\neq 1$, we have partial information.
This model has a clear interpretation for SDNs, describing the  situation where only a subset of individuals can be contacted at each time $k$ (e.g.\ random interviews).

In this section, we review the approach described  in \cite{RAVAZZI2020} where the objective is: Given the sequence of observation $\{\mathbf{z}(k)\}_{k=1}^{t}$ estimate of the matrix $\mathbf{W}$, call it $\widehat{\mathbf{W}}_t$. In \cite{RAVAZZI2020}, theoretical conditions are also provided on the number of samples that are sufficient to have an error not larger than a fixed tolerance $\epsilon$ with high probability. For clarity of exposition, these theoretical results are not reviewed here.


\subsection{Overview of the proposed approach to influence estimation}
To reconstruct the influence matrix, we start by recalling the  definition of  opinions' cross-correlation matrix
$$
\boldsymbol{\boldsymbol{\Sigma}}^{[\ell]}(k):=\mathbb{E}\left[\mathbf{x}(k)\mathbf{x}(k+\ell)^{\top}\right].
$$
We have seen  that the evolution of the covariance matrix  $\boldsymbol{\boldsymbol{\Sigma}}^{[\ell]}(k)$ is described by
	\begin{equation}\label{eq:Sigmak}
	\mathbf{\Sigma}^{[\ell+1]}(k)=\mathbf{\Sigma}^{[\ell]}(k)\overline{\mathbf{\Gamma}}^{\top}+\mathbb{E}[\mathbf{x}(k)]\overline{\boldsymbol{b}}^{\top}.
	\end{equation}
	Moreover $\mathbf{\Sigma}^{[\ell]}(k)$  converges to $\mathbf{\Sigma}^{[\ell]}({\infty})$ for all non-negative integer $\ell$ which satisfy
	\begin{equation}\label{eq:YW}
	\mathbf{\Sigma}^{[\ell+1]}(\infty)=\mathbf{\Sigma}^{[\ell]}(\infty)\overline{\mathbf{\Gamma}}^{\top}+\mathbb{E}[\mathbf{x}(\infty)]\overline{\boldsymbol{b}}^{\top}.
	\end{equation}

The simple linear equation above provides the motivation for the approach described in this section. This approach can be summarized as follows: First, from the partial random measurements $\{\mathbf{z}(k)\}_{k=1}^t$, estimate the expected terminal state $\mathbb{E}[\mathbf{x}(\infty)]$ and the terminal covariance matrices $\boldsymbol{\boldsymbol{\Sigma}}^{[0]}(\infty)$ through $\boldsymbol{\boldsymbol{\Sigma}}^{[\ell]}(\infty)$ for some $\ell$. Given these estimates and using~\eqref{eq:YW}, estimate the matrix~$\overline{\mathbf{\Gamma}}$. Finally, to estimate the influence matrix $\mathbf{W}$ by exploiting the relation between $\overline{\mathbf{\Gamma}}$ and $\mathbf{W}$.


\subsection{Estimating the expected opinion profile and the cross-correlation matrices}
We now show how we can exploit the model of the observations and the data collected to estimate the opinion's expectation an covariance.
In order to estimate the expected opinion profile $\mathbb{E}[\mathbf{x}(\infty)]$, we start with time averages of the  observations $\mathbf{z}(k)$. It can be shown that
$$
\mathbb{E}[\mathbf{z}(k)]=\mathbf{\pi}\circ\mathbb{E}[\mathbf{x}(k)]
$$
where $\mathbf{\pi}=\mathbb{E}[\mathbf{p}(k)]$ and $\circ$ denotes the entrywise product.

This allows us to estimate the expectation of the opinions from available data. More precisely, we start by estimating  $\mathbb{E}[\mathbf{z}(k)]$ using time averages
$$
\overline{\mathbf{z}}(t)=\frac{1}{t}\sum_{k=1}^{t}\mathbf{z}(k)
$$
and obtain
\begin{equation}
\label{x_estimation}
\widehat{x}_i(t)=\frac{\overline{z}_i(t)}{\pi_i}.
\end{equation}

Estimating covariance matrices can be done in a similar way. More precisely,  the cross correlation matrices $\boldsymbol{\boldsymbol{\Sigma}}^{[\ell]}(\infty)$ are estimated from  the empirical covariance matrix of the observations $\mathbf{z}(k)$. Let us denote
$$
\mathbf{S}^{[\ell]}(k):=\mathbb{E}[\mathbf{z}(k)\mathbf{z}(k+\ell)^{\top}].
$$
Then,
$$\mathbf{S}^{[\ell]}(k)=\mathbf{\Pi}^{[\ell]}(k)\circ\boldsymbol{\boldsymbol{\Sigma}}^{[\ell]}(k)$$
where $\mathbf{\Pi}^{[\ell]}=\mathbb{E}[\mathbf{p}(k)\mathbf{p}(k+\ell)^{\top}]$ and $\circ$ denotes the Hadamard product.
Since $\mathbf{S}^{[\ell]}(k)$ is unknown we estimate $\mathbf{S}^{[\ell]}(k)$ using time averages
$$
\widehat{\mathbf{S}}^{[\ell]}(t)=\frac{1}{t-\ell}\sum_{k=1}^{t-\ell}\mathbf{z}(k)\mathbf{z}(k+\ell)^{\top}
$$
from which we get
\begin{equation}
\label{Sigma_estimation}
\widehat{\boldsymbol{\boldsymbol{\Sigma}}}_{ij}^{[\ell]}(t)={\widehat{\mathbf{S}}_{ij}^{[\ell]}(t)}/{\mathbf{\Pi}_{ij}^{[\ell]}}.
\end{equation}

Although these seem rather ad-hoc estimates of the needed quantities, It can be shown that they converge to the desired values as the number of measurements tend to infinity. More precisely, in \cite{RAVAZZI2020} a careful analysis of the procedures developed above shows that the estimates converge to the true values at a rate of $O(1/\sqrt{t})$ where $t$ is the number of measurements used.
	
As an example of the estimation procedure above, consider first the case of \emph{independent homogeneous random sampling}. In this case, we have $\pi=\rho$
\begin{align*}
 \mathbf{\Pi}^{[0]}&=\mathbb{P}(i,j\in\V_k)=\rho \mathbf{I}_n+\rho^2(\1\1^{\top}-\mathbf{I}_n) \\
\mathbf{\Pi}^{[\ell]}&=\mathbb{P}(i\in\V_k,j\in\V_{k+\ell})=\rho^2\1\1^{\top} \quad \text{if }\ell\neq 0
\end{align*}
from which $\widehat{x}(t)={\overline{z}(t)}/{\rho}$ and
\begin{gather}
\widehat{\boldsymbol{\boldsymbol{\Sigma}}}_{ij}^{[\ell]}(t)=\frac{1}{\rho^2}\widehat{\mathbf{S}}^{[\ell]}(t)-\left(\frac{1-\rho}{\rho^2}\widehat{\mathbf{S}}^{[\ell]}(t)\circ \mathbf{I}_n\right) \mathbf{1}(\ell=0)
\end{gather}

As a second example, let us consider the case of \emph{intermittent observations}. In this case, we have $\pi=\rho$
\begin{align*}
 \mathbf{\Pi}^{[0]}&=\rho\1\1^{\top}\quad\text{and}\quad
\mathbf{\Pi}^{[\ell]}=\rho^2\1\1^{\top} \quad \text{if }\ell\neq 0
\end{align*}
from which $\widehat{\mathbf{x}}(t)={\overline{\mathbf{z}}(t)}/{\rho}$ and
$\widehat{\boldsymbol{\boldsymbol{\Sigma}}}_{ij}^{[\ell]}(t)=\widehat{\mathbf{S}}^{[\ell]}(t)/{\rho^2}.
$

\subsection{Estimating the influence matrix}
In principle, the estimators  $\widehat{\mathbf{\Sigma}}^{[1]}(t)$ and $\widehat{\mathbf{\Sigma}}^{[0]}(t)$ of $\mathbf{\Sigma}^{[1]}(t)$ and $\mathbf{\Sigma}^{[0]}(t)$ together with \eqref{eq:YW} can be used to estimate the dynamics matrix $\overline{\mathbf{\Gamma}}$. However, there is a significant obstacle that one needs to deal with when using such a ``naive'' approach. Given the fact that one has random observations, it is likely that the procedure described above produces ``poor'' estimates of   $\mathbf{\Sigma}^{[1]}(t)$, due to the fact that, for several $k$,  many of the entries of $\mathbf{z}(k)\mathbf{z}(k+\ell)^{\top}$ might be zero.

To circumvent this,  we start by choosing a number $N_\Sigma$ of covariance matrices that are going to be considered in the estimation of dynamics and use a combination of these covariance matrices. More precisely, given estimates~$\widehat{\mathbf{\Sigma}}^{[\ell]}(t)$, we compute
\begin{equation}
\widehat{\mathbf{\Sigma}}_{-}(t)  \doteq \frac{1}{N_{\mathbf{\Sigma}}}\sum_{\ell=0}^{N_\Sigma-1} \widehat{\mathbf{\Sigma}}^{[\ell]}(t); \qquad
\widehat{\mathbf{\Sigma}}_{+}(t) \doteq \frac{1}{N_{\mathbf{\Sigma}}}\sum_{\ell=1}^{N_\Sigma} \widehat{\mathbf{\Sigma}}^{[\ell]}(t)
\end{equation}
and note that these matrices (approximately) satisfy \eqref{eq:YW}. Hence, can be used to estimate the structure of the network.

With this at hand, let us consider two types of networks. Let us first assume that one knows  in advance that the network is dense. In this case a possible estimator of $\overline{\mathbf{\Gamma}}$ can be obtained by directly solving the set of linear equations~\eqref{eq:YW}. In other words, the estimator is
\begin{equation}\label{eq:est_Gamma}
\widehat{\overline{\mathbf{\Gamma}}}(t)^{\top}=\widehat{\boldsymbol{\Sigma}}_{-}(t)^{\dag}(\widehat{\boldsymbol{\Sigma}}_{+}(t)-\overline{\mathbf{x}}(t)\overline{\boldsymbol{b}}^{\top})
\end{equation}

\medskip

In the case of networks that are known to be sparse, then one can solve a sparsity inducing optimization problem aimed at finding the sparsest graph that is compatible with available information. More precisely, in this case the estimator can be obtained by solving
\begin{align*}
&\qquad\qquad\widehat{\overline{\mathbf{\Gamma}}}(t)^{\top}=\argmin{\mathbf{M}\in\R^{\V\times\V}}\sum_{i,j,i\neq j}|\mathbf{M}_{ij}|\\
&\text{s.t.}\quad\|\widehat{\boldsymbol{\Sigma}}_{-}(t)\mathbf{M}-(\widehat{\boldsymbol{\Sigma}}_{+}(t)-\overline{\mathbf{x}}(t)\overline{\boldsymbol{b}}^{\top})]\|_{\max}\leq \eta
\end{align*}

\begin{tcolorbox}[title= \sf \textbf{Performance of influence estimation: asynchronous gossip-based FJ model},colframe=carmine!10,colback=carmine!10,coltitle=black,]
	\sf \small
	The estimation error on matrix $\overline{\mathbf{\Gamma}}(t)$ is based on the previous estimation of the cross-correlation matrices. In particular, using \eqref{eq:est_Gamma} we have to invert $\hat{\Sigma}_+$ and the estimation error depends on the singular values of $\hat{\Sigma}_\pm$. More precisely, it can be shown that,
with probability at least $1-\delta$, we have
\begin{align}
&\|\overline{\mathbf{\Gamma}}(t)-\widehat{\overline{\mathbf{\Gamma}}}(t)\|_2\nonumber\\
&\ =O\left(\frac{n(\sigma^{+}_{\max}+n)}{
{(\underline{\sigma}^-_{\min})}^2\Pi^{\star}\sqrt{\delta(t+1)\beta(1-\lambda_{\max})}}\right)
\end{align}
where $\sigma^{+}_{\max}=\|\Sigma_+\|_2$ and ${\underline{\sigma}^-_{\min}\doteq{\min({\sigma}^-_{\min},\widehat{\sigma}^-_{\min})}}$, being
$\sigma^-_{\min}$, $\widehat{\sigma}^-_{\min}$ the minimum singular value of $\Sigma_-$ and $\widehat{\Sigma}_-$, respectively.

\end{tcolorbox}

\subsection{Estimating the network topology and the influence matrix} \label{sec:est_topology}
Once an estimate of the average transition matrix $\overline{\mathbf{\Gamma}}(t)$ has been obtained, we can retrieve the topology of the influence network  in a straightforward manner, by noticing that $\mathrm{supp}(\overline{\mathbf{\Gamma}})=\mathrm{supp}(\mathbf{W})$. Hence, we can reconstruct the support of   $\mathbf{W}$  by taking the elements of the estimated matrix $\widehat{\overline{\mathbf{\Gamma}}}$ that are significantly larger than zero.

The estimation of the intensity of the influence can be done by exploiting previously developed results. More precisely, the following equality holds
$$
\widehat{\mathbf{W}}(t)=\widehat{\mathbf{D}}\mathbf{\Lambda}^{-1}\left[\overline{\mathbf{\Gamma}}(t)-(1-\beta)\mathbf{I}_n-\beta\mathbf{\Lambda}\left(\mathbf{I}_n-\widehat{\mathbf{D}}^{-1}\right)\right],
$$
where $\widehat{\mathbf{D}}$ represents an estimate of the degree matrix $\mathbf{D}$ obtained from the reconstructed support. That is, $\widehat{\mathbf{D}}$  is the diagonal matrix with elements
\[
\widehat{\mathbf{D}}_{i,i}=\|\mathrm{supp}(\boldsymbol{\gamma}_i)\|_0,
\]
with $\boldsymbol{\gamma}_i^\top$ being the $i$-th row of matrix $\widehat{\overline{\mathbf{\Gamma}}}$.

\subsection{Influence estimation in multiplex networks}
Also for the model described in the box~\boxref{F\&J model on multiplex networks} we can proceed by estimating the cross-correlation matrices and then use relations \eqref{eq:Sigmak3} and \eqref{eq:Sigmak4} for each dynamical system replacing the theoretical covariances $\mathbf{\Sigma}^{(s)}_{[\ell]}({ t})$ with estimated value $ \widehat{\mathbf{\Sigma}}^{(s)}_{[\ell]}(t)$. (see the methodology summarized in Figure \ref{fig:Flusso})

Leveraging on estimation of VAR processes \cite{Yonina-1} and on the ergodicity of the dynamical systems it can be shown that with probability at least $1-\delta$
$$
\|\mathbf{W}^{(s)}-\widehat{\mathbf{W}}^{(s)}(t)\|_F\leq \frac{C(n,\|\mathbf{Q}_{\eta}\|)}{(1-\sigma_{\max})^4\sqrt{t}\rho}
$$
where $C(n,\|\mathbf{Q}_{\eta}\|)$ is a constant independent of $t$.
This bound can be improved by imposing new constraints in the recovery by exploiting correlations among different dynamical systems (see models $\mathcal{M}_cc$ and $\mathcal{M}_{cs}$ in the box~\boxref{F\&J model on multiplex networks}).
If the correlations are not known among influence matrices the idea proposed in \cite{8796302} is to leverage on global properties of the local processes to correct the local estimates of $\mathbf{S}^{(s)}_{[0]}(\infty)$. Moreover, the reconstruction performance suffer in case the sample size is not large, the number of observed data must be larger than the number of unknowns in order to have a full rank estimation of $\hat{\mathbf{\Sigma}}_{[0]}^{( s)}(\infty)$.
The Bayesian approach is a powerful estimation framework since it combines prior probabilistic information, parametrized by some unknown hyperparameters, and gathered observations.

\begin{tcolorbox}[title= \sf \textbf{\color{black}Bayesian estimation of $\mathbf{S}_{[0]}^{( s)}(\infty)$}, colframe=carmine!10,
colback=carmine!10,
coltitle=black,
]\sf \small
In the absence of additional information on the model, the selection of the prior distribution is quite delicate. A commonly used approach is to consider the conjugate prior of the multivariate normal distribution.
More precisely, we consider the inverse-Wishart with matrix $\mathbf{\Psi}$ and  $\nu > n+1$ degrees, i.e. 
or, equivalently,
$$
\hat{\mathbf{S}}_{[0]}^{( s)}(\infty) = \gamma^{{ (s)}} \bar{\mathbf{S}} + (1-\gamma^{{ (s)}}) \hat{\mathbf{S}}_{\text{\tiny SCM}}^{( s)}(\infty)
$$
where\smallskip
\begin{itemize}
\item $\hat{\mathbf{S}}_{\text{\tiny SCM}}^{( s)}(\infty)  $ is the sample covariance matrix\medskip
\item $\bar{\mathbf{S}} = \frac{\mathbf{\Psi}}{\nu-(n+1)}$ is the prior mean/mode\medskip
\item $ \gamma^{{ (s)}} =  \frac{\nu  - (n+1)}{\nu + T^{{ (s)}} - (n+1)} \in(0,1)$\medskip
is a term balancing the two contributions according to the sample size $T^{{ (s)}}$ and informative level of the prior (degrees of freedom $\nu$).
\end{itemize}
\bigskip

Then the Inverse-Wishart {parameters estimation} are obtained via {\textbf{alternating minimization}}
\begin{align*}
&(\hat{\mathbf{\Psi}},\hat{\nu}) \! \\
&= \!\!\!\!\argmin{\mathbf{\Psi}>0, \,\nu>n+1} \!\! -\sum_{s=1}^m \log \frac{\det^{\frac{\nu}{2}} (\Psi) \, \Gamma_n\!\! \left( \frac{\nu+T^{{  (s)}}+n}{2}\right)}{\pi^{\frac{nT^{{  (s)}}}{2}} \det^{\frac{\nu+T^{{  (s)}}}{2}} (\Psi+\mathbf{Z}^{(s)} ({\mathbf{Z}^{(s)}})^\top) }
\end{align*}

\end{tcolorbox}


In\cite{8796302}, the performance of the proposed estimators are tested within $\mathcal{M}_{cc}$ and $\mathcal{M}_{cs}$.
The simulations show that the approach based on the  Bayesian method achieves better performance in the estimation.  In particular, for both the considered models the variance of the reconstruction error is much lower for the proposed approach compared to the conventional ML estimator.
It is worth remarking that the recovery of the transition matrices depends significantly on the conditioning of the estimated covariance matrices. Although the matrices are invertible, the reconstruction performance suffer in case the sample size is not large.
In this sense, the Bayesian method acts as a regularizer of the covariance estimation in an adaptive fashion, i.e., with automatic selection of the regularization parameter.
In fact, by putting a prior distribution on the covariance matrix,  the reconstruction formula will be a combination of a sample statistic (computed from the observed data) and a function of the hyperparameters (prior information). The latter can indeed help in case of scarce data, while its effect vanishes asymptotically as posterior estimates converge to the ML counterparts for large samples (Bernstein-von Mises theorem), thus converging to the classical SCM. This is a ``natural weighting'' mechanism, which automatically regulates (through the parameters $\gamma^{(s)}$) the relative importance of prior model and data according to the sample size, automatically switching to a non-informative prior (retrieved for limit values of the hyperparameters) if conversely the sample size is large.

%



\section{Concluding remarks}

Although the phenomenon of social influence has been long studied in social and behavioral sciences,
mathematical characterization of influence between individuals is not a trivial task.
How to understand which connections between people are most essential and
who are the genuine leaders of the group?

Granovetter~\cite{Granovetter73,Granovetter83} proposed the theory of ``strong'' and ``weak'' ties connecting, respectively,
close friends and acquaintances. Strong ties build densely connected subgraphs (communities) in a network, whereas weak ties build bridges between these densely knit communities. This principle has led to a number of mathematical characteristics~\cite{Sun2011} measuring influence between two individuals as a function of their positions in a network. At the same time, Granovetter argued that some ``weak'' ties not only have a strong impact on an individual but, in fact \emph{are actually vital for an individual's integration into modern society}~\cite{Granovetter83}. ``Weak'' ties facilitate exchange of information between closed communities, enabling, in particular, the mobility of labour and integration of individuals into political movements. Hence, ``static'' characteristics considering only links between individuals and ignoring the specific features of their interactions can be misleading.
Alternative methods are needed that consider a social network as a dynamical system.

In this survey, we focus on two novel directions of research, concerned with dynamical networks of social influence. The statistical approach adopted in machine learning considers a social network as a probabilistic graphical model and treating social influence as a measure of statistical correlation between some data produced by individuals (e.g. information which events they attend and which goods they consume). The approach of social influence network theory~\cite{Friedkin:Johnsen:2011} considers social influence as a process altering opinions of the individuals; to find the parameters of these models, methods of identification theory should be used. Even for a parsimonious opinion formation model, proposed by Friedkin and Johnson, the problem of parameter identification appears to be non-trivial and is closely related to compressed sensing and other rapidly growing branches of signal processing theory.

Many problems related to recovery of influence networks' structure remain beyond the scope of this survey and are still waiting for solutions. Identification problems become quite challenging when a dynamical model nonlinearly depends on unknown parameters as e.g. bounded confidence models surveyed in~\cite{ProTempo:2018}. Along with continuous (real-valued) measurements, models can deal with discrete (finite valued) data as e.g. cellular automata considered in physical literature~\cite{Castellano:2009} or CODA (continuous opinion - discrete action) model~\cite{MARTINS:2014}. Even more complicated for analysis is the case of \emph{temporal} social network where both nodes and arcs can emerge and disappear. Such models are vital to understand online social networks dynamics where individuals can easily create and delete user profiles. System theory lacks tools to cope with such temporal models, a promising framework of \emph{open} multi-agent systems has recently been proposed in~\cite{Hendrickx:2017,FranceschelliFrasca:2018}.

Perhaps, the most challenging problem lying at the frontier between computer science, social sciences and systems theory is to extract the structure of an online social network from big data produced by users. Unlike simplified mathematical models, in real life people do not broadcast numbers and communicate via webforums, microblogs, mobile apps and other social media. The numbers thus have to be extracted from textual and multimedia information, which requires advanced tools for video and language processing, big data analytics and efficient numerical methods that are able to deal with large-scale dynamical systems. We hope that this survey will help to recruit young talented researchers to the vibrant and fascinating area of dynamical social network analysis.







\end{multicols}

\bibliographystyle{IEEEtran}

\end{document}